# Bipartite Fluctuations as a Probe of Many-Body Entanglement

H. Francis Song,[1] Stephan Rachel,[1] Christian Flindt,[2] Israel Klich,[3] Nicolas Laflorencie,[4] and Karyn Le Hur[1]

[1]*Department of Physics, Yale University, New Haven, CT 06520*
[2]*Département de Physique Théorique, Université de Genève, CH-1211 Genève, Switzerland*
[3]*Department of Physics, University of Virginia, Charlottesville, VA 22904*
[4]*Laboratoire de Physique Théorique, Université de Toulouse, UPS, (IRSAMC), Toulouse, France*
(Dated: September 4, 2011)

We investigate in detail the behavior of the bipartite fluctuations of particle number $\hat{N}$ and spin $\hat{S}^z$ in many-body quantum systems, focusing on systems where such U(1) charges are both conserved and fluctuate within subsystems due to exchange of charges between subsystems. We propose that the bipartite fluctuations are an effective tool for studying many-body physics, particularly its entanglement properties, in the same way that noise and Full Counting Statistics have been used in mesoscopic transport and cold atomic gases. For systems that can be mapped to a problem of non-interacting fermions we show that the fluctuations and higher-order cumulants fully encode the information needed to determine the entanglement entropy as well as the full entanglement spectrum through the Rényi entropies. In this connection we derive a simple formula that explicitly relates the eigenvalues of the reduced density matrix to the Rényi entropies of integer order for any finite density matrix. In other systems, particularly in one dimension, the fluctuations are in many ways similar but not equivalent to the entanglement entropy. Fluctuations are tractable analytically, computable numerically in both density matrix renormalization group and quantum Monte Carlo calculations, and in principle accessible in condensed matter and cold atom experiments. In the context of quantum point contacts, measurement of the second charge cumulant showing a logarithmic dependence on time would constitute a strong indication of many-body entanglement.

## I. INTRODUCTION

The study of quantum many-body systems has traditionally involved the analysis of ground-state and excited-state energies, correlation functions, and symmetry-breaking order parameters to characterize different states of matter [1, 2]. Recently, however, there has been great interest in using a different feature to understand the properties of a quantum many-body system, namely entanglement. Entanglement is believed to be particularly relevant at zero temperature, especially for topologically ordered states for which conventional order parameters are not sufficient to fully define the state of the system [3–5].

The degree of entanglement in a system is often quantified using the bipartite entanglement entropy between a subsystem and the remainder of the system, most commonly using the scaling of the entanglement entropy with subsystem size. Specifically, given a wave function $|\Psi\rangle$—usually the ground state but often a time-evolved wave function—the entanglement entropy is defined as the von Neuman entropy of the reduced density matrix $\hat{\rho}_A = \text{Tr}_B |\Psi\rangle\langle\Psi|$ of subsystem $A$ obtained by tracing out the degrees of freedom in the remainder of the system $B$. Thus

$$\mathcal{S}(\hat{\rho}_A) = -\text{Tr}(\hat{\rho}_A \ln \hat{\rho}_A). \quad (1.1)$$

More generally, the $\alpha$-Rényi entropies are defined as

$$\mathcal{S}_\alpha(\hat{\rho}_A) = \frac{1}{1-\alpha} \ln[\text{Tr}(\hat{\rho}_A^\alpha)] \quad (1.2)$$

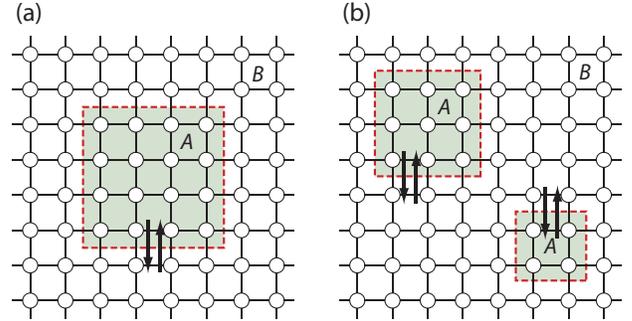

FIG. 1: (color online). Quantum many-body systems. We can divide the system into two parts $A$ and the remainder $B$, where $A$ can be (a) a single connected region or, more generally, (b) multiple disconnected regions. The result is the formation of a boundary, across which two subsystems of a local Hamiltonian interact, for example by exchanging particles (arrows).

and reduce to the von Neumann entropy in the limit $\alpha \to 1$. For a pure state $\mathcal{S}_\alpha(\hat{\rho}_A) = \mathcal{S}_\alpha(\hat{\rho}_B)$ and it makes no difference which subsystem is analyzed; indeed, this symmetry is an important feature of entanglement entropy. The scaling of entanglement entropy with subsystem size has yielded a wealth of interesting results for both gapped and critical systems [3, 6–19]. Alternatively, detailed information about the structure of entanglement can be obtained by studying the full spectrum of eigenvalues of the reduced density matrix, specifically through the eigenvalues of the "entanglement Hamiltonian" $\hat{H}_A$ defined by $\hat{\rho}_A = e^{-\hat{H}_A}/\text{Tr}(e^{-\hat{H}_A})$ [20–32].

A distinguishing feature of using entanglement entropy to study many-body systems is the introduction of a boundary between the two subsystems $A$ and $B$ as in Fig. 1, which represents a change in perspective from the study of point-point correlators and local order parameters. The special role played by the area of this boundary has long been recognized, and indeed was an early motivation for the study of entanglement entropy [33, 34], because most Hamiltonians of interest contain local interactions which in turn lead to correlations localized along the entire boundary. In many situations this is formalized in *area laws* for the entanglement entropy [35], while important violations have been studied since then [8, 12]. The important role played by the boundary, moreover, is not unique to entanglement entropy—all proposed measures of many-body entanglement are based on it. In systems where particles are exchanged between parts of the system, for example, all particle exchange occurs through the boundary and therefore one expects, and can confirm, that particle number fluctuations will similarly be sensitive to the area. Indeed, as we have shown previously [36] and further elaborate on in this work, for non-interacting systems of fermions in which entanglement originates precisely from the uncertainty of the position of particles, particle number fluctuations are equivalent to the entanglement entropy.

Fundamentally, entanglement is what distinguishes quantum systems from classical ones, and the belief that studying the entanglement properties of a quantum system, whether by analyzing the scaling of the entanglement entropy with subsystem size or by investigating the more detailed entanglement spectrum, has led to many striking and beautiful insights. In particular, the recent interest in many-body entanglement was in large part motivated by the work in Refs. [33, 34] showing that the entanglement entropy of a subsystem in a quantum field theory is generally expected to scale as the area of the subsystem, and the work of Refs. [7, 37] showing that in conformally invariant one-dimensional gapless systems the entanglement entropy of an interacting many-body system exhibits universal logarithmic scaling with subsystem size, with the prefactor controlled by the central charge of the underlying conformal field theory (CFT). These results have been confirmed by many numerical, and sometimes analytical, calculations for microscopic Hamiltonians, and indeed the question of which problems are tractable by classical simulation methods has itself been an important part of the investigation of many-body entanglement [38].

Except for systems that can be mapped to non-interacting fermions [36] as suggested in Refs. [39, 40] and discussed extensively in the present work, however, it has not been clear how entanglement entropy can be measured experimentally (similar ideas are discussed in Ref. [41]; a method for determining the lower bound is proposed in Ref. [35]; see Ref. [42] for a completely different, promising approach). Even theoretically, entanglement entropy is difficult to calculate analytically and in dimensions greater than one there are no good numerical methods for obtaining this quantity due to the fact that the entanglement entropy is a non-trivial function of the reduced density matrix and hence of the full ground-state wave function (however, some significant progress has been made recently for the Rényi entropies in quantum Monte Carlo (QMC) [43–46]). This has prompted a search for other measures of entanglement that behave similarly to the entanglement entropy yet are easier to determine, primarily in QMC calculations as they can be applied to a wide variety of systems in any dimension.

In this work we show that using the same setup shown in Fig. 1 but measuring instead the fluctuations of a conserved U(1) charge such as particle number $\hat{N}$ and spin $\hat{S}^z$ reveals important features of quantum many-body systems including their entanglement properties, quite similarly to the way in which Full Counting Statistics (FCS), the study of charge transfer across mesoscopic conductors, has been intensely analyzed in mesoscopic transport [47–50] and in cold atom systems [51, 52]. Indeed, our result for non-interacting fermions unequivocally demonstrates the importance of studying the full set of cumulants, beyond the fluctuations (noise) encoded in the second cumulant. The present work, moreover, may be understood as applying some of the same principles underlying the field of FCS to the study of the ground states of many-body Hamiltonians, with the boundary introduced by the bipartition playing the role of the scattering or interaction region (e.g., quantum dots) in mesoscopic systems.

To be precise, suppose the total particle number is conserved and that the pure state $|\Psi\rangle$ is an eigenstate of the particle number operator $\hat{N}_\text{tot}$, $\hat{N}_\text{tot}|\Psi\rangle = N_\text{tot}|\Psi\rangle$ where $N_\text{tot}$ (without the caret) is a positive, real integer. Since we will almost always be concerned with pure states at zero temperature in this work, we only consider pure states here. We then define the fluctuations as [53]

$$\mathcal{F} = \langle(\hat{N}_A - \langle\hat{N}_A\rangle)^2\rangle \quad (1.3)$$

where the expectation value is taken with respect to $|\Psi\rangle$. We can generalize the fluctuations to the cumulants defined by

$$C_n = (-i\partial_\lambda)^n \ln\chi(\lambda)|_{\lambda=0}, \quad (1.4)$$

where for the ground state of a many-body Hamiltonian we might define

$$\chi(\lambda) = \langle\exp(i\lambda\hat{N}_A)\rangle \quad (1.5)$$

as the generating function for particle number fluctuations within subsystem $A$, with $\mathcal{F} = C_2$. In the context of two-terminal mesoscopic transport the time-dependent generating function is defined as [47]

$$\chi(\lambda, t) = \sum_m P_m(t) e^{i\lambda m}, \quad (1.6)$$

where $P_m(t)$ is the probability that $m$ charges are transferred from the source lead to the drain lead. As we will see below, in the framework of Fig. 1 the *even* cumulants are the appropriate generalization of the fluctuations because of their invariance under exchange of subsystems at zero temperature. We note that the fluctuations (and cumulants) can always be expressed as a sum over the subsystem of the appropriate correlation functions of density operators [cf. Eq. (3.23)], which shows how the fluctuations may be considered as the quantity that bridges local correlation functions with the highly non-local and non-linear entanglement entropy. As such the fluctuations are expected to share some, but not all, essential features with the entanglement entropy.

The fluctuations at zero temperature (and more generally all even cumulants) share with the entanglement entropy two important properties:

1. The fluctuations (and all cumulants) are zero for a product state $|\Psi\rangle = |\Psi_A\rangle \otimes |\Psi_B\rangle$. For the product state, $|\Psi_A\rangle$ and $|\Psi_B\rangle$ are separately eigenstates of $\hat{N}_A$ and $\hat{N}_B$, respectively, which can be seen by noting that if either $|\Psi_A\rangle$ or $|\Psi_B\rangle$ is a superposition of states with different numbers of particles then so must the total state $|\Psi\rangle$ since $\hat{N}$ is additive. Thus there are no particle number fluctuations in $A$ or $B$.

   Unlike the entanglement entropy, the converse of this statement is not always true: there exist states with zero fluctuations which are *not* product states. This can occur because one can construct a Hamiltonian where $\hat{N}_A$ and $\hat{N}_B$ are separately conserved from the start. Then all eigenstates of the Hamiltonian will have zero fluctuations even though the entanglement entropy is nonzero. We note that if the fluctuations, i.e., the second cumulant, are nonzero, then all cumulants are zero and implies that $\hat{N}_A = \langle\hat{N}_A\rangle$ and $\hat{N}_B = \langle\hat{N}_B\rangle$. However, unlike total particle number conservation, particle number conservation within subsystems is rare (essentially, only density-density interactions are allowed between $A$ and $B$ as in Ref. [54]) and we can generally expect the converse of this statement to hold in realistic systems.

2. At zero temperature the fluctuations are *symmetric* between $A$ and the remainder of the system $B$. Since $\hat{N}_A + \hat{N}_B = N_{\text{tot}}$ we have $\hat{N}_A - \langle\hat{N}_A\rangle = -(\hat{N}_B - \langle\hat{N}_B\rangle)$ and therefore

$$\mathcal{F}_A = \mathcal{F}_B. \quad (1.7)$$

   More generally, switching $A$ and $B$ is equivalent to changing $\lambda \to -\lambda$ in the generating function (1.5). Thus the even cumulants are invariant under this switch and $C_n$ for even $n$ are symmetric in $A$ and $B$ as well.

   This property is crucial to understanding why the fluctuations are generally expected to behave similarly to the entanglement entropy, in the following sense. In some of the earliest works on the scaling of the entanglement entropy with the area of the subsystem, it was suggested [34] that this could be understood from the fact that the entanglement entropy of a pure state is symmetric between the subsystem and the remainder of the system, which implies that the relevant correlations lie at the boundary shared by the two systems and not in the "volume" of the respective subsystems. We now know from many violations of the area law that symmetry is not sufficient to guarantee an area law, but symmetry is a property shared by all proposed entanglement measures.

The fluctuations $\mathcal{F} = C_2$, in addition, share with the entanglement entropy the property of being subadditive, in the sense that (at zero temperature)

$$\mathcal{F}_A + \mathcal{F}_B \geq 0, \quad (1.8)$$

which must be true because $\mathcal{F}_A = \mathcal{F}_B \geq 0$. The higher-order cumulants and Rényi entropies for $\alpha > 1$ are not subadditive.

Thus with the exception noted under Property 1 the fluctuations can, in a broad sense, be considered an entanglement measure similar to previously proposed measures (disussed below), while the question of how *similarly* the fluctuations behave to the entanglement entropy itself is a separate and interesting question addressed in the following sections. It is, however, important to note that the fluctuations are not an entanglement measure in the strict quantum information sense, which is a non-negative function of a state which cannot increase under local operations and classical communications and which vanishes for separable states. On the other hand, given the prior knowledge that the ground state is an eigenstate of the particle number operator, for example, the fluctuations contain (perhaps incomplete) information about entanglement in the system.

The advantage of the fluctuations, meanwhile, is that they are much more accessible—analytically, numerically, and experimentally. Computing the fluctuations involves trivial adjustments to existing density matrix renormalization group (DMRG) [38, 55] and QMC codes [in DMRG the reduced density matrix is almost always stored in block-diagonal form according to the U(1) quantum number, so that the sum of the eigenvalues within each block can be used to easily determine cumulants of any order]. There are, moreover, existing experimental techniques that can in principle be used to measure the fluctuations in real systems [56–60]. This is the main motivation of the present work: to explore the behavior of the fluctuations in a wide variety of systems in anticipation of such experimental measurements.

The outline of this paper is as follows. In Sec. II we show that for systems that can be mapped to a problem of non-interacting fermions the full set of even cumulants is completely equivalent to the entanglement entropy by showing how the von Neumann entanglement entropy [36], and more generally the Rényi entanglement entropies, can be written exactly in terms of the even cumulants. In particular, this usually implies that the dominant scaling of any given cumulant with subsystem size is identical to that of the entanglement entropy. We present applications of the formulas, as well as a (completely general) method for obtaining the full entanglement spectrum from the Rényi entropies, or in the case of non-interacting fermions, from the cumulants. In Sec. III we explore the generalization for one-dimensional interacting systems as suggested in Ref. [53], including the case of disjoint intervals (Fig. 1b). Although the exact relation found in systems of non-interacting fermions no longer holds we show how many similarities exist between fluctuations and the entanglement entropy. We then establish that for gapped quantum many-body systems an area law holds for the fluctuations. Finally, in Sec. IV we compare bipartite fluctuations to previously proposed measures of many-body entanglement and show that fluctuations possess almost all of the same properties while being much easier to access both theoretically and experimentally. We conclude by suggesting possible experimental realizations and future work.

Appendices provide essential but somewhat technical calculations, while the Supplementary Material contains longer, more detailed computations for several of the results presented in the main text.

Unless stated otherwise, we work in units where the Planck constant $\hbar$ and Boltzmann constant $k_B$ are $\hbar = k_B = 1$. For lattice systems we assume that the lattice spacing $a = 1$, while for continuum systems we will explicitly include the short-distance cutoff.

## II. EXACT RELATIONS BETWEEN CHARGE STATISTICS AND ENTANGLEMENT ENTROPIES

In this section we show how, for non-interacting fermions (and systems that can be mapped to a problem of non-interacting fermions), two important measures of many-body entanglement can be expressed exactly in terms of the statistics of charge fluctuations. Specifically, we derive series for the von Neumann and the Rényi entanglement entropies in terms of the full set of (even) cumulants of the charge fluctuations. Similar ideas have previously been suggested in Ref. [39] in the context of ground-state entanglement entropy and in Ref. [40] in relation to electron transport through a quantum point contact (QPC). Here, we extend the results of Ref. [36] to the computation of the generalized Rényi entropies and demonstrate that such relations in fact hold in general for non-interacting fermionic systems in any dimension. Importantly, the intimate connections between the entanglement entropies and the charge fluctuations facilitate a novel approach to the experimental determination of many-body entanglement entropy in non-interacting systems where charge fluctuations may be measured.

The restriction to non-interacting fermions is not as limiting as it first appears; aside from the intrinsic importance of free fermions in various dimensions, in one dimension the Jordan-Wigner transformation maps the spin-1/2 XY model to a model of non-interacting fermions, and a similar mapping can be carried out for bosons with infinite on-site repulsion (hard-core bosons) to a model of free fermions. In two dimensions the integer quantum Hall effect is described by a model of non-interacting fermions, as are the related topological insulators. The case of hard-core bosons in one dimension seems particularly promising for the experimental detection of many-body entanglement, since the number of particles in a given region is already accessible in present-day experiments with cold atoms in optical lattices [61–63].

Before going through the detailed derivations, we first present the central results of this section. As we demonstrate below, the von Neumann entanglement entropy $\mathcal{S}$ can be expressed in terms of the charge statistics as [36]

$$\mathcal{S} = \lim_{K \to \infty} \sum_{n=1}^{K+1} \alpha_n(K) C_n, \qquad (2.1)$$

where $C_n$ are the cumulants defined in Eqs. (1.4),(1.5). The cutoff-dependent coefficients $\alpha_n(K)$ are given by

$$\alpha_n(K) = \begin{cases} 2 \sum_{k=n-1}^{K} \frac{S_1(k, n-1)}{k! k} & \text{for } n \text{ even}, \\ 0 & \text{for } n \text{ odd}, \end{cases} \qquad (2.2)$$

where $S_1(n, m)$ are the unsigned Stirling numbers of the first kind.

A similar, but lengthier, relation also exists for the Rényi entanglement entropies of integer order,

$$\mathcal{S}_n = \lim_{R \to \infty} \sum_{k=1}^{nR} \beta_k(n, R) C_k. \qquad (2.3)$$

Here, the coefficients depend on the cutoff $R$ as

$$\beta_k(n, R) = \begin{cases} \frac{1}{1-n} \sum_{r=1}^{R} \sum_{m=0}^{r} \sum_{s=k}^{nr} (-1)^{r+s+nr+nm} \\ \quad \times \frac{1}{r} \binom{R}{r} \binom{r}{m} \binom{nm}{nr-s} \frac{S_1(s,k)}{(s-1)!} & \text{for } k \text{ even}, \\ 0 & \text{for } k \text{ odd}. \end{cases} \qquad (2.4)$$

Interestingly, only even-order cumulants contribute in the series (2.1) and (2.3). Physically, this is a consequence of the entanglement entropies being symmetric between the subsystem and its complement for pure states. One may for example think of a QPC at zero temperature, Eq. (1.6), where $P_m$ describes the probability of transmitting $m$ charges from one subsystem, the left lead, to its complement, the right lead: charge conservation implies that the number of electrons $m$ collected in the right lead is equal to $-m$ charges collected in the left. Only even-order cumulants are then symmetric in the two leads, and the entanglement entropies for the QPC can therefore only depend on the even cumulants.

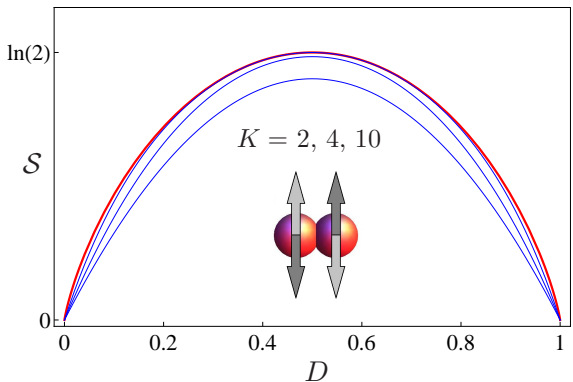

FIG. 2: (color online). Entanglement entropy of a pair of spins. The two spins are in the state $|D\rangle$ in Eq. (2.5), which is a product state for $D = 0, 1$ and maximally entangled for $D = 1/2$. The red line shows the exact result given by Eq. (2.6), while the blue lines show the series (2.1) with increasing cutoff number $K$ from bottom to top. For $K = 10$, the series has basically converged to the exact result.

Since the series (2.1) and (2.3) are universally valid for *any* non-interacting system, the entanglement entropies can only depend on the even-order cumulants, including for systems where the even cumulants are not necessarily symmetric under interchange of the two subsystems. We consider an example of such a system in Sec. II C 2.

The two series above yield a sequence of approximations to the entanglement entropies that is an increasingly better lower bound to the exact results: if the series (2.1) for the von Neumann entanglement entropy is truncated at a finite $K$, the approximation only becomes better as $K$ is increased. Thus, with only a finite number of cumulants, the series provides a converging approximation of the exact von Neumann entropy. This is also true for the Rényi entropies. For infinite cutoff $K$, our series (2.1) is moreover *formally* equivalent to the series suggested in Ref. [40], which can be written as $S = \sum_{n=1}^{\infty} \alpha_n C_n$, where $\alpha_n = 2\zeta(n)$ for $n$ even with $\zeta(n)$ being the Riemann zeta function, and zero for $n$ odd. This is shown explicitly in the Supplementary Material. As we show below, however, the convergence properties of the series (2.1) and the one presented in Ref. [40] are very different.

As a simple illustration of how the series can be applied, consider two electron spins in the state

$$|D\rangle_{AB} = \sqrt{1-D}|\uparrow\rangle_A|\downarrow\rangle_B - \sqrt{D}|\downarrow\rangle_A|\uparrow\rangle_B \quad (2.5)$$

with $0 \leq D \leq 1$. Clearly, for $D = 1/2$ the spins are in the maximally entangled singlet state, while for $D = 0, 1$ the spin state is factorized. It is easy to calculate the entanglement entropy of spin $A$ with respect to spin $B$ (or vice versa) directly from the definition (1.1),

$$\mathcal{S} = -[D \ln D + (1-D)\ln(1-D)], \quad (2.6)$$

which is zero for $D = 0, 1$ and the maximal $\ln 2$ for $D = 1/2$. The entanglement entropy can also be obtained from the fluctuations of the spins using the series (2.1). For $D = 0, 1$ the two spins are clearly in a well-defined state and no fluctuations occur. In contrast, in the spin singlet state $D = 1/2$ the two spin states of spin $A$ are equally probable and the $z$-component of each spin fluctuates. We can think of the two spins as sharing a spin excitation: for the state $|\uparrow\rangle_A|\downarrow\rangle_B$ spin $A$ is excited, while for the state $|\downarrow\rangle_A|\uparrow\rangle_B$ the excitation has been transferred to spin $B$.

Now, the distribution $P(m)$ for the number $m$ of transferred excitations between spin $A$ and spin $B$ is

$$P(m) = \begin{cases} 1-D & \text{for } m=0, \\ D & \text{for } m=1, \\ 0 & \text{otherwise.} \end{cases} \quad (2.7)$$

The corresponding generating function is

$$\chi(\lambda) = \sum_m P(m)e^{i\lambda m} = 1 + (e^{i\lambda} - 1)D, \quad (2.8)$$

from which the cumulants of $m$ easily follow as $C_n = (-i\partial_\lambda)^n \ln \chi(\lambda)|_{\lambda=0}$. In Fig. 2 we show the entanglement entropy calculated from the cumulants using the series (2.1) with increasing cutoff number $K$ together with the exact result, Eq. (2.6). Already with $K = 2$, where only the second cumulant is included, the series yields a good approximation of the exact entanglement entropy, while basically full convergence is obtained for $K = 10$. We note that the convergence is not uniform over all values of $D$. This simple example illustrates how the entanglement entropy can be obtained from the (correlated) fluctuations of the spins.

It is instructive to attempt the same calculation using the series presented in Ref. [40]. The terms of this series are $2\zeta(n)C_n$, where $2\zeta(n) \sim 2$ for large $n$. However, the cumulants corresponding to Eq. (2.8) go as $|C_n| \sim (n-1)!/\pi^n$ for large $n$, so that the series of Ref. [40] diverges under normal summation methods. Since the factorial divergence of the $|C_n|$ is typical for non-gaussian fluctuations [60], moreover, the problem is quite generic. As explained in the Supplementary Material, this is due to the existence of singularities in the complex plane possessed by almost all cumulant generating functions and hence the problematic step of integrating its series expansion term by term. In particular, for any experimental situation where only a finite number of cumulants is available, the series from Ref. [40] is ill-suited for the determination of the von Neumann entanglement entropy. Instead the series (2.1) should be used as it converges for any distribution of the fluctuations.

## A. Derivation

We now present the detailed derivation of the series (2.1) for the von Neumann entanglement entropy.

The series (2.3) for the Rényi entropies follows from a similar but lengthier derivation, an outline of which is given in Appendix A. In Sec. II C we illustrate the use of the series with several different applications, and the reader may at this point wish to proceed directly to the concrete examples which can be understood without going through the derivation below.

The derivation takes as starting point the expressions for the entanglement entropy

$$\mathcal{S} = -\text{Tr}\{M \ln M + (1-M)\ln(1-M)\} \quad (2.9)$$

and the cumulant generating function

$$\ln \chi(\lambda) = \ln \det\{[1 + (e^{i\lambda} - 1)M]e^{-i\lambda Q}\} \quad (2.10)$$
$$= \text{Tr} \ln\{[1 + (e^{i\lambda} - 1)M]e^{-i\lambda Q}\} \quad (2.11)$$

in terms of the correlation matrix $M$ and a phase factor $e^{-i\lambda Q}$ which eventually drops out of the problem. For lattice models involving non-interacting fermions, the entries of $M$ are the Green's functions $M_{ij} = \langle \hat{a}_j^\dagger \hat{a}_i \rangle$ of the fermionic degrees of freedom [64] as described in the Supplementary Material, while for a generic two-terminal scattering situation $M$ is determined by the scattering matrix of the problem [65, 66]. Indeed, Eq. (2.11) is the famous Levitov-Lesovik determinant formula [47]. In both cases $M$ has real eigenvalues in the range 0 to 1 due to fermionic statistics. The similarity between Eq. (2.8) and Eq. (2.11) highlights the fundamental result that in a two-terminal (bipartite) setup charge transfer can be understood in terms of simple elementary processes [65, 66].

To relate the von Neumann entropy to the cumulants of charge fluctuations, we first expand the logarithms of Eq. (2.9) around $M = 0, 1$ to obtain

$$\mathcal{S} = \sum_{k=1}^{\infty} \frac{A_k}{k} \quad (2.12)$$

with coefficients

$$A_k = \text{Tr}[M(1-M)^k + M^k(1-M)]. \quad (2.13)$$

In the next step we relate $\text{Tr}(M^k)$ to the charge statistics described by Eq. (2.11). To this end the *factorial cumulants* (see Supplementary Material)

$$F_k = \partial_\lambda^k \ln \chi_f(\lambda)|_{\lambda=0} \quad (2.14)$$

turn out to be particularly useful, where

$$\ln \chi_f(\lambda) = \ln \chi(-i \ln(\lambda + 1)) \quad (2.15)$$

is the factorial cumulant generating function. The factorial cumulants corresponding to the generating function (2.11) are given by [50]

$$F_k = (-1)^{k-1}(k-1)![\text{Tr}(M^k) - Q]. \quad (2.16)$$

Solving for $\text{Tr}(M^k)$ and substituting into Eq. (2.13) after having used the binomial theorem gives

$$A_k = (-1)^{k-1}\left[\frac{F_k}{(k-1)!} + \frac{F_{k+1}}{k!}\right] + \sum_{r=0}^{k} \binom{k}{r}\frac{F_{r+1}}{r!}, \quad (2.17)$$

or, from Eq. (2.12),

$$\mathcal{S} = \sum_{k=1}^{\infty} \left\{ \frac{(-1)^{k-1}}{k}\left[\frac{F_k}{(k-1)!} + \frac{F_{k+1}}{k!}\right] + \sum_{r=0}^{k}\binom{k}{r}\frac{F_{r+1}}{r!k} \right\}. \quad (2.18)$$

The phase factor $Q$ does not contribute to the sum.

Eq. (2.18) expresses the von Neumann entropy purely in terms of factorial cumulants, which in principle are measurable quantities. It is nevertheless convenient, and somewhat more elegant, to express $\mathcal{S}$ directly in terms of ordinary cumulants. This is simply a matter of algebra, but we present here some important intermediate steps that would allow the interested reader to reproduce the derivation.

The factorial cumulants can be expressed as a linear combination of the ordinary cumulants as

$$F_{k \geq 1} = \sum_{n=1}^{k}(-1)^{k-n}S_1(k,n)C_n, \quad (2.19)$$

where $S_1(k,n)$ are the unsigned Stirling numbers of the first kind. We can therefore write Eq. (2.17) as

$$A_k = \sum_{n=1}^{k+1}(-1)^{n-1}\left[\frac{S_1(k,n)}{(k-1)!} - \frac{S_1(k+1,n)}{k!}\right]C_n$$
$$- \sum_{r=0}^{k}\sum_{n=1}^{r+1}(-1)^{r-n}\binom{k}{r}\frac{S_1(r+1,n)}{r!}C_n. \quad (2.20)$$

We note that $S_1(k, k+1) = 0$. Using the recursion relation [67] $S_1(k+1, n) = S_1(k, n-1) + kS_1(k, n)$ in the first term and switching the order of sums in the second, we can simplify $A_k$ to

$$A_k = \sum_{n=1}^{k+1}\left\{(-1)^n\left[\frac{S_1(k, n-1)}{k!}\right.\right.$$
$$\left.\left. - \sum_{r=n-1}^{k}(-1)^r\binom{k}{r}\frac{S_1(r+1,n)}{r!}\right]C_n\right\}. \quad (2.21)$$

Substituting into Eq. (2.12) gives (we can exclude $n = 1$ since it contributes zero to the sum)

$$\mathcal{S} = \sum_{k=1}^{\infty}\sum_{n=2}^{k+1}\left\{(-1)^n\left[\frac{S_1(k, n-1)}{k!k}\right.\right.$$
$$\left.\left. - \sum_{r=n-1}^{k}(-1)^r\binom{k}{r}\frac{S_1(r+1,n)}{r!k}\right]C_n\right\}, \quad (2.22)$$

which we can write as a series in terms of the $C_n$ by introducing a cutoff $K$ in the outer sum and switching the order of the sums to get

$$\mathcal{S} = \lim_{K \to \infty}\sum_{n=2}^{K+1}\alpha_n(K)C_n, \quad (2.23)$$

where the cutoff-dependent coefficients $\alpha_n(K)$ are

$$\alpha_n(K) = (-1)^n \sum_{k=n-1}^{K} \left\{ \frac{1}{k} \left[ \frac{S_1(k, n-1)}{k!} - \sum_{r=n-1}^{k} (-1)^r \binom{k}{r} \frac{S_1(r+1, n)}{r!} \right] \right\}. \quad (2.24)$$

It can be shown numerically for arbitrary integers $n, k \geq n-1$ that the second term in square brackets is simply

$$\sum_{r=n-1}^{k} (-1)^r \binom{k}{r} \frac{S_1(r+1, n)}{r!} = (-1)^{n-1} \frac{S_1(k, n-1)}{k!}, \quad (2.25)$$

so that we obtain Eq. (2.2). To the best of our knowledge Eq. (2.25) is a new identity for the Stirling numbers and a rigorous proof is desirable. As a check of the algebra, it is known that [68]

$$\sum_{k=n-1}^{\infty} \frac{S_1(k, n-1)}{k! k} = \zeta(n), \quad (2.26)$$

so that the coefficients (2.2) agree with the series proposed in Ref. [40] in the limit of infinite cutoff, $K \to \infty$. The similar series (2.3) for the Rényi is derived in Appendix A. The convergence of both the von Neumann entanglement entropy (2.1) and Rényi entanglement entropies (2.3) is proved in Appendix B.

### B. Rényi Entropies and the Entanglement Spectrum

As mentioned in the Introduction, many recent works [20, 21, 24, 25] have focused on the entanglement spectrum, or the full set of eigenvalues of the reduced density matrix. The Rényi entropies of integer order are known to be fully equivalent to the entanglement spectrum [69], which implies that, remarkably, for the non-interacting fermion systems considered here the charge statistics also encodes the entanglement spectrum.

In this connection we present a very simple method for computing the full entanglement spectrum of a (finite) reduced density matrix from the corresponding Rényi entropies of integer order. Specifically, for $D$ degrees of freedom in the reduced density matrix only the Rényi entropies of order $2, \ldots, D$ are required. Although our interest here is on non-interacting fermions, the method is in fact completely general and is applicable to *any* density matrix and hence of considerable interest to studies where the Rényi entropies but not the von Neumann entropy can be computed [43, 46]. The mathematical basis for this result, the Newton-Girard formulas [70], is long-established but to the best of our knowledge is not well-known in the physics literature. The method presented here should be compared with the standard method of analytically continuing the Rényi entropies $\mathcal{S}_\alpha$ to $\alpha \to 1$ (see below) and the method presented in Ref. [69].

Suppose we are given a $D \times D$ reduced density matrix $\rho$. The Rényi entropies are defined for general $\alpha > 0$ as

$$\mathcal{S}_\alpha = \frac{1}{1 - \alpha} \ln[\text{Tr}(\rho^\alpha)]. \quad (2.27)$$

It is usually noted that the von Neumann entropy can be obtained by taking the limit as $\alpha \to 1$,

$$\mathcal{S} = -\text{Tr}(\rho \ln \rho) = \lim_{\alpha \to 1} \mathcal{S}_\alpha. \quad (2.28)$$

For convenience let $R_\alpha$ be the trace of the $\alpha$-th power of the reduced density matrix,

$$R_\alpha = \text{Tr}(\rho^\alpha) = e^{(1-\alpha)\mathcal{S}_\alpha}. \quad (2.29)$$

Note that $R_1 = 1$ for a properly normalized density matrix.

Define the $D \times D$ matrix

$$E = \begin{pmatrix} 1 & 1 & 0 & \cdots & \\ R_2 & 1 & 2 & 0 & \cdots \\ \vdots & \ddots & \ddots & \ddots & 0 \\ R_{D-1} & R_{D-2} & \cdots & 1 & D-1 \\ R_D & R_{D-1} & \cdots & R_2 & 1 \end{pmatrix}, \quad (2.30)$$

i.e., a quasi-lower triangular matrix with $R_1 = 1$ on the main diagonal, $R_2$ on the sub-diagonal, $R_n$ on the $(n-1)$-th sub-diagonal, $1, 2, 3, \ldots, D-1$ on the super-diagonal, and zero everywhere else. For example, for $D = 5$ we have

$$E = \begin{pmatrix} 1 & 1 & 0 & 0 & 0 \\ R_2 & 1 & 2 & 0 & 0 \\ R_3 & R_2 & 1 & 3 & 0 \\ R_4 & R_3 & R_2 & 1 & 4 \\ R_5 & R_4 & R_3 & R_2 & 1 \end{pmatrix}. \quad (2.31)$$

Denote the matrix obtained by taking the first $n \times n$ submatrix of $E$ as $E_n$. Then we can form the polynomial

$$P(x) = \sum_{n=0}^{D} \frac{(-1)^n}{n!} (\det E_n) x^{D-n}, \quad (2.32)$$

with the understanding that $\det E_0 = 1$.

The polynomial $P(x)$ is actually the characteristic polynomial of the density matrix $\rho$: $P(x) = \det(xI - \rho)$. If the individual eigenvalues of the reduced density matrix are desired then we can simply find all the roots of $P(x)$. For example, in the simple case where the reduced density matrix represents a pure state the Rényi entropies would be zero and hence $R_2 = R_3 = \cdots = R_D = 1$. Then (written here for the case $D = 5$)

$$E = \begin{pmatrix} 1 & 1 & 0 & 0 & 0 \\ 1 & 1 & 2 & 0 & 0 \\ 1 & 1 & 1 & 3 & 0 \\ 1 & 1 & 1 & 1 & 4 \\ 1 & 1 & 1 & 1 & 1 \end{pmatrix}. \quad (2.33)$$

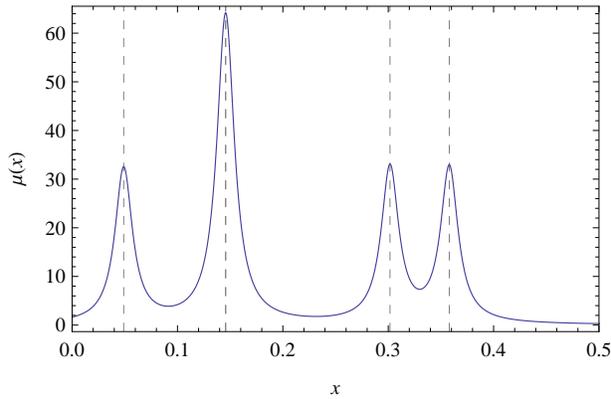

FIG. 3: (color online). Spectral density function corresponding to the spectrum in Eq. (2.37), using Eq. (2.36) with $\epsilon = 0.01$.

It can be shown that, aside from $\det E_0 = 1$, only $\det E_1 = 1$ is nonzero. The corresponding characteristic polynomial is therefore

$$P(x) = x^{D-1}(x-1), \qquad (2.34)$$

so that the eigenvalues are 1 (multiplicity 1) and 0 (multiplicity $D-1$) as expected. In contrast, for the fully mixed state where $\rho$ has $1/D$ on the diagonal and zero everywhere else, we have $R_n = D^{1-n}$ so that the characteristic polynomial is

$$P(x) = \left(x - \frac{1}{D}\right)^D \qquad (2.35)$$

and all of the eigenvalues are $1/D$ (multiplicity $D$). The essential point, of course, is that we can obtain this not from knowing the full density matrix but from only the Rényi entropies and Eq. (2.32).

For analytical purposes, it may be advantageous to write down the spectral density function

$$\mu(z) = \frac{1}{\pi} \lim_{\epsilon \to 0^+} \operatorname{Im} \partial_z \ln P(z - i\epsilon), \qquad (2.36)$$

where $\epsilon$ is a small positive number that can be taken arbitrarily close to zero for increased accuracy. As an example, suppose the eigenvalues of the reduced density matrix $\rho$ are

$$0.0489694,\ 0.145703,\ 0.145703,\ 0.301666,\ 0.357959 \qquad (2.37)$$

where we have included a pair of degenerate eigenvalues to illustrate the effects of degeneracy. Then by performing the above calculation for $P(x)$ and using the resulting polynomial in Eq. (2.36) with $\epsilon = 0.01$, the spectral density function appears as shown in Fig. 3.

For a long time it has been possible to determine the von Neumann entanglement entropy only with exact diagonalization (ED) for small systems and DMRG for one-dimensional and quasi-one-dimensional systems. For the

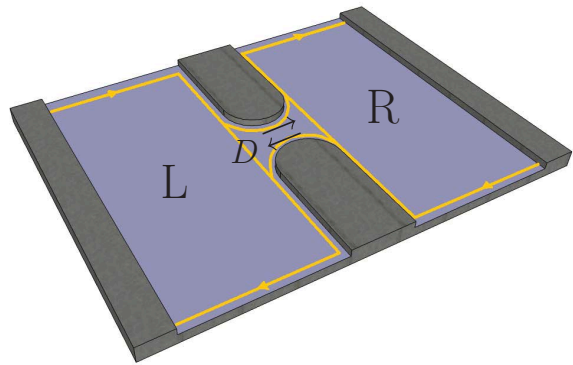

FIG. 4: (color online). Quantum point contact (QPC) with chiral edge states (yellow lines) running along the edges of a sample containing a two-dimensional electron gas in a strong perpendicular magnetic field. The sample is divided into a left (L) and a right (R) region by a split gate acting as a QPC. Electrons in the two in-coming edge states are transmitted through the QPC with probability $D$ or reflected with probability $1-D$. The number of electrons that have been transmitted, say, from the left to the right region, is denoted as $m$, which can be both positive and negative depending on the direction of the tunneling events.

Rényi entanglement entropies of integer order $n > 1$ it was recently shown that they can also be determined in quantum Monte Carlo (QMC) calculations in two dimensions using an elegant trick [43, 46]. The von Neumann entanglement entropy, however, has remained inaccessible outside of ED and DMRG, although it has been assumed that the Rényi entropies could be analytically continued to obtain the desired result. With the method presented here it is now possible to obtain not only the von Neumann entanglement entropy but also the full entanglement spectrum, assuming the Rényi entropies can be obtained to sufficiently high order.

### C. Applications

We now illustrate the use of our formulas on a number of different examples. In one dimension, we expand on the analysis given in Ref. [36] of the entanglement entropy generated in a QPC under both equilibrium and non-equilibrium conditions, as well as of the spin-1/2 XX model. For the latter, we discuss the influence of finite temperature and strong disorder in the chain. For non-interacting fermions in two dimensions we show how the cumulants reproduce the violation of the strict area laws for free fermions, while, in contrast, for bulk states of integer quantum Hall systems the area law is strictly obeyed.

### 1. Quantum Point Contact

We first consider the entanglement entropy generated between two electronic reservoirs connected via a QPC, Fig. 4. In the situation discussed here, the chiral edge states of a two-dimensional electron gas in a strong perpendicular magnetic field constitute the one-dimensional leads from which incoming electrons are either transmitted through the QPC and appear in the other reservoir or are reflected back into the out-going edge channel. We analyze the entanglement entropy between the two Fermi seas separated by the QPC.

Without a bias difference between the electronic reservoirs, the cumulant generating function for the number of charges $m$ that have been transferred through the QPC (say, from left to right) is well-known and reads [71]

$$\ln \chi(\lambda) = \frac{(i\lambda_*)^2}{2\pi^2} G(t) \qquad (2.38)$$

with

$$G(t) = \ln\left\{\frac{\tau_\beta}{\tau_c} \sinh \frac{t}{\tau_\beta}\right\} \qquad (2.39)$$

determining the time dependence. Here $\tau_\beta = h/(\pi k_B T)$ is the time scale associated with the temperature $T$ and $\tau_c$ is an ultraviolet short-time cutoff. Additionally, the counting field $\lambda$ enters via the relation

$$\sin \frac{\lambda_*}{2} = \sqrt{D} \sin \frac{\lambda}{2} \qquad (2.40)$$

involving the QPC transmission $D$. It is assumed that the QPC is initially pinched off and instantaneously opened at time $t = 0$ when the counting of transmitted electrons begins. A more detailed description of the opening of the QPC and its influence on the charge statistics can be addressed using the numerical procedure described in Ref. [72], but is not essential here.

Interestingly, for perfect transmission $D = 1$, Eq. (2.40) immediately yields $\lambda = \lambda_*$, implying that the fluctuations described by Eq. (2.38) are purely gaussian with zero mean, variance $C_2 = G/\pi^2$, and all higher cumulants equal to zero. In this case, we can directly take the limit of infinite cutoff in Eq. (2.1) such that the coefficients $\alpha_n(K \to \infty) = 2\zeta(n)$ are given by the Riemann zeta function (cf. the generalization to the case with fractional quantum Hall reservoirs in Ref. [73]). Since only the second cumulant is nonzero, we find $\mathcal{S} = G/3$ having used $\zeta(2) = \pi^2/6$. Considering now the zero temperature limit $\tau_\beta \to \infty$ for which $G(t) = \ln(t/\tau_c)$, we readily obtain the simple result

$$\mathcal{S} = \frac{1}{3} \ln \frac{t}{\tau_c}. \qquad (2.41)$$

Remarkably, this expression agrees with the prediction

$$\mathcal{S} = \frac{c}{3} \ln \frac{\ell}{\alpha} \qquad (2.42)$$

for a window in space of size $\ell$ of a conformal field theory with central charge $c = 1$ and short-distance cutoff $\alpha$ [6, 7, 37, 74]. In our case, an analogous window $[0, t]$ in *time* is used during which particles can delocalize among the two reservoirs, thereby making them entangled. It should be noted that this problem of local quantum *quenches* and the analogy between Eqs. (2.41) and (2.42) is quite non-trivial and has been addressed extensively in Ref. [75].

In principle, the logarithmic dependence on time of the gaussian equilibrium quantum noise of a QPC is measurable. Indeed, measurement of the second charge cumulant showing a logarithmic dependence on time alone would constitute a strong indication of many-body entanglement. However, measurements of the equilibrium quantum noise should be contrasted with several other sources of noise that we discuss in turn below. Firstly, an imperfect transmission of the QPC, $D < 1$, leads to non-gaussian fluctuations which modify the result above. Additionally, any nonzero temperature causes classical thermal fluctuations, such that the von Neumann entropy eventually becomes linear in time. This is also the effect of a finite bias voltage, which gives rise to shot noise due to individual binomial charge transfer events that contribute only classically to the entropy increase without generating any entanglement between the reservoirs.

In the case of imperfect transmission, the entanglement entropy maintains its logarithmic growth with time but the prefactor changes. This is illustrated in Fig. 5a, showing the entanglement entropy as a function of time for a QPC with transmission $D = 0.5$ obtained from the series (2.1) with increasing cutoff number $K$. For $K = 2$, where only the second cumulant is included, we immediately find $\mathcal{S} = [5/(2\pi)^2] \ln(t/\tau_c) \simeq 0.13 \ln(t/\tau_c)$. This is already a good approximation of the exact result to which the series basically converges for $K \simeq 30$. Importantly, the result obtained for $K = 2$ provides a *lower* bound on the full entanglement entropy, implying that a measurement of the second cumulant showing a logarithmic dependence on time would constitute a strong indication of many-body entanglement. For comparison, we also show Eq. (2.41) corresponding to a perfectly transmitting QPC with $D = 1$. The entanglement entropy is maximal for $D = 1$, since the entanglement is generated by charge-neutral processes in which the two regions exchange particles [40]. For imperfect QPCs where incoming electrons may reflect back on the QPC, the entanglement entropy decreases with decreasing $D$ and, ultimately, when the QPC is completely pinched off with $D = 0$ no electrons are transmitted and no entanglement is generated.

We next discuss the influence of a nonzero electron temperature. Due to charge conservation, the even cumulants are the same if we count the number of electrons that have been transferred from right to left instead of from left to right. The von Neumann entropy of the two regions with respect to each other is therefore identical even at finite temperatures, where the full system is no longer in a pure state. In Fig. 5b we show the von Neumann entropy as a function of time at a realistic temperature of $T = 10$ mK and associated time scale $\tau_\beta \simeq 1.5$ ns. For short times $t < \tau_\beta$, the time dependence remains logarithmic since $G(t) \simeq \ln(t/\tau_c)$ and the behavior is nearly identical to the zero temperature result indicated by a red line. In contrast, at long times $t > \tau_\beta$, temperature effects become important and the linear-in-time dependence $G(t) \simeq t/\tau_\beta + \ln[\tau_\beta/(2\tau_c)]$ (shown as a dashed line) for high temperatures eventually prevails. Importantly, the zero-temperature behavior observed on sub-nanosecond time scales may be experimentally accessible with current high-frequency noise measurement techniques that operate at bandwidths in the gigahertz regime [56, 58, 76–78].

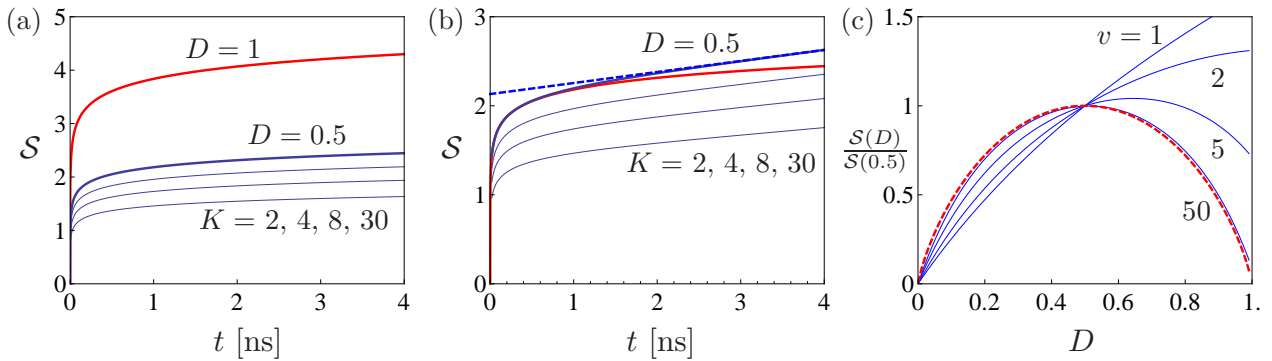

FIG. 5: (color online). Entanglement entropy in a quantum point contact (QPC). (a) Zero-temperature results for the time dependent entanglement entropy with different QPC transmissions $D$. The $D = 1$ result shown in red is given by Eq. (2.41). For $D = 0.5$, results were obtained from the series (2.1) with increasing cutoff number $K$ from bottom to top. The thick blue line is the converged result for $K = 30$. The ultraviolet short-time cutoff is $\tau_c = 10^{-5}$ ns. (b) Finite-temperature results for $T = 10$ mK (or $\tau_\beta \simeq 1.5$ ns), $D = 0.5$, and $\tau_c = 10^{-5}$ ns. Blue lines show results obtained with increasing cutoff number $K$ and the thick blue line is the converged result. For comparison, zero and high temperature limits are indicated with a red and a blue dashed line, respectively. For short times $t < \tau_\beta$ the time-dependence is logarithmic, while for long time the high-temperature behavior eventually prevails and the entropy grows linearly with time. (c) Results for a biased QPC as a function of the transmission $D$. Here $v = eV/(2k_B T)$ is the ratio of the applied voltage $V$ over temperature $T$. In the long-time limit $t \gg \tau_\beta$, the ratio $\mathcal{S}(D)/\mathcal{S}(0.5)$ does not depend on time. The cutoff number is $K = 10$. As the voltage is increased, the entanglement entropy changes from a nearly linear dependence on $D$ to that of a binomial event with success probability $D$, given by Eq. (2.45) and shown with a red dashed line.

Finally, it is instructive to consider the effect of a finite bias between the two Fermi seas. Our series are also applicable to such non-equilibrium situations. Focusing only on the thermal noise and the shot noise due to the applied bias, the cumulant generating function reads

$$\ln \chi(\lambda) = -\frac{t}{\pi \tau_\beta} u_+ u_-, \qquad (2.43)$$

where

$$u_\pm = v \pm \cosh^{-1}\{D\cosh(v + i\lambda) + (1-D)\cosh(v)\} \qquad (2.44)$$

and $v = eV/(2k_B T)$ is the ratio of the applied voltage over temperature [71]. Fig. 5c shows the von Neumann entropy as a function of the QPC transmission $D$ for several applied voltage differences. For low voltages, the entropy generated is mainly due to thermal fluctuations and depends nearly linearly on $D$. As the bias voltage is increased, shot noise-dominated processes become increasingly important and in the regime, where the applied bias voltage is much larger than temperature, the generated entropy is solely due to individual binomial charge transfer events occurring with probability $D$. In that case, the von Neumann entropy is that of a classical binomial process involving $eVt/h$ attempts for which

$$\mathcal{S} = -\frac{eVt}{h}\left[D\ln D + (1-D)\ln(1-D)\right]. \qquad (2.45)$$

This expression is shown with a dashed red line and is in excellent agreement with our results for the high-bias limit. Eq. (2.45) is also the electron-hole entanglement predicted for a biased tunnel junction [79], and in both cases the Renyi entropies for small $D$ are

$$\mathcal{S}_n = \frac{n}{n-1} N, \qquad (2.46)$$

where $N = (eVt/h)D$ is the number of transferred particles [80]. We note that in the finite-bias situation considered in the last part, the generated von Neumann entropy is only due to classical fluctuations and does not reflect true many-body entanglement.

### 2. The Spin-1/2 XX Chain

In our next example, we consider the one-dimensional spin-1/2 XX model. We first analyze the situation where the coupling between neighboring spins is constant along the chain and discuss in this case the influence of a finite temperature on the spin fluctuations and the entanglement entropy in the chain. Secondly, we turn to the situation where the coupling between neighboring spins has random magnitude. Importantly, the relation (2.1) between the von Neumann entanglement entropy and the cumulants is linear, such that the equation also holds when averaged over the disorder. The series then becomes a linear relation between the *disorder-averaged* entanglement entropy and the *disorder-averaged* cumulants.

The spin-1/2 XX model in one dimension is described by the Hamiltonian

$$\hat{H} = \sum_i J_i (\hat{S}_i^x \hat{S}_{i+1}^x + \hat{S}_i^y \hat{S}_{i+1}^y), \quad (2.47)$$

where $\hat{S}_i^\alpha$, $\alpha = x, y, z$, is the operator for the $\alpha$-component of the spin on site $i$. We first consider $J_i = J$ being constant along the chain and discuss analytically the zero-temperature and high-temperature behaviors of both the von Neumann entanglement entropy and the cumulants of the spin fluctuations. These results serve as important limiting check points for our numerical calculation with intermediate temperatures. Unless otherwise stated, in this section all system sizes count the number of lattice sites, i.e., lengths are in terms of the lattice spacing and are dimensionless.

The zero-temperature result for the entanglement entropy was determined by Calabrese and Cardy who found [7]

$$\mathcal{S}(L/2, L; T=0) = \frac{1}{3} \ln \frac{L}{\pi} + s_1, \quad (2.48)$$

where $L$ is the total length of the spin chain (in units of the lattice spacing) and the subsystem is taken to be half the chain of length $L/2$. The constant $s_1$ was obtained analytically by Jin and Korepin and can be found in Ref. [81].

Although it is difficult to compute the full set of cumulants analytically, it is worth obtaining analytical results for the fluctuations for the spin-1/2 XX chain [53]. For periodic boundary conditions (PBCs) the computation in the Supplementary Material gives

$$\pi^2 \mathcal{F}_{\text{XX}}^{\text{PBC}}(\ell) = \ln \ell + f_1 \quad (2.49)$$

plus $O(\ell^{-2})$ corrections, with $f_1 = 1 + \gamma + \ln 2$. For open boundary conditions (OBCs) the computation in the Supplementary Material leads to

$$\mathcal{F}_{\text{XX}}^{\text{OBC}}(\ell) = \frac{1}{2} \mathcal{F}_{\text{XX}}^{\text{PBC}}(2\ell) + \frac{1}{2\pi^2(2\ell)}$$
$$- \frac{(-1)^\ell}{\pi^2(2\ell)} [\ln(2\ell) + \gamma + \ln 2]$$
$$+ \frac{(-1)^\ell}{\pi^2(2\ell)^2} [\ln(2\ell) - \ln 2] \quad (2.50)$$

plus $O(\ell^{-2})$ corrections. In both cases the linear term vanishes, and for the case of OBCs subleading oscillations exist.

It is interesting to derive Eq. (2.49) for PBCs in a way different from the computation presented in the Supplementary Material that illustrates the similarity to the computation of the entanglement entropy for the spin-1/2 XX chain in Ref. [81], although it is less controlled than the previous method. The method also suggests how one might be able to compute all of the cumulants. In principle, the full set of cumulants can be calculated from the generating function

$$\chi(\lambda) = \left\langle \exp\left(i\lambda \sum_{j=1}^\ell \hat{S}_j^z\right) \right\rangle \quad (2.51)$$

$$= e^{-i\lambda \ell/2} \left\langle \exp\left(i\lambda \sum_{j=1}^\ell \hat{a}_j^\dagger \hat{a}_j\right) \right\rangle \quad (2.52)$$

where $\hat{a}_j^\dagger, \hat{a}_j$ are the Jordan-Wigner fermion creation and annihilation operators (cf. Supplementary Material). We have

$$\chi(\lambda) = \det\left[\left(\cos\frac{\lambda}{2}\right) I + \left(i \sin\frac{\lambda}{2}\right) \tilde{M}\right], \quad (2.53)$$

where $\tilde{M}$ is the $\ell \times \ell$ matrix with elements

$$\tilde{M}_{ij} = \begin{cases} \frac{2}{\pi}(-1)^{(i-j+1)/2} \frac{1}{i-j} & \text{for } i-j \text{ odd}, \\ 0 & \text{for } i-j \text{ even}. \end{cases} \quad (2.54)$$

For $\ell \to \infty$ the determinant (2.53) can be evaluated using the generalized Fisher-Hartwig conjecture for Toeplitz determinants to obtain the result (we assume $\lambda < \pi$ since we desire the derivatives at $\lambda = 0$ for the cumulants) [82]

$$\ln \chi(\lambda) \simeq -\frac{\lambda^2}{2\pi^2} \ln(2\ell) + 2 \ln\left[G\left(1 + \frac{\lambda}{2\pi}\right) G\left(1 - \frac{\lambda}{2\pi}\right)\right], \quad (2.55)$$

where $G(z)$ is the Barnes-G function defined by

$$G(1+z) = (2\pi)^{z/2} \exp\left[-\frac{1}{2}z - \frac{1}{2}(1+\gamma)z^2\right]$$
$$\times \prod_{k=1}^\infty \left[\left(1 + \frac{z}{k}\right)^k \exp\left(-z + \frac{z^2}{2k}\right)\right]. \quad (2.56)$$

Using

$$\ln[G(1+z)G(1-z)] = -(1+\gamma)z^2 + O(z^4) \quad (2.57)$$

we find that $\mathcal{F}_{\text{XX}}^{\text{PBC}} = C_2 = (-i\partial_\lambda)^2 \ln \chi(\lambda)|_{\lambda=0}$ agrees with Eq. (2.49).

At finite temperature the von Neumann entropy can be found from the correlation matrix $M$ of the full chain given in the Supplementary Material (see also Ref. [83]). In the high-temperature limit $\beta J \ll 1$ the correlation

matrix is

$$M = \frac{1}{2} \begin{pmatrix} 1 & -\beta J/4 & & & -p\beta J/4 \\ -\beta J/4 & 1 & -\beta J/4 & & \\ & \ddots & \ddots & \ddots & \\ & & -\beta J/4 & 1 & -\beta J/4 \\ -p\beta J/4 & & & -\beta J/4 & 1 \end{pmatrix}, \quad (2.58)$$

with $p$ determined by the choice of boundary conditions (Supplementary Material); here we restrict to $p = \pm 1$ for PBCs. Considering a subsystem of length $\ell < L$, the eigenvalues of the correlation matrix restricted to the subsystem $M^{(\ell)}$ are easily found to be

$$\nu_n = \frac{1}{2}\left(1 - \frac{\beta J}{2}\cos\frac{\pi n}{\ell+1}\right) \quad (2.59)$$

for $n = 1, \ldots, \ell$. Using the expansion of the binary entropy function $H_2(x) = -x\ln x - (1-x)\ln(1-x)$

$$H_2\left(\frac{1-\epsilon}{2}\right) = \ln 2 - \frac{\epsilon^2}{2} + O(\epsilon^4) \quad (2.60)$$

for $\epsilon \ll 1$ and the identity

$$\sum_{n=1}^{\ell} \cos^2\frac{\pi n}{\ell+1} = \frac{\ell-1}{2}, \quad (2.61)$$

the von Neumann entropy (2.9) becomes

$$\mathcal{S}(\ell, L; \beta J \ll 1) \simeq \left(\frac{\beta J}{4}\right)^2 + \left[\ln 2 - \left(\frac{\beta J}{4}\right)^2\right]\ell. \quad (2.62)$$

For comparison, if we consider the full chain $\ell = L$ we can similarly demonstrate that the von Neumann entropy is given by

$$\mathcal{S}(L, L; \beta J \ll 1) \simeq \left[\ln 2 - \left(\frac{\beta J}{4}\right)^2\right]L. \quad (2.63)$$

Eq. (2.63) is of course nothing but the thermodynamic entropy of the XX model at high temperature, which is purely extensive in the system size $L$. Eq. (2.62), meanwhile, is *almost* extensive in $\ell$, and the constant contribution becomes negligible with increasing temperature, i.e., as $\beta J$ decreases. This is in contrast to the zero-temperature behavior (2.48) where the logarithmic dependence on the subsystem size is a clear signature of true quantum correlations, i.e., entanglement.

We now compute the cumulants $C_n$ for the total spin $\hat{S}^z$ in the subsystem. Since we know the eigenvalues of $M^{(\ell)}$ we can write for $\ell < L$

$$\ln \chi(\lambda) = \sum_n \ln\{1 + (e^{i\lambda}-1)\nu_n]e^{-i\lambda/2}\} \quad (2.64)$$

$$= \left(\ln\cos\frac{\lambda}{2}\right)\ell - i\tan\frac{\lambda}{2}\sum_n\left(\frac{\beta J}{2}\cos\frac{\pi n}{\ell+1}\right)$$

$$+ \frac{1}{2}\tan^2\frac{\lambda}{2}\sum_n\left(\frac{\beta J}{2}\cos\frac{\pi n}{\ell+1}\right)^2 + O((\beta J)^3). \quad (2.65)$$

The second term of Eq. (2.65) is an odd function of $\lambda$ and does not contribute to the even cumulants, so we may write the effective cumulant generating function to leading order in $\beta J$ as

$$\ln \chi(\lambda) \simeq -\left(\frac{\beta J}{4}\right)^2 \tan^2\frac{\lambda}{2}$$

$$+ \left[\ln\cos\frac{\lambda}{2} + \left(\frac{\beta J}{4}\right)^2\tan^2\frac{\lambda}{2}\right]\ell. \quad (2.66)$$

Similarly, for $\ell = L$

$$\ln\chi(\lambda) \simeq \left[\ln\cos\frac{\lambda}{2} + \left(\frac{\beta J}{4}\right)^2\tan^2\frac{\lambda}{2}\right]L. \quad (2.67)$$

It is easily shown numerically that the cumulants arising from Eqs. (2.66), (2.67) lead to the entanglement entropies given in Eqs. (2.62), (2.63), respectively.

In Fig. 6 we illustrate how the series (2.1) converges to the von Neumann entropy at various temperatures as the number of cumulants is increased. Exact results were obtained by numerically diagonalizing the correlation matrix (see Ref. [84] and the Supplementary Material). The results interpolate nicely between the zero and high temperature limits given by Eqs. (2.48) and (2.62), respectively. In particular, at low temperature where the entanglement entropy is similar to that of a pure state, the entanglement is essentially equal for the subsystem and its complement, such that $\mathcal{S}(\ell, L) \simeq \mathcal{S}(L-\ell, L)$ as seen in Fig. 6a. In the high-temperature regime, Fig. 6d, this symmetry is broken and the von Neumann entropy for the subsystem and its complement are now very different. This is clearly an effect of the finite temperature, which causes random and independent spin flips on each site. This effect is extensive in the subsystem size $\ell$ and causes the even cumulants of the spin fluctuations in the subsystem and its compliment to be different.

For subsystem size $\ell = L/2$ the temperature $T^*$ where thermal effects become important can be estimated from CFT arguments as follows. For an infinite system and subsystem size $x$ at finite temperature $T = 1/\beta$, we can use the mapping [7]

$$x \to \frac{v\beta}{\pi}\sinh\frac{\pi x}{v\beta} \quad (2.68)$$

where $v$ is the effective velocity, in this case $v = J$. The sinh becomes important when $\pi x \sim v\beta$, or with $x = L/2$ [83],

$$T^* \sim \frac{2J}{\pi L}. \quad (2.69)$$

This can already be seen in Figs. 6b and 6c with temperatures around $T^*$. For $T = 2T^*$, Fig. 6c, the von Neumann entropy is clearly not symmetric. The crossover between zero-temperature and high-temperature behavior is further illustrated in Fig. 7 showing the entanglement entropy as a function of temperature for different system sizes $L$. Here the partition is taken in the middle of the chain and periodic boundary conditions have been imposed. The solid lines show the analytical zero- and high-temperature limits, while the vertical dashed lines correspond to the crossover temperature $T^*$ defined by Eq. (2.69). The crossover temperature $T^*$ separates the von Neumann entropy into a low-temperature regime, governed by zero-temperature behavior with logarithmic dependence on the subsystem size $\ell$, and a high-temperature limit, where the von Neumann entropy depends linearly on $\ell$. In the low-temperature regime, the generated von Neumann entropy is mostly due to quantum correlations or *entanglement*, whereas in the high-temperature regime the von Neumann entropy mainly reflects classical, thermal fluctuations. This is important for actual experiments, where one wishes to measure the entanglement and not classical fluctuations.

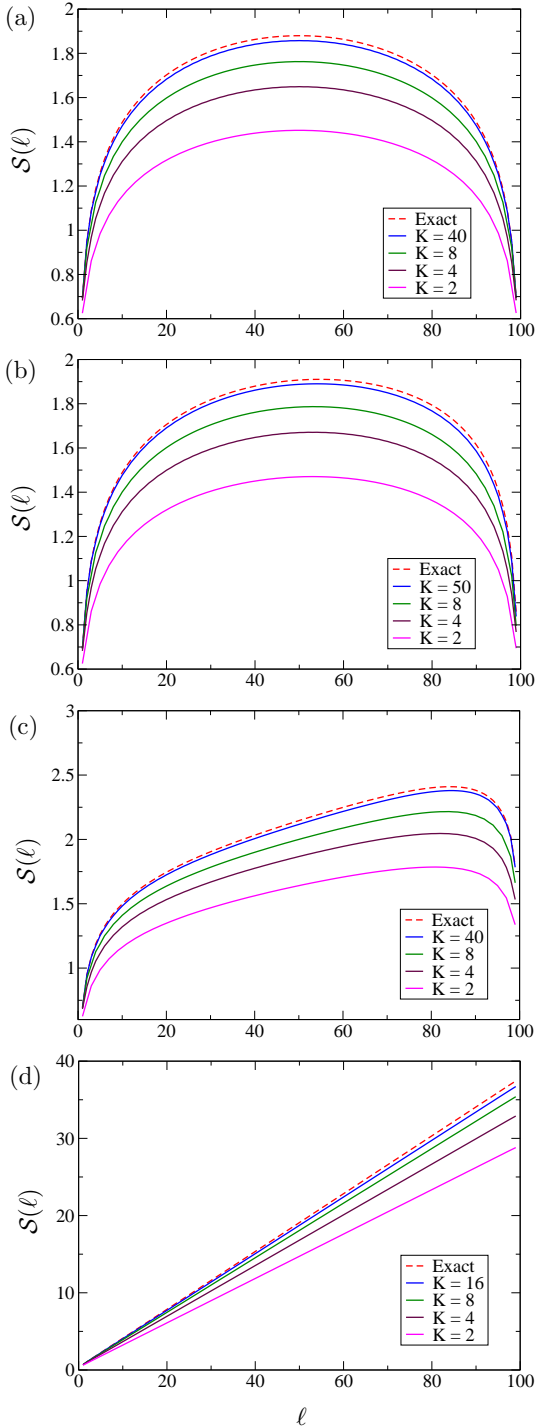

FIG. 6: (color online). Exact entanglement entropy (dashed) and approximation by cumulants (solid lines) of the spin-1/2 XX chain, $L = 100$, as a function of subsystem size, at temperature (a) $T = 0.5T^*$, (b) $T = T^*$, (c) $T = 2T^*$, (d) $T = 50T^*$ where $T^* = 2J/(\pi L)$ is the temperature at which thermal effects become significant. The cutoff $K$ increases from bottom to top. Note the differences in scale on the vertical axes.

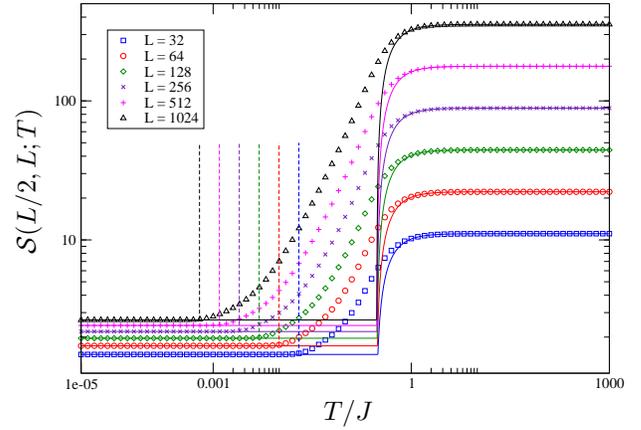

FIG. 7: (color online). Entanglement entropy of the spin-1/2 XX chain as a function of temperature $T$, for several system sizes $L$. The partition is taken in the middle at $\ell = L/2$. Solid lines show analytical results for the zero- and high-temperature limits, and dashed lines indicate the crossover temperature $T^*$, Eq. (2.69), where the entanglement entropy begins to deviate significantly from the zero-temperature behavior.

Finally, we briefly discuss the effect of disorder. We thus allow the coupling strength $J_i$ between sites $i$ and $i+1$ to vary along the chain with probability distribution $P(J_i)$. For essentially any distribution $P(J_i)$, the fixed point of the model is known to be described by the random singlet phase (RSP) in which the ground state is a product of singlets between spins at arbitrarily large distances [85]. Each singlet $|ij\rangle = (|\uparrow_i\downarrow_j\rangle - |\downarrow_i\uparrow_j\rangle)/\sqrt{2}$ that crosses the boundary between a subsystem of length $\ell$ and the remainder of the system contributes $\ln 2$ to the entanglement entropy. It is thus clear that we can write [74]

$$\mathcal{S}_{\mathrm{RSP}}(\ell) = \bar{n} \ln 2, \qquad (2.70)$$

where $\bar{n}$ is the disorder-averaged number of singlets that cross the boundary between the two blocks. It has been shown that $\bar{n} \sim (1/3) \ln \ell$ [74, 86, 87].

For a factorized state like the RSP, it is also easy to find the cumulants of the total spin $\hat{S}^z$ in the subsystem. For a subsystem of length $\ell$ the generating function for $\hat{S}^z$ is

$$\chi(\lambda) = \langle \Psi_{\text{RSP}} | \exp\left(i\lambda \sum_{j=1}^{\ell} \hat{S}_j^z\right) | \Psi_{\text{RSP}} \rangle, \qquad (2.71)$$

where $|\Psi_{\text{RSP}}\rangle = \prod_{\langle i,j \rangle} |ij\rangle$ is a product of non-overlapping singlets covering the entire system. Singlets between two spins within the subsystem ("in-in") or two spins outside the subsystem ("out-out") are eigenstates of $\hat{S}_i^z + \hat{S}_j^z$ with eigenvalue 0 and contribute a factor of $e^0 = 1$ to the expectation value. In contrast, a singlet between one spin inside the subsystem and one outside ("in-out") contributes $\langle ij | \exp[i\lambda(\hat{S}_i^z + \hat{S}_j^z)] | ij \rangle = \cos(\lambda/2)$ to the product, so if there are $n$ in-out singlets then

$$\chi(\lambda) = \cos^n \frac{\lambda}{2}. \qquad (2.72)$$

The disorder-averaged cumulant generating function is therefore

$$\overline{\ln \chi(\lambda)} = \bar{n} \ln \cos \frac{\lambda}{2}. \qquad (2.73)$$

Eq. (2.1) applied to Eq. (2.73) then leads to Eq. (2.70). As can be shown using exact diagonalization, moreover, the formula holds exactly for *each separate realization of the disorder* so that the equivalence between the cumulants and the entanglement entropy is a true equivalence.

It is worth noting that the previous discussion does not apply if the disorder is *locally correlated* [88]. In this case the fixed point is not a RSP and the prefactor of the logarithmic growth of the entanglement entropy $\mathcal{S}$ with subsystem size is not universal. Since the model is still that of non-interacting fermions, however, Eq. (2.1) still holds. Perhaps more suggestively [cf. Eq. (3.33)], $\mathcal{S}/C_2 \sim \pi^2/3$ for locally correlated disorder.

3. *Entanglement Entropy of Fermions in Two Dimensions*

We now consider systems in dimension higher than one and focus in particular on the area laws obeyed by systems in two dimensions. It has already been well-established through explicit arguments [89, 90] and CFT arguments on the Fermi surface [12] that for free fermions the entanglement entropy of a subsystem of linear size $L$ in $d$-dimensions typically obeys an area law with multiplicative logarithmic corrections:

$$\mathcal{S} \sim L^{d-1} \ln L, \qquad \text{quasi-area law.} \qquad (2.74)$$

For the particular case of free fermions on a square lattice with PBCs at half-filling as in Fig. 8, a numerical fit shows that the prefactor of the leading $L \ln L$ term is almost certainly exactly $1/3$ as predicted by a formula based on the Widom conjecture [90, 91]. Of course, unlike the one-dimensional case the prefactor is not universal and depends, for example, on the chemical potential. The behavior of the cumulants is necessarily similar as illustrated in Fig. 8. The approximation of the exact entanglement entropy using the series (2.1) improves, and we eventually find the quasi-area law (2.74) for the entanglement entropy with prefactor 0.333 in agreement with Ref. [91]. The similar behavior of the cumulants can be understood from the CFT nature of the problem as we discuss in further detail in Sec. III A 4.

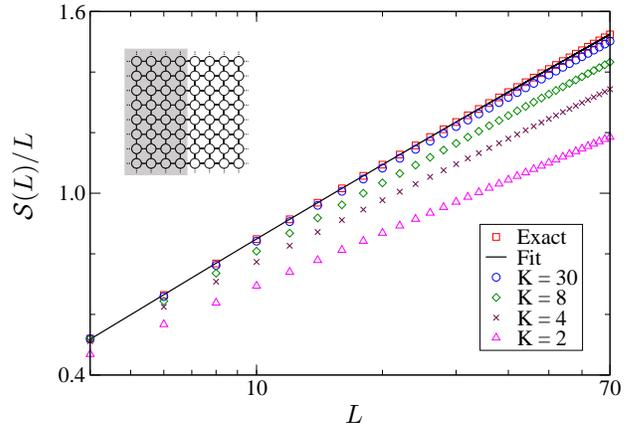

FIG. 8: (color online). Entanglement entropy and approximation by cumulants for free fermions on an $L \times L$ square lattice with PBCs, where the partition is taken to be half the system. The solid line shows a fit to the generic form $\mathcal{S} = aL\ln L + bL + c\ln L + d + e/L$, where, in particular, $a \simeq 0.333$.

Of course, there also exist systems of non-interacting fermions in $d$ dimensions which obey the usual *strict* area law

$$\mathcal{S} \sim L^{d-1}, \qquad \text{strict area law}, \qquad (2.75)$$

without multiplicative logarithmic corrections. Ref. [91] has explored this behavior as a function of the dimensionality and shape of the Fermi surface, showing that entanglement entropy (and hence the charge statistics) contains important information about the geometry of the Fermi surface. The argument of Ref. [12] elegantly illustrates the role of the Fermi surface in the appearance of the multiplicative logarithmic correction for Fermi liquids: one essentially breaks the Fermi surface into $O(L^{d-1})$ "patches," with each patch described by a $c = 1$ CFT and hence contributing $(1/3) \ln L$ to the entanglement entropy.

Here we consider an important example of a two-dimensional system of fermions that obeys a strict area law: the integer quantum Hall effect (IQHE) at filling factor $\nu = 1$. The real-space entanglement entropy for this system was computed for various geometries in Ref. [92] and found to obey a strict area law. Following Ref. [92], we thus consider the lowest Landau level of electrons in a cylinder of size $L_x \times L_y$ in the gauge $\mathbf{A} = B(0, x)$ where $B$ is the magnetic field, i.e., the magnetic field is perpendicular to the surface of the cylinder. The cylinder is periodic in the $y$-direction and infinite in the $x$-direction. We will work in units where the magnetic length is equal to 1 [92]. In the limit $L_x, L_y \gg 1$ the ground-state correlation matrix $M_{\mathbf{rr'}} = \langle \hat{c}^\dagger_\mathbf{r} \hat{c}_{\mathbf{r'}} \rangle$, where $\hat{c}_\mathbf{r}$ is the fermion annihilation operator at position $\mathbf{r}$, is given by

$$M_{\mathbf{rr'}} = \frac{1}{2\pi} \exp\left[-\frac{1}{4}(x-x')^2 - \frac{1}{4}(y-y')^2 - \frac{i}{2}(x+x')(y-y')\right]. \quad (2.76)$$

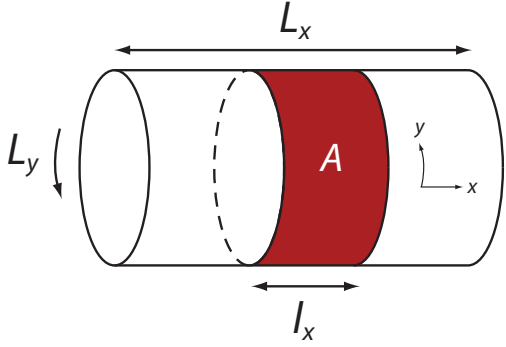

FIG. 9: (color online). Cylinder geometry for the integer quantum Hall effect. The "area" of subsystem $A$ is $2L_y$.

We are interested in the entanglement entropy of the region $A$ defined by

$$-\frac{\ell_x}{2} \leq x \leq \frac{\ell_x}{2}, \qquad 0 \leq y \leq L_y \quad (2.77)$$

as shown in Fig. 9. It was found in Ref. [92] that for $\ell_x \gg 1$ the eigenvalues of $M_{\mathbf{rr'}}$ are

$$\eta(\mu, \ell_x) = \frac{1}{2}\left[\mathrm{Erf}\left(\mu + \frac{\ell_x}{2}\right) - \mathrm{Erf}\left(\mu - \frac{\ell_x}{2}\right)\right], \quad (2.78)$$

where $\mu$ is a continuous index ranging from $-\infty$ to $\infty$ with measure $L_y/(2\pi)$ and $\mathrm{Erf}(x) = (2/\sqrt{\pi})\int_0^x dt\, e^{-t^2}$ is the error function. Each eigenvalue contributes $L_y H_2(\eta(\mu, \ell_x))/(2\pi)$ to the entanglement entropy, where $H_2(x) = -x \ln x - (1-x)\ln(1-x)$ is the binary entropy function. The total entanglement entropy is then given by the area law

$$\mathcal{S}(\ell_x, L_y) = 2\alpha(\ell_x) L_y, \quad (2.79)$$

with

$$\alpha(\ell_x) = \frac{1}{4\pi} \int_{-\infty}^\infty d\mu\, H_2(\eta(\mu, \ell_x)). \quad (2.80)$$

In this geometry the "area" is given by the perimeter length $2L_y$, so the result is expected to depend only on $L_y$ for long cylinders $\ell_x \gg 1$. Indeed, in this limit the prefactor $\alpha(\ell_x)$ approaches the constant value $\alpha_0 \simeq 0.203$.

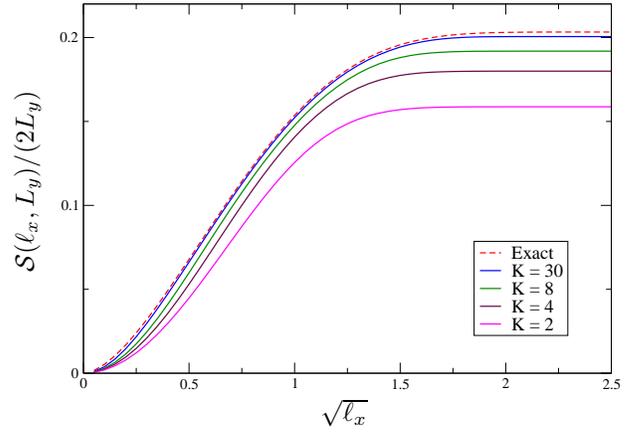

FIG. 10: (color online). The exact (dashed line) entanglement entropy, under the assumptions that lead to Eq. (2.79), and approximation by cumulants for different cutoffs $K$, increasing from bottom to top, for the integer quantum Hall effect on a cylinder at filling factor $\nu = 1$.

The factor of 2 in the area law (2.79) is due to the particular choice of geometry, where the subsystem has *two* boundaries of length $L_y$ at $x = \pm \ell_x/2$ connecting it to its complement. For a subsystem consisting of a semi-infinite cylinder with only a single boundary of length $L_y$, the factor of 2 in Eq. (2.79) would be replaced by unity. The absence of a constant term, moreover, indicates the absence of topological order [92]. The "strip" geometry has also recently been used to study entanglement and Rényi entropies in the spin-1/2 Heisenberg model on a square lattice [46].

These results also follow directly from the fluctuations of charge across the boundaries between the subsystem and its compliment. To illustrate this, we evaluate the generating function for charge fluctuations (2.11) using the correlation matrix in Eq. (2.76). We find

$$\ln \chi(\lambda, \ell_x) = \frac{L_y}{2\pi}\int_{-\infty}^\infty d\mu\, \ln[1 + (e^{i\lambda}-1)\eta(\mu, \ell_x)], \quad (2.81)$$

with corresponding cumulants

$$C_n(\ell_x) = 2\beta_n(\ell_x) L_y \quad (2.82)$$

where

$$\beta_n(\ell_x) = \frac{1}{4\pi}\int_{-\infty}^\infty d\mu\, (-i\partial_\lambda)^n \ln[1+(e^{i\lambda}-1)\eta(\mu,\ell_x)]|_{\lambda=0}. \quad (2.83)$$

Fig. 10 shows the approximation by cumulants to the entanglement entropy [specifically, the prefactor $\alpha(\ell_x)$ in Eq. (2.80)], plotted as a function of $\sqrt{\ell_x}$ following Ref. [92]. As mentioned previously, the entanglement entropy (and also the cumulants) are independent of $\ell_x$ for large $\ell_x$, and depends only on $L_y$. We note that similar calculations can be carried out for the disk and spherical geometries [92].

## III. FLUCTUATIONS IN ONE-DIMENSIONAL INTERACTING SYSTEMS

As shown in Sec. II, for systems that can be mapped to a problem of non-interacting fermions the complete set of even cumulants of the charge fluctuations fully determines the von Neumann and Rényi entanglement entropies, as well as the entanglement spectrum. This equivalence appears to be unique to non-interacting fermions [73]. For many one-dimensional conformally invariant critical systems with a conserved U(1) charge, however, the fluctuations can be shown to scale logarithmically with the system size, just as for entanglement entropy. We can also determine the prefactor which depends not on the central charge of the theory but on the compressibility (susceptibility) of the U(1) charge. In the following sections we show how this comes about and discuss when and how the behavior of the fluctuations and entanglement entropy differ. The area-law behavior of fluctuations in gapped systems is also discussed.

In the remainder of the paper we will be concerned exclusively with the second cumulant $C_2$, which we refer to as the "fluctuations" $\mathcal{F}$.

### A. One-Dimensional Critical Models

The low-energy properties of many one-dimensional systems are captured by Luttinger liquid (LL) theory [93–95] and are described by the Hamiltonian

$$H_{\text{LL}} = \frac{v}{2\pi} \int dx \left[ K(\partial_x \theta)^2 + \frac{1}{K}(\partial_x \phi)^2 \right] \quad (3.1)$$

where $v$ and $K$ are the effective velocity and Luttinger parameter, respectively. We recall that for such a LL, a free bosonic CFT with central charge $c = 1$, the entanglement entropy of a region of size $x$ in an infinite system scales as

$$\mathcal{S}_{\text{LL}}(\ell) = \frac{1}{3} \ln \ell + s_1 \quad (3.2)$$

where $s_1$ is a non-universal constant and is sometimes combined into the logarithm term through a short-distance cutoff $a$ as $\ln(\ell/a)$. Eq. (3.2) is for a system with PBCs; for OBCs the prefactor is divided by 2, $1/3 \to 1/6$, while $\ell \to 2\ell$. For systems with finite size $L$, moreover, conformal mapping leads to the simple substitution

$$\ell \to \frac{L}{\pi} \sin \frac{\pi \ell}{L}. \quad (3.3)$$

#### 1. Luttinger Liquids

We now consider the fluctuations. In the following we measure distances in physical units and thus restore the short-distance cutoff to emphasize the ultraviolet divergence of the fluctuations and its regularization. The long wavelength ($|k| \ll \rho_0$, $\rho_0$ is the mean density) part of the density fluctuations in a LL is given by

$$\rho(x) = \rho_0 + \frac{1}{\pi} \partial_x \phi(x), \quad (3.4)$$

so that for a block of length $\ell$ extending from $x = 0$ to $x = \ell$ we have

$$\hat{N}_A - \langle \hat{N}_A \rangle = \frac{1}{\pi}[\phi(x) - \phi(0)], \quad (3.5)$$

where we have used the fact that $\langle \phi(x) - \phi(y) \rangle = 0$. The field $\phi$ has the mode expansion

$$\phi(x) = \phi_0 + \sqrt{\frac{\pi K}{2L}} \sum_{q \neq 0} \frac{e^{-\alpha |q|/2}}{\sqrt{|q|}}(e^{iqx}\hat{b}_q + e^{-iqx}\hat{b}_q^\dagger) \quad (3.6)$$

where $\alpha$ is a short-distance cutoff, not necessarily the same as $a$ for the entanglement entropy, and $\hat{b}_q^\dagger, \hat{b}_q$ are bosonic creation and annihilation operators. For $L \to \infty$ we can convert the momentum sum into an integral to find

$$\langle [\phi(\ell) - \phi(0)]^2 \rangle = 2K \int_0^\infty dq \, \frac{e^{-\alpha q}}{q} \sin^2 \frac{q\ell}{2} \quad (3.7)$$

$$= \frac{K}{2} \ln \frac{\ell^2 + \alpha^2}{\alpha^2}. \quad (3.8)$$

Thus for $\ell \gg \alpha$

$$\pi^2 \mathcal{F}_{\text{LL}}(\ell) = \langle [\phi(\ell) - \phi(0)]^2 \rangle \sim K \ln \frac{\ell}{\alpha}. \quad (3.9)$$

As for the entanglement entropy, the same result with $K \to K/2$ and $\ell \to 2\ell$ is obtained when there is a boundary, due to the constraint $\phi(0) = $ constant. This can be seen from

$$\phi(x) = \phi_0' + i\sqrt{\frac{\pi K}{L}} \sum_{q > 0} \frac{e^{-\alpha q/2}}{\sqrt{q}} \sin(qx)(\hat{b}_q - \hat{b}_q^\dagger) \quad (3.10)$$

for OBCs. Then

$$\langle [\phi(\ell) - \phi(0)]^2 \rangle = K \int_0^\infty dq \, \frac{e^{-\alpha q}}{q} \sin^2(q\ell) \quad (3.11)$$

$$= \frac{K}{4} \ln \frac{(2\ell)^2 + \alpha^2}{\alpha^2}. \quad (3.12)$$

Assuming $\ell \gg \alpha$ we obtain

$$\pi^2 \mathcal{F}_{\text{LL}}^{\text{OBC}}(\ell) \sim \frac{K}{2} \ln \frac{2\ell}{\alpha}. \quad (3.13)$$

We see that for both PBCs and OBCs in the limit $\ell \to \infty$ the leading terms of the entanglement entropy and fluctuations obey the relation

$$\mathcal{S}(\ell) \sim K \frac{\pi^2}{3} \mathcal{F}(\ell), \quad (3.14)$$

which generalizes the prefactor $2\zeta(2) = \pi^2/3$ for free fermions [36, 40]. As a simple example, for the fermionic Hubbard model with spin-charge separation into two LLs [94] the ratio of the entanglement entropy to the fluctuations is controlled by $K_c$ for the charge degree of freedom and $K_s = 1$ for the spin degree of freedom.

### 2. Case of Disjoint Intervals

For the entanglement entropy, there has been growing interest in the case where the subsystem $A$ is composed of *disjoint* intervals [7, 96, 97]. In particular, consider the case where $A$ extends from $x_1$ to $x_2$ and also $x_3$ to $x_4$ (we consider PBCs once more). It was originally predicted in Ref. [7] that for a free boson the entanglement entropy is given by

$$\mathcal{S}(x_1, x_2, x_3, x_4) \stackrel{?}{=} \frac{1}{3} \ln \frac{x_{12} x_{34} x_{14} x_{23}}{x_{13} x_{24}} + 2s_1, \quad (3.15)$$

cf. Eq. (3.2). It has been shown [96, 97] that this is not quite correct, due to the fact that for disjoint intervals the topology of the Riemann surface used in the derivation is highly non-trivial, i.e., non-local. It is interesting that, as we show below, the particle number fluctuations within LL theory, i.e., for a free boson, retains the simpler form of Eq. (3.15). This is a case where the entanglement entropy is clearly more sensitive to the full operator content of a CFT.

For the two-interval case considered above the number operator for subsystem $A$ is

$$\hat{N}_A - \langle \hat{N}_A \rangle = \frac{1}{\pi}\{[\phi(x_2) - \phi(x_1)] + [\phi(x_4) - \phi(x_3)]\}, \quad (3.16)$$

so that

$$\begin{aligned}
\pi^2 &\mathcal{F}_{\mathrm{LL}}(x_1, x_2, x_3, x_4) \\
&\sim K \ln \frac{x_{12} x_{34}}{\alpha^2} \\
&\quad + 2\langle [\phi(x_2) - \phi(x_1)][\phi(x_4) - \phi(x_3)]\rangle \quad (3.17)\\
&= K \ln \frac{x_{12} x_{34}}{\alpha^2} \\
&\quad + \frac{K}{2} \int_{-\infty}^{\infty} dq \left[ \frac{e^{-\alpha|q|}}{|q|}(e^{iqx_2} - e^{iqx_1}) \times \right.\\
&\qquad\qquad \left. (e^{-iqx_4} - e^{-iqx_3}) \right] \quad (3.18)\\
&\sim K \ln \frac{x_{12} x_{34} x_{14} x_{23}}{x_{13} x_{24} \alpha^2}, \quad (3.19)
\end{aligned}$$

where $x_{ij} = x_j - x_i$. Thus Eq. (3.19) is essentially identical to the naïve result for the entanglement entropy in Eq. (3.15).

### 3. Subleading Corrections

It is possible to compute subleading corrections to the fluctuations within LL theory. To illustrate, we consider the spin-1/2 XXZ model with PBCs at zero magnetic field (i.e., $\langle \hat{S}_i^z \rangle = 0$), which is described by the Hamiltonian

$$\hat{H}_{\mathrm{XXZ}} = J\sum_i (\hat{S}_i^x \hat{S}_{i+1}^x + \hat{S}_i^y \hat{S}_{i+1}^y + \Delta \hat{S}_i^z \hat{S}_{i+1}^z). \quad (3.20)$$

For $-1 < \Delta \leq 1$ Hamiltonian (3.20) is gapless and described by LL theory. The isotropic point $\Delta = 1$ is somewhat special due to the effects of a marginally relevant operator and its infrared behavior is best considered in the context of the Haldane-Shastry model described below, while $\Delta = -1$ is not conformally invariant because of the quadratic spectrum. At $\Delta = 0$ the model reduces to the exactly solvable spin-1/2 XX chain where the exact results of Sec. II apply, but it is instructive to analytically compute $\mathcal{F}$, Sec. II C 2.

From LL theory, for $x \gg \alpha$ the spin-spin correlation function in the ground state of Hamiltonian (3.20) is given by

$$\begin{aligned}
\langle &\hat{S}_{i+r}^z \hat{S}_i^z \rangle - \langle \hat{S}_{i+r}^z \rangle \langle \hat{S}_i^z \rangle \\
&= -\frac{K}{2\pi^2} \frac{1}{r^2} + \frac{2^{2K-2} A_1}{\pi^2} \frac{(-1)^r}{r^{2K}} \\
&\quad + \sum_{m \geq 2} \frac{2^{2m^2 K - 2} A_m}{\pi^2} \frac{\cos(\pi m r)}{r^{2m^2 K}}, \quad (3.21)
\end{aligned}$$

where from the Bethe ansatz solution of the XXZ model

$$K = \frac{1}{2}\left(1 - \frac{\cos^{-1}\Delta}{\pi}\right)^{-1} \quad (3.22)$$

while the $A_m$ are non-universal coefficients (interestingly, there also exists a conjectured formula for $A_1$, see Refs. [98, 99]). For the XX model, for example, $K = 1$ and $A_1 = 1/(2\pi^2)$, $A_{m\geq 2} = 0$. Strictly speaking, of course, Eq. (3.21) is an asymptotic expansion. However, we will see that summing this expression gives results in excellent agreement with the numerical results, neglecting the linear term arising from the short-distance physics not taken into account within LL theory. With precise knowledge of the microscopic physics this linear term vanishes, as illustrated by calculations for the Haldane-Shastry chains described below and detailed in the Supplementary Material. It is also important to note that Eq. (3.21) is not correct for the isotropic point $\Delta = 1$ where marginal operators induce multiplicative logarithmic corrections to correlation functions.

To find the fluctuations, we first rewrite Eq. (1.3) as

$$\mathcal{F}_A = \sum_{i,j \in A} [\langle \hat{S}_i^z \hat{S}_j^z \rangle - \langle \hat{S}_i^z \rangle \langle \hat{S}_j^z \rangle]. \quad (3.23)$$

As shown in Ref. [53] and the Supplementary Material, the fluctuations corresponding to Eq. (3.21) are then given by

$$\pi^2 \mathcal{F}_{\mathrm{XXZ}}(\ell) = K \ln \ell + f_2 - A_1 \frac{(-1)^\ell}{\ell^{2K}} \quad (3.24)$$

plus $O(\ell^{-2})$ corrections. This provides an alternative method for finding the LL parameter $K$ in numerical calculations instead of using the spin-spin correlation function [53]. We note that for $K \leq 1$ the $r^{-2K}$ term dominates at long distances over the Fermi liquid-like term proportional to $r^{-2}$. The logarithmic divergence, in contrast, originates from the universal Fermi liquid-like term which decays as $r^{-2}$, i.e., from the short-distance correlations. This is of course an important feature of the entanglement entropy, whose divergence is due to short-distance correlations at the boundary between two subsystems.

Since the SU(2)-symmetric point $\Delta = 1$ is an important example of a model where the full symmetry group is actually larger than, but contains, U(1), it is worth checking that the above considerations are also valid for a closely-related system known as the Haldane-Shastry model [100, 101]. This model is in the same universality class as the Heisenberg model and has the peculiar feature that its finite-size behavior is the same as its thermodynamic limit, so that all logarithmic corrections are absent. The Haldane-Shastry model is described by the Hamiltonian

$$H_{\rm HS} = \sum_{i<j} J_{ij} \hat{\mathbf{S}}_i \cdot \hat{\mathbf{S}}_j, \qquad (3.25)$$

where $J_{ij} = 1/d(i-j)^2$ and $d(i,j)$ is the chordal distance

$$d(x) = \frac{L}{\pi}\left|\sin\frac{\pi x}{L}\right|. \qquad (3.26)$$

The exact spin-spin correlation is known to be [102]

$$\langle \hat{S}^z_{i+r}\hat{S}^z_i\rangle - \langle \hat{S}^z_{i+r}\rangle\langle \hat{S}^z_i\rangle = \frac{1}{4}(-1)^r \frac{\operatorname{Si}(\pi r)}{\pi r}, \qquad (3.27)$$

where

$$\operatorname{Si}(x) = \int_0^x dt\, \frac{\sin t}{t}. \qquad (3.28)$$

We note that the asymptotic expansion of Eq. (3.27) yields the same two leading terms as the LL expression in Eq. (3.21) with $K=1/2$, but the exponents differ for the remaining terms. As shown in the Supplementary Material the fluctuations are given by

$$\pi^2 \mathcal{F}_{\rm HS} = \frac{1}{2}\ln\ell + f_{\rm HS} - \frac{\pi^2}{16}\frac{(-1)^\ell}{\ell} \qquad (3.29)$$

plus $O(\ell^{-2})$ corrections, where $f_{\rm HS}$ is related to an integral whose numerical value is $f_{\rm HS}/\pi^2 \simeq 0.197217$. Fig. 11 shows the excellent agreement between DMRG results and the analytical expressions, with the usual mapping $\ell \to (L/\pi)\sin(\pi\ell/L)$ for finite systems. The calculation in the Supplementary Material also illustrates how explicitly demonstrating the vanishing of the linear term is non-trivial and depends very much on the exact microscopic details of the theory. Since the vanishing of the linear, "volume" term is nevertheless universal, it is interesting that the short-distance behavior is not arbitrary but rather constrained by this necessity. The example of the Haldane-Shastry chain also shows that the leading logarithmic scaling of the fluctuations can persist even when long-range interactions are introduced.

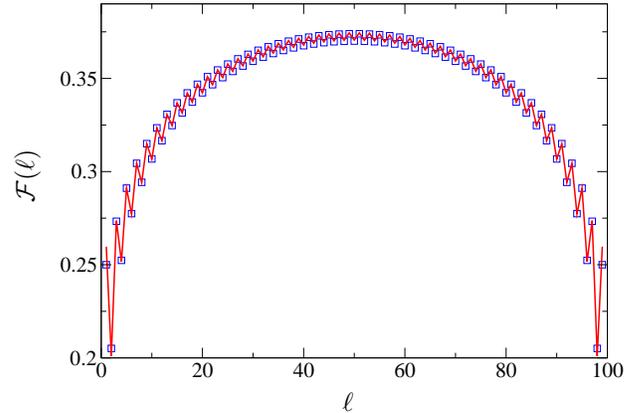

FIG. 11: (color online). Fluctuations of the Haldane-Shastry model, $L = 100$. Numerical results from DMRG (squares) are in excellent agreement with the analytical result (solid line) in Eq. (3.29) with finite-size mapping.

### 4. Extension to CFTs with a Conserved U(1) Charge

For critical one-dimensional systems described by CFT, a more general argument may be given for the scaling of the fluctuations that does not rely on traditional LL theory. Briefly reviewing the argument given in Ref. [53], this is because a conserved U(1) charge is always described by a free bosonic generator in the theory, and consequently the generating function is given by

$$\chi(\lambda) = \langle e^{i\lambda(\hat{N}_A - \langle \hat{N}_A\rangle)}\rangle = \left(\frac{x}{\alpha}\right)^{-g\lambda^2/(2\pi^2)}. \qquad (3.30)$$

Then $\pi^2 \mathcal{F}_A = \pi^2(-i\partial_\lambda)^2 \ln\chi(\lambda)|_{\lambda=0} = g\ln(x/\alpha)$. The prefactor $g$ can always be fixed by considering the physical meaning of the conserved charge, but we can give a heuristic argument for its value as follows. At finite temperature $1/\beta$ (we will set the effective velocity $v=1$ for now) the mapping $z \to z' = [\beta/(2\pi)]\ln z$ in Eq. (3.30) gives

$$\pi^2 \mathcal{F}(x,\beta) = g\ln\left(\frac{\beta}{\pi\alpha}\sinh\frac{\pi x}{\beta}\right). \qquad (3.31)$$

For sufficiently large $x \gg \beta$ such that interactions across the boundary can be neglected (which is possible since correlations decay exponentially), we may consider the subsystem $A$ to be a grand canonical ensemble in equilibrium with a bath consisting of the remainder of the system [103]. This is of course only possible if the total particle number is fixed. Then from standard statistical mechanics one has $\mathcal{F}(x,\beta) \sim \kappa x/\beta$ where $\kappa = \partial n/\partial \mu$ is the compressibility (susceptibility $\chi = \partial m/\partial B$ for spins), so that by matching Eq. (3.31) for $x \gg \beta, \alpha$ we find

$$g = \pi v \kappa. \qquad (3.32)$$

We have put in the velocity $v$ for completeness. Note that this is consistent with the LL expression since $K = \pi v \kappa$.

Thus for such systems we observe that

$$\mathcal{S}(\ell) \sim \frac{c}{\pi v \kappa} \frac{\pi^2}{3} \mathcal{F}(\ell), \qquad \ell \to \infty, \qquad (3.33)$$

which again generalizes the free-fermionic prefactor $2\zeta(2) = \pi^2/3$ [36, 40].

An important class of models is described by the SU(2)-symmetric Wess-Zumino-Witten non-linear $\sigma$-model with topological coupling $k$ [SU(2)$_k$ WZW model]. They are CFTs with central charge $c = 3k/(k+2)$, so that the SU(2)$_1$ WZW model, with $c = 1$, is simply a free-boson theory. Furthermore, the spin susceptibility is given by $\pi v \chi = k/2$ where $v$ is the effective velocity [53, 104]. A microscopic realization of the SU(2)$_k$ WZW model for general $k$ was found by Takhtajan [105] and Babujian [106] (TB) governed by the Hamiltonian

$$\mathcal{H}_{\mathrm{TB}}^S = \sum_i \sum_{\nu=0}^{2S} \left( \sum_{j=1}^{\nu} \frac{1}{j} \right) P_{i\,i+1}^{(\nu)} \qquad (3.34)$$

where $P^{(\nu)}$ is a projector onto the subspace with total spin $\nu$ of a bond $(i, i+1)$. The TB Hamiltonian is soluble by Bethe ansatz and $k = 2S$. We performed DMRG calculations of the Hamiltonian (3.34) for $k = 2$ and 3 ($k = 1$ being the spin-1/2 Heisenberg chain) and found the predicted result $\pi v \chi = k/2$ within a considerable error since the susceptibility is renormalized (like for the spin-1/2 Heisenberg chain). For $S > 1/2$ there are no models analogous to the Haldane-Shastry chain which do not have logarithmic corrections and renormalization effects (see below). The central charges are not affected by the renormalization and we obtained the expected values $c = 3/2$ and $c = 9/5$, respectively. Since $k = 2S$ for the TB models, one might ask how we can in general know that we really measure $k/2$ and not just the spin $S$. In fact we can prove that we measure $k/2$: the half–integer spin-$S$ Heisenberg models with nearest-neighbor interactions are all known to be described by the same SU(2)$_1$ WZW model. That is, from the CFT point of view the $S = 1/2, 3/2, 5/2$, etc. Heisenberg models are all equivalent (so $k \neq 2S$). While we have found within DMRG for the $S = 3/2$ TB chain that $\pi v \chi$ is in agreement with $k = 3$, we find for the $S = 3/2$ Heisenberg model $k = 1$ (as well as $c = 1$).

These findings can be further generalized to SU($N$), SP($N$), or SO($N$)-symmetric WZW models; for instance, we expect that $\pi v \chi = k(N^2 - 1)/6$ for SU($N$)$_k$ WZW models [107, 108] but leave this topic for future work.

## B. Numerical Results

### 1. Zero Temperature

One can check the previous analytical results in large-scale numerical simulations using DMRG or QMC. DMRG results have already be presented in Ref. [53] for the spin-1/2 XXZ chain in the critical regime, and Appendix C gives a detailed account of how cumulants of charge fluctuations can be determined in DMRG. Here we show QMC data obtained at zero and finite temperature using the Stochastic Series Expansion (SSE) method [109, 110]. As in DMRG, it is rather straightforward to compute the fluctuations of the magnetization of a subsystem $A$ embedded in a spin chain with PBCs. We choose the simplest partition, cutting a length-$L$ ring into two halves as depicted in Fig. 12, and compute

$$\mathcal{F}(L/2) = \left\langle \left( \sum_{i=1}^{L/2} \hat{S}_i^z \right)^2 \right\rangle - \left\langle \sum_{i=1}^{L/2} \hat{S}_i^z \right\rangle^2. \qquad (3.35)$$

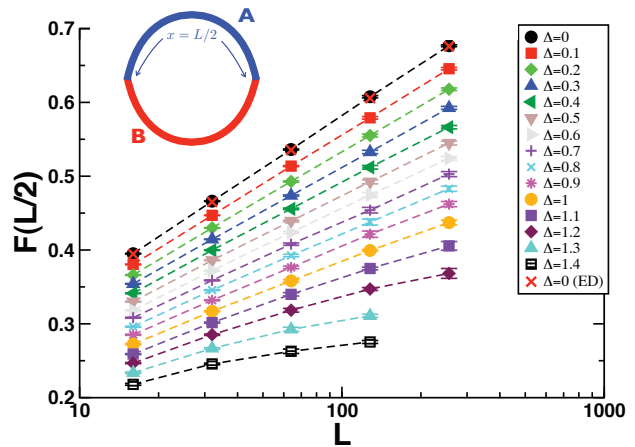

FIG. 12: Quantum Monte Carlo (QMC) results for the fluctuations $\mathcal{F}(L/2)$ versus the total system size $L$ for periodic chains of lengths $L = 16, 32, 64, 128, 256$. Different colored symbols are for different values of the Ising anisotropy of the XXZ chain. For the XX point $\Delta = 0$, QMC data (black circles) are compared to exact diagonalization results (red crosses). Logarithmic scaling is observed in the critical regime $\Delta \leq 1$ whereas for $\Delta > 1$ $\mathcal{F}(L/2)$ tends to saturate at large $L$. Inset: schematic picture for the periodic ring where subsystem A is taken with $x = L/2$ sites.

QMC results for the ground-state expectation value $\mathcal{F}(L/2)$ obtained using $T \to 0$ extrapolations (see below) are shown in Fig. 12 for various anisotropies $\Delta$ and sizes $L = 16, 32, 64, 128, 256$. For $\Delta = 0$ we successfully compared QMC results with exact diagonalization performed using the free-fermion representation. In the entire antiferromagnetic critical regime $\Delta \in [0, 1]$ the scaling of $\mathcal{F}$ is logarithmic, while $\mathcal{F}$ tends to saturate once $\Delta \geq 1$, as clearly visible in Fig. 12. Assuming a logarithmic scaling of the form

$$\pi^2 \mathcal{F}(L/2) = K_{\mathrm{eff}} \ln L + f_1 \qquad (3.36)$$

where $f_1$ is a non-universal constant and $K_{\mathrm{eff}}(L)$ is the effective (size-dependent) LL parameter used to fit the

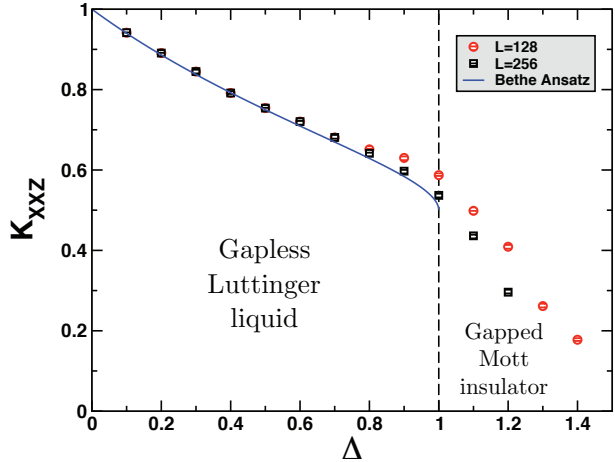

FIG. 13: Prefactor $K_{\text{eff}}$ of the logarithmic scaling Eq. (3.36) for the fluctuations of the 1D XXZ chain computed with QMC, see Fig. 12. $K_{\text{eff}}$ was extracted by fitting QMC data for $L \in [2^{p-1}, 2^p]$ with $L = 128$ ($p = 7$) and $L = 256$ ($p = 8$). The blue line shows the exact Bethe Ansatz result for $K(\Delta)$ Eq. (3.22).

QMC data for $L \in [2^{p-1}, 2^p]$. Results for $K_{\text{eff}}$ as a function of $\Delta$ are shown in Fig. 13 for $L = 128$ ($p = 7$) and $L = 256$ ($p = 8$), and are compared with the Bethe Ansatz result in Eq. (3.22). There we see three important features: (1) In the gapped regime ($\Delta > 1$) the logarithmic scaling is of course not present and we find that the effective LL parameter rapidly vanishes with $L$. (2) For $\Delta < 0.8$, the finite-size effects on $K_{\text{eff}}$ appear to be very small since numerical estimates and Bethe Ansatz results almost coincide. This turns out to be natural because we know from the renormalization group (RG) equations [111] that $K_{\text{eff}}(L)$ goes to its infinite value $K$ as $L^{-\alpha}$ with exponent [112] $\alpha = 8(K - 1/2)$ which becomes less than 1 if $K < 5/8$, i.e., $\Delta \gtrsim 0.9$. (3) At the isotropic Heisenberg point $\Delta = 1$, a marginal operator in the effective field theory induces finite size logarithmic corrections to the prefactor $K_{\text{eff}}$ which makes the convergence to the thermodynamic limit $K = 1/2$ slower. The same fixed point without such finite-size effects, the Haldane-Shastry model, has already been discussed in the previous section.

2. *Finite-Temperature QMC Results*

Fluctuations at finite temperature can be compared to the Curie constant of the uniform susceptibility,

$$C(T) = T\chi(T), \quad (3.37)$$

where the uniform susceptibility is defined as usual by

$$\chi(T) = \left(\frac{\partial m_z}{\partial h}\right)_{h=0} \quad (3.38)$$

$$= \frac{1}{LT}\left(\sum_{i,j=1}^{L} \langle \hat{S}_i^z \hat{S}_j^z \rangle - \sum_{i=1}^{L} \langle \hat{S}_i^z \rangle^2\right). \quad (3.39)$$

The fluctuations within a subsystem $A$ containing $\ell$ spins reads

$$\mathcal{F}(\ell) = \sum_{i,j=1}^{\ell} \langle \hat{S}_i^z \hat{S}_j^z \rangle - \sum_{i=1}^{\ell} \langle \hat{S}_i^z \rangle^2, \quad (3.40)$$

which can be rewritten as

$$\mathcal{F}(\ell, T) = \ell C(T) - \sum_{i \in A}\sum_{j \in B} \langle \hat{S}_i^z \hat{S}_j^z \rangle. \quad (3.41)$$

The above expression contains useful information. At high temperatures, the second term on the righthand side of Eq. (3.41) vanishes because there are no correlations between distant spins. In the limit $T \to \infty$, therefore, the fluctuations $\mathcal{F}$ coincide with $\ell C$, as can be seen in Fig. 14 where finite-$T$ QMC results are presented for $\mathcal{F}(L/2, T)$ and $\chi(T)$. More precisely, the main panel of Fig. 14 shows QMC results for $\mathcal{F}(L/2, T)$ and $L/2 \times C(T)$ obtained for isotropic Heisenberg rings of various lengths $L = 32, 64, 128, 256$. The extensive Curie constant $L/2 \times C(T)$ starts to deviate from $\mathcal{F}(L/2, T)$ for $T \sim J$. Below this energy scale $C$ goes to zero, as expected for an antiferromagnet, and $\mathcal{F}$ decreases smoothly down to very low temperatures where it saturates to its ground-state expectation value (3.36) when the temperature $T$ becomes less than the finite-size gap $\Delta(L) \sim J/L$. To be more quantitative, let us define the "quantum contribution"

$$\mathcal{F}_Q(\ell, T) = \mathcal{F}(\ell, T) - \ell C(T) \quad (3.42)$$

$$= -\sum_{i \in A}\sum_{j \in B} \langle \hat{S}_i^z \hat{S}_j^z \rangle \quad (3.43)$$

which measures the quantum part in the fluctuations since thermal contributions are subtracted. Of course $\mathcal{F}_Q$ and $\mathcal{F}$ coincide at $T = 0$, as visible in Fig. 14. At high temperature, correlations between $A$ and $B$ are very short-ranged, being governed essentially by $\xi(T) \sim 1/T$. In this limit, we therefore expect

$$\mathcal{F}_Q(\ell, T) \sim \frac{J}{T}, \quad T \to \infty, \quad (3.44)$$

which is indeed observed numerically as shown in the inset of Fig. 14. When $T$ decreases, $\mathcal{F}_Q(L/2, T)$ increases logarithmically:

$$\mathcal{F}_Q(L/2, T) = -\frac{1}{2\pi^2}\ln\frac{T}{J} + \text{const.} \quad (3.45)$$

up to saturation when the temperature becomes smaller than the finite size gap $\Delta(L) \sim J/L$. This will be further discussed in the experimental section below.

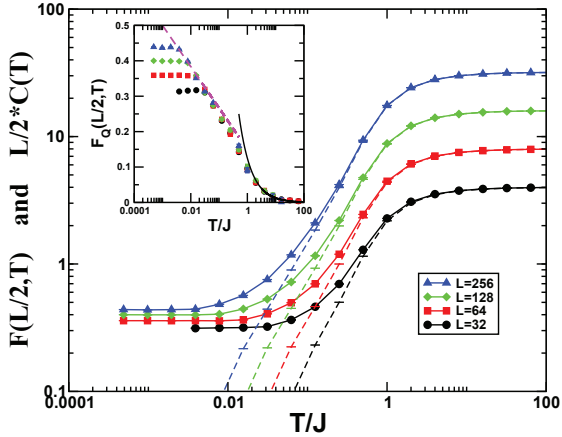

FIG. 14: Finite temperature QMC results for SU(2) Heisenberg rings. Main panel: $\mathcal{F}(L/2,T)$ vs. $T/J$ for various lengths $L = 32, 64, 128, 256$ (symbols) plotted together with the subsystem Curie constant $L/2 \times C(T)$ (dashed lines) over a wide temperature range. Inset: Difference $\mathcal{F}_Q(L/2,T)$ from Eq. (3.43) plotted versus $T/J$. The dashed pink line is $\gamma - (\ln T)/(2\pi^2)$ with $\gamma = 0.15$, and the full black line is the high-$T$ limit $\simeq 0.125/T$.

### C. Gapped Systems

Until now we have only considered gapless models where we could find universal results for the fluctuations. Gapped models are, in contrast, non-universal since a new length scale is introduced by the gap. Nevertheless, gapped models possess interesting aspects from an entanglement point of view. In gapped one-dimensional systems the entanglement entropy obeys a strict area law [113], and we expect the same for the fluctuations due to the finite correlation length and exponentially decaying correlation functions. In one dimension, the scaling behavior of a strict area law is a constant since the region at the cut consists of a single point (or two for PBCs). For a large enough subsystem $A$ ("large" being governed by the size of the gap) it is therefore sufficient to consider the cut at the boundary between both subsystems.

As a paradigm for gapped systems we consider the spin-$S$ Affleck-Kennedy-Lieb-Tasaki (AKLT) chain [114] because we can evaluate all the quantities rigorously. The AKLT states, also called valence bond solid (VBS), serve as model wave functions for Haldane-gap spin chains [115]. Most recently, the AKLT models have attracted some interest since they can be seen as topologically non-trivial states of matter [116, 117]. The spin $S$ AKLT wave function can be most conveniently written as

$$|\psi_S\rangle = \prod_i \left(a_i^\dagger b_{i+1}^\dagger - b_i^\dagger a_{i+1}^\dagger\right)^S |0\rangle \qquad (3.46)$$

using Schwinger bosons [118, 119]: $|\uparrow\rangle = a^\dagger |0\rangle$,

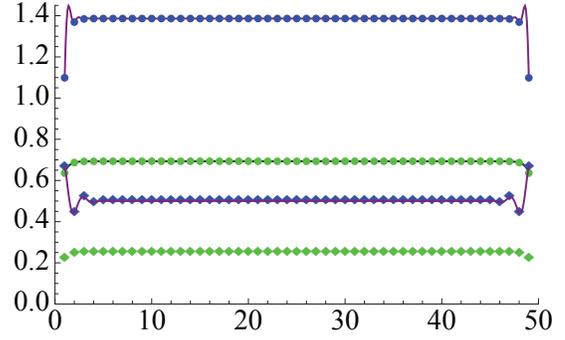

FIG. 15: (Color online) Fluctuations (diamonds) and entanglement entropy (dots) for the spin 1 AKLT chain with $L = 50$. Blue (green) data points correspond to PBCs (OBCs). The curve fitting the EE for PBCs (blue dots) corresponds to Eq. (3.48), the black curve fitting the EE for OBCs (green dots) is the analytic expression Eq. (3.47). The purple line fitting the fluctuations for PBC (blue diamonds) corresponds to Eq. (3.51). Green data points are obtained for OBCs and $S^z_{\text{tot}} = 1$, thus providing an example for an AKLT state with nonzero magnetization. The bulk values of the curves are (from top to bottom) $2\ln 2$, $\ln 2$, $1/2$, and $1/4$.

$|\downarrow\rangle = b^\dagger |0\rangle$ while the spin 1 states are defined as $|1,1\rangle = 1/\sqrt{2} (a^\dagger)^2 |0\rangle$, $|1,0\rangle = a^\dagger b^\dagger |0\rangle$, and $|1,-1\rangle = 1/\sqrt{2}(b^\dagger)^2 |0\rangle$. A singlet bond reads $(a_i^\dagger b_j^\dagger - b_i^\dagger a_j^\dagger)|0\rangle$. If we perform the partitioning of the system into $A$ and $B$, the cut inbetween is nothing than an intersected singlet bond. The entanglement entropy of an intersected singlet bond is $\ln 2$, the fluctuations are $1/4$. Thus we might expect the entanglement entropy (fluctuations) to saturate to the value $\ln 2$ ($1/4$) for OBC and to the value $2\ln 2$ ($1/2$) for PBC. In the following, we will see that for OBC this guess is too naive. For clarity we restrict the discussion for the moment to the spin 1 case. For PBC the AKLT state is a spin singlet, for OBC, however, free edge spins (called *dangling* spins) appear [114]. The dangling spins with $S = 1/2$ couple either into a singlet or a triplet, yielding a four-fold degenerate ground state. The intersected valence bond at the cut contributes always $\ln 2$ ($1/4$) to the entanglement entropy (fluctuations). In addition, there might be a contribution from the physical boundary (i.e., the other edge of subsystem $A$) if the dangling spins form a $S^z_{\text{tot}} = 0$ state, i.e., either $(|\uparrow\downarrow\rangle - |\downarrow\uparrow\rangle)/\sqrt{2}$ or $(|\uparrow\downarrow\rangle + |\downarrow\uparrow\rangle)/\sqrt{2}$. For the channels with finite magnetization $S^z_{\text{tot}} = \pm 1$, i.e., either $|\uparrow\uparrow\rangle$ or $|\downarrow\downarrow\rangle$, the physical edges do not contribute to either entanglement entropy or fluctuations.

Exact expressions for the von Neumann entropy have already been derived [120–123]. For OBCs the entanglement entropy is given by

$$\mathcal{S}^{\text{OBC}}(\ell) = \ln 2 - \frac{e^{-2\ell/\xi}}{2} \qquad (3.47)$$

where $\xi^{-1} = \ln 3$ is the correlation length of the AKLT state. This expression, the black curve in Fig. 15, per-

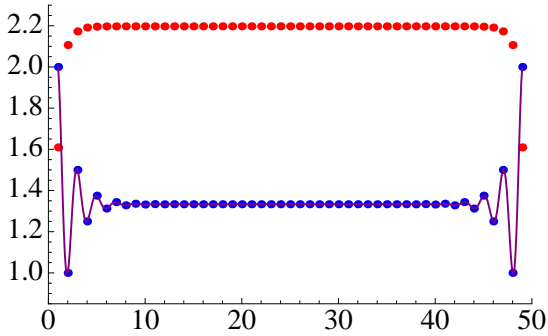

FIG. 16: (Color online) Fluctuations (blue dots, bottom) and entanglement entropy (red points, top) for the spin 2 AKLT chain. The purple line corresponds to Eq. (3.51), all data points obtained within DMRG ($L = 50$ and PBC). $\mathcal{F}$ saturates to $4/3$ and $\mathcal{S}$ to $2\ln 3$.

fectly matches the DMRG data (green dots). For PBCs, the von Neumann entropy is given by [121, 122]

$$\mathcal{S}^{\text{PBC}}(\ell) = -3\lambda_A \ln \lambda_A - \lambda_B \ln \lambda_B ,\quad (3.48)$$

where $\lambda_A = \frac{1}{4}\frac{(1-p^\ell)(1-p^{L-\ell})}{1-p^{L-1}}$ and $\lambda_B = \frac{1}{4}\frac{(1+3p^\ell)(1+3p^{L-\ell})}{1-p^{L-1}}$; $L$ is the system size and $p = -1/3$. Again, the agreement between this anaytical expression and the DMRG data is excellent (see top curve in Fig. 15). In what follows we compute the fluctuations for general spin $S$. We start with the static spin-spin correlation function [118, 124]

$$\langle \hat{S}_i^z \hat{S}_j^z \rangle - \langle \hat{S}_i^z \rangle \langle \hat{S}_j^z \rangle = \frac{(S+1)^2}{3}(-1)^{i-j}e^{-|i-j|/\xi} \quad (3.49)$$

with correlation length $\xi^{-1} = \ln(1+2/S)$. The sum over $i,j$ can be carried out by writing

$$\mathcal{F}_{\text{AKLT}}(\ell) = \frac{S(S+1)}{3}\ell + 2\sum_{k=1}^{\ell}(\ell-k)(-e^{-1/\xi})^k , \quad (3.50)$$

which is a geometric series and therefore easily summed:

$$\mathcal{F}_{\text{AKLT}}(\ell) = \frac{S(S+2)}{6}[1-(-1)^\ell e^{-\ell/\xi}]. \quad (3.51)$$

The last term in brackets vanishes for $\ell \to \infty$, so the fluctuations reach a constant. In Fig. 15, $\mathcal{F}_{\text{AKLT}}(\ell)$ is plotted together with the DMRG data. We still find a constant ratio $\mathcal{S}/\mathcal{F} = 4\ln 2$ for the spin-1 VBS state. The ratio depends purely on the cuts (intersected valence bonds) rather than on microscopic details. Remarkably, the ratio is the same for PBCs and OBCs as it is the case for conformally invariant models.

For the generalization to higher-spin AKLT chains one deals with $S$ intersected singlet bonds per cut rather than a single one. The bonds are, however, not independent and it is not simply a factor $S$ which renders the correct entanglement entropy or fluctuations, respectively.

The simplest way is to consider the reduced density matrix, and one will find for the von Neumann entropy $\mathcal{S} = \ln(1+S)$ per cut [122]; similarly, we find that the fluctuations are given by $\mathcal{F} = S(S+2)/12$ per cut in agreement with the asymptotic value of (3.51) as well as our DMRG findings, see Fig. 16.

These exact results might be extended for various generalized AKLT states [119, 125, 126] where the SU(2) symmetry of the electron spin is replaced by SU($n$), SP($n$), or SO($n$) symmetry.

## IV. CONCLUDING REMARKS ON BIPARTITE FLUCTUATIONS

### A. Comparison to Several Proposed Measures of Many-Body Entanglement

As noted in the Introduction, the difficulty of computing the entanglement entropy (1.1) for most many-body systems described by microscopic Hamiltonians has led to the search for alternative measures of entanglement which behave similarly to the entanglement entropy. Most of the measures behave quite similarly in one dimension to the entanglement entropy and to each other, while in higher dimensions they generally differ.

One such quantity is the valence bond (VB) entanglement entropy for SU(2) quantum spin systems [127, 128], which generalizes the idea, first used in the context of random spin chains [74, 86, 87], that for a pure valence bond state whose wave function is a product of disjoint singlet bonds between two spin-1/2s the entanglement entropy of a subsystem is essentially the number of singlet bonds that cross the boundary between the subsystem and the remainder of the system, with each singlet $|\Psi_s\rangle = (|\uparrow_A\downarrow_B\rangle - |\downarrow_A\uparrow_B\rangle)/\sqrt{2}$ contributing $\ln 2$ to the entanglement entropy. For more general wave functions which are superpositions of pure VB states the number of such crossings can be averaged to obtain a well-defined (with certain restrictions [127]) measure of entanglement. In one-dimensional critical systems the scaling of the VB entanglement entropy $\mathcal{S}^{\text{VB}}$ has been shown to be logarithmic with subsystem size, though without the central charge-dependence of the von Neumann entanglement entropy [129]. It is also now clear that in higher dimensions the VB entanglement entropy does not necessarily scale exactly as the von Neumann entanglement entropy; for the spin-1/2 Heisenberg antiferromagnet on the square lattice in two dimensions, for example, the von Neumann entanglement entropy obeys a strict area law with an additive logarithmic correction while the VB entanglement entropy contains a multiplicative logarithmic correction [45, 130]. Interestingly, however, the strict area law can be recovered if a modified definition of entropy based on loops of the transition graph is used in the valence-bond basis [131].

For pure VB states the fluctuations can also be understood in a similar manner: for each singlet we can consider the spin fluctuation $\mathcal{F}_A = \langle(\hat{S}_A^z)^2\rangle = 1/4$. Then for a pure VB state with $n$ bonds crossing the boundary, $\mathcal{S} = \mathcal{S}^{\text{VB}} = (4\ln 2)\mathcal{F} = n\ln 2$. This idea is explored further below.

Another suggested measure of entanglement is the logarithmic bipartite fidelity [132] $F = -\ln(|\langle A \cup B|A \otimes B\rangle|^2)$, where the state $|A \cup B\rangle$ is the ground state of a many-body Hamiltonian and $|A \otimes B\rangle = |A\rangle \otimes |B\rangle$ is the ground state of the same Hamiltonian with all interactions between the two subsystems removed. The argument of the logarithm, $|\langle A \cup B|A \otimes B\rangle|^2$, has an interpretation as the probability of finding the ground state energy if the energy is measured just after connecting two previously disconnected subsystems. What makes the logarithmic bipartite fidelity particularly interesting is the fact that not only is the scaling with system size logarithmic in one-dimensional gapless systems but the prefactor is proportional to the central charge of the underlying CFT just as for the entanglement entropy. Thus the appearance of the central charge is not unique to the entanglement entropy. On the other hand, unlike the entanglement entropy the finite-size result cannot be obtained from the infinite result by a simple substitution arising from a conformal transformation, and the quantity remains difficult to both measure experimentally and compute in numerical simulations beyond one dimension.

### B. Relation to the Valence Bond Entanglement Entropy

#### 1. Basic Notions

The VB entanglement entropy was introduced as an alternative measure of bipartite entanglement in SU(2) symmetric quantum spin systems [127, 128, 133]. While conceptually distinct from von Neumann and Rényi entropies [43, 45], VB entanglement entropy (and alternative definitions in the VB basis [131]) has been shown to be an efficient entanglement witness and a powerful tool to detect quantum phase transitions [127, 134] in SU(2) spin systems. Among several interesting features, the VB entanglement entropy was computed analytically for the critical spin-1/2 Heisenberg chain in Ref. [129] by Jacobsen and Saleur, who found a logarithmic scaling of the form

$$\mathcal{S}^{\text{VB}}_{\text{1DHAF}}(x) = \frac{4\ln 2}{\pi^2}\ln x + r_1, \quad (4.1)$$

where $r_1$ is a constant. Such a scaling turns out to be accidentally close to the CFT result for the entanglement entropy [7] where the prefactor $1/3$ is similar to the $4(\ln 2)/\pi^2 \simeq 0.28$ above. In two dimensions, the Néel-ordered spin-1/2 antiferromagnetic Heisenberg model was shown to exhibit a log-modified area law [45, 127, 128]. If an enforced bond dimerization leads to a gapped VB solid state, a strict area law was found instead:

$$\mathcal{S}^{\text{VB}}_{\text{2DHAF}}(x) \sim \begin{cases} ax\ln x + bx & \text{(Néel)}, \\ b'x & \text{(VBS)}. \end{cases} \quad (4.2)$$

If $x$ denotes the perimeter of the boundary between the two subsystems, QMC estimates are $a \simeq 0.1$ and $b \simeq 0.02$ [127] for the Heisenberg antiferromagnet on the square lattice.

For a single VB state $|\varphi\rangle$, each virtual bond $b = [ij]$ connecting two lattice sites $i$ and $j$ can be assigned by its occupation number $n_b$ such that

$$n_b(\varphi) = \begin{cases} 1 & \text{if } b \text{ is occupied by a singlet}, \\ 0 & \text{otherwise}. \end{cases} \quad (4.3)$$

Then the VB entanglement entropy for such a state $|\varphi\rangle$ between two subsystems $A$ and $B$ is defined as follows

$$\mathcal{S}^{\text{VB}}(\varphi) = \ln 2 \times \sum_{b_A} n_{\text{b}}(\varphi), \quad (4.4)$$

where the sum is performed over *all* the bonds $b_A$ connecting the two subsystems $A$ and $B$. Of course, only the occupied bonds will contribute 1. This obviously puts an upper bound on $\mathcal{S}^{\text{VB}}$ for a subsystem of $N$ sites: $\mathcal{S}^{\text{VB}}_{\max} = N\ln 2$.

#### 2. Links with $\mathcal{F}$

For a SU(2)-symmetric system, if the singlet ground state is a *unique* dimer covering $|\Phi_0\rangle = |\varphi\rangle$, the two-point correlation is simply given by

$$\langle\Phi_0|\hat{S}_i^z\hat{S}_j^z|\Phi_0\rangle = -\frac{1}{4}n_{[ij]}. \quad (4.5)$$

Using the fact that at zero temperature (see Eq. 3.41)

$$\mathcal{F} = -\sum_{i\in A}\sum_{j\in B}\langle\hat{S}_i^z\hat{S}_j^z\rangle \quad (4.6)$$

we have the simple result

$$\mathcal{F} = \frac{\mathcal{S}^{\text{VB}}}{4\ln 2}. \quad (4.7)$$

In the random singlet state where $\mathcal{S}^{\text{vN}}(x) = \mathcal{S}^{\text{VB}}(x) = [(\ln 2)/3]\ln x$, it yields $\mathcal{F}(x) = (1/12)\ln x$.

However, a general singlet state can be a complicated superposition involving a large number of dimer coverings, such that one can expand any singlet state on a (overcomplete) valence bond basis $\{|\varphi_i\rangle\}$:

$$|\Phi_0\rangle = \sum_p \lambda_p|\varphi_p\rangle. \quad (4.8)$$

In such a case the VB entanglement entropy has been defined in Refs. [127, 128] as

$$\mathcal{S}^{\text{VB}}(\Phi_0) = \frac{\sum_p \lambda_p \mathcal{S}^{\text{VB}}(\varphi_p)}{\sum_p \lambda_p}. \quad (4.9)$$

From such a definition, one sees immediately that the equality (4.7) is not valid anymore. Indeed, if one defines

$$\mathcal{F}(\varphi_p, \varphi_{p'}) = -\sum_{i \in A}\sum_{j \in B} \langle \varphi_p | \hat{S}_i^z \hat{S}_j^z | \varphi_{p'} \rangle, \quad (4.10)$$

from Eq. (4.9) we get

$$\mathcal{S}^{\mathrm{VB}}(\Phi_0) = 4\ln 2 \frac{\sum_p \lambda_p \mathcal{F}(\varphi_p, \varphi_p)}{\sum_p \lambda_p}, \quad (4.11)$$

whereas

$$\mathcal{F}(\Phi_0) = \sum_p \sum_{p'} \lambda_p \lambda_{p'} \mathcal{F}(\varphi_p, \varphi_{p'}). \quad (4.12)$$

Clearly we see that, except for trivial cases, a general superposition of the form Eq. (4.8) yields $\mathcal{F} \neq \mathcal{S}^{\mathrm{VB}}/(4\ln 2)$. An illustration based on a simple 4-site example is provided in Appendix D.

Despite their fundamental differences, it is quite interesting to notice that for the Heisenberg chain, when comparing the exact expression of Jacobsen and Saleur [129] Eq. (4.1) for $\mathcal{S}^{\mathrm{VB}}$ with $\mathcal{F}$, one sees that the leading logarithmic term of

$$\mathcal{F}_{\mathrm{1DHAF}}(\ell) = \frac{1}{2\pi^2} \ln \ell + f_1, \quad (4.13)$$

is directly proportional with a factor $8\ln 2$ (to be compared with $4\ln 2$ for a pure valence bond state). However, this coefficient does not hold when moving away from the Heisenberg point along the XXZ critical line [129].

### C. Non-Equivalence of Fluctuations and Entanglement Entropy

As noted in the Introduction, the equivalence of the charge statistics with the entanglement entropy for non-interacting fermions, while suggestive, is particular to the nature of non-interacting particles: here entanglement entropy arises solely from the uncertainty of the position of particles and therefore particle-number fluctuations within a given volume encodes the full information needed to determine the entanglement entropy. We have shown that many essential features are retained in the fluctuations for interacting systems, especially when the theory is essentially gaussian as in one-dimensional Luttinger Liquids. It is worth noting, however, that, even in principle, it is unlikely that there exists a universal series similar to Eq. (2.1) for interacting systems with more complicated coefficients. This can be seen by observing the simple fact that when calculating the fluctuations (or cumulants) from the reduced density matrix, one essentially averages over the eigenvalues in each particle-number sector, so that the individual eigenvalues cannot be discerned. Stated another way, the full counting statistics cannot, in general, contain the same amount of information as the entanglement entropy. The example for DMRG in Appendix C makes this clear.

To see this more explicitly, it is already known that for an ideal Bose gas the relation between entanglement entropy and fluctuations is actually $\mathcal{S} \sim \ln\sqrt{\mathcal{F}}$ [39, 135]. In the absence of particle number conservation in the full system, moreover, the arguments given in the Introduction for the symmetry between subsystems no longer holds and the fluctuations are not expected to behave as the entanglement entropy. For the quantum Ising chain in transverse field, for example, the fluctuations of $\hat{S}^z$ *do not* scale logarithmically [73]. Of course, for this particular case one must instead relate the entanglement entropy to fluctuations of the appropriate *quasi-particles* (see Supplementary Material).

### D. Accessible Entropy and Charge Fluctuations

Charge conservation restricts the possible local operations available when using fermionic modes for quantum information processes purposes. A modification of the usual entanglement entropy, which quantifies the useful, accessible entanglement was proposed in Ref. [136]. There, it was suggested to quantify the accessible entanglement entropy $S_A^{\mathrm{res}}$ by averaging entanglement entropy over super-selection sectors (see also Refs. [79, 137, 138]). In Ref. [139] it was suggested that constraints on local operations may be treated as a resource for hiding information in correlations which are blocked from local probing.

In Ref. [140], charge fluctuations have been used as an estimate of the difference between accessible and inaccessible entropy via the inequality:

$$S_A - \Delta S \leq S_A^{\mathrm{res}} \leq S_A, \quad (4.14)$$

$$\Delta S = \tfrac{1}{2}\log\left[2\pi e\left(C_2 + \tfrac{1}{12}\right)\right] \quad (4.15)$$

where

$$\rho_{n,m} = \frac{1}{p_{n,m}} \Pi_n^A \otimes \Pi_m^B \rho \Pi_n^A \otimes \Pi_m^B, \quad (4.16)$$

$$p_{n,m} = \mathrm{Tr}(\Pi_n^A \otimes \Pi_m^B \rho \Pi_n^A \otimes \Pi_m^B), \quad (4.17)$$

where $\Pi_n^A$ project onto sectors with fixed particle number $n$ in $A$ (i.e., on states $\psi$ in $A$, such that $\hat{N}_A \psi = n\psi$). Similarly, $\Pi_m^B$ projects on sectors with $N_B = m$ in $B$.

An immediate consequence of the examples we have considered in the previous sections is that the $C_2$ scaling can only lead to sub-leading logarithmic (or log log in one dimension) corrections to the accessible entanglement entropy.

### E. Experiments

In principle, bipartite fluctuations are accessible in experiment, which was part of the motivation for studying them in place of entanglement entropy, which remains the quintessential measure of "quantum-ness" in a many-body system. The latter, however, has always suffered from lack of measurability (but see Ref. [42]) despite the great theoretical interest stemming from fundamental results. In this work we have shown how for non-interacting fermions the fluctuations are equivalent to entanglement entropy [36, 40], while for interacting systems we trade non-equivalence for measurability. It remains to show how this can be accomplished in real systems, and below we propose and discuss three different systems where $\mathcal{F}$ might be measured.

### 1. Quantum Point Contacts

As discussed in Sec. II C 1, a quantum point contact (QPC) is a beam splitter with tunable transmission and reflection that serves as a "door" between electron reservoirs [36, 40]. The current fluctuations $\mathcal{F}(t)$ measured at this QPC are exactly our bipartite fluctuations when replacing the temporal window $t$ by the spatial extent $x$. Due to space-time duality, such a replacment is always allowed for conformally invariant sytems [36], even when they are interacting. Although the two situations are not identical, the QPC measurements being more closely related to the concept of local quantum quenches [75, 141], QPCs with free-electron reservoirs offer the opportunity to unambiguously measure entanglement entropy using the formalism of Sec. II. The first few cumulants, in particular $C_2$, have already been successfully measured [58].

### 2. Cold Atoms

The second system where the fluctuations are available are cold atomic quantum gases loaded into optical lattices. To be more specific, we consider the recent experiments where site-resolved images of atoms in opical lattices have been taken for the first time with a *quantum gas microscope* [57, 61, 62, 142]. In these experiments, it was possible to measure the mean and variance (i.e., the fluctuations) of the particle number *per site*. Strictly speaking, current technology allows for the measurement of parity number, which corresponds to particle number only for free fermions and hard-core bosons (in the Tonks regime). It is possible, however, that in future experiments actual particle number at the subsystem level may be measured.

### 3. Quantum Antiferromagnets

Third, bipartite fluctuations can also be measured in quantum magnetic systems, with the $z$-component of spin $\hat{S}^z$ (or the quantization axis, more generally) playing the role of particle number in QPCs and cold atom systems [143]. We note that Eq. (3.23) essentially defines the Curie constant of the partial uniform susceptibility, $\mathcal{F}_A = T \times d\langle S^z_A \rangle / dh_A|_{h_A \to 0}$ where $h_A$ is a small external uniform magnetic field applied only to region $A$. Such a setup could be realized by applying the magnetic field over the entire sample and covering the region $B$ (the remainder of the system) with superconducting Meissner screens [130]. Such screens would eliminate the external field as well as the magnetic response outside the region $A$. By varying the size of $A$, the scaling of $\mathcal{F}_A$ could be measured. A setup is sketched in Fig. 17

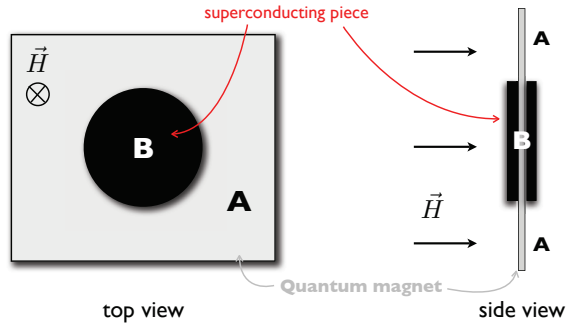

FIG. 17: Experimental setup proposed to extract the fluctuation $\mathcal{F}_A$ of the magnetization within a subregion $A$. A superconducting device (black disk) is placed on top of the quantum antiferromagnet (on region $B$) so that the Meissner effect will induce $h_B = 0$ in the antiferromagnet whereas $h_A = H$. The superconducting device is also placed on the other side such that only the $A$ contribution of the field-induced magnetization will be measured.

### F. Conclusion

We have presented a detailed study of the behavior of bipartite fluctuations of conserved U(1) charges such as particle number $\hat{N}$ and spin $\hat{S}^z$ in various quantum many-body systems and proposed that the study of such fluctuations is a useful tool for characterizing such systems. For systems that can be mapped to a problem of non-interacting fermions we have shown that the fluctuations and higher-order cumulants encode all of the information needed to determine the commonly used measures of entanglement, the von Neumann entanglement entropy and Rényi entanglement entropies. As we have shown, the full entanglement spectrum can also be determined in this manner, demonstrating for the first time that, in principle, the entanglement spectrum is an accessible quantity. In other systems the behaviors of the fluctuations and entanglement entropy differ, but the fluctuations still reveal important features of the system while having the advantage of being tractable analytically in many cases, computable numerically in both DMRG and QMC, and in principle experimentally accessible. We therefore hope that this work will motivate further work into understanding more fully the behavior of the fluctuations as well as their relation to the entanglement entropy in interacting systems in one and higher dimensions.

### Acknowledgements

The work by HFS and KLH was supported by NSF Grant No. DMR-0803200, the Yale Center for Quantum Information Physics (DMR-0653377). KLH also acknowledges support from DOE Grant DE-FG02-08ER46541. HFS acknowledges support from the facilities and staff of the Yale University Faculty of Arts and Sciences High Performance Computing Center. SR acknowledges support from the Deutsche Forschungsgemeinschaft under Grant No. RA 1949/1-1 and partially from NSF DMR-0803200, and thanks Peter Schmitteckert for many discussions and use of his DMRG code. The work by CF was supported by the Swiss National Science Foundation. I. K. acknowledges financial support from NSF Grant No. DMR-0956053. IK and KLH thank the Aspen Center for Physics for its kind hospitality. NL acknowledges M. Mambrini and J. Jacobsen for discussions regarding some subtleties on valence bond physics.

### APPENDIX A: DERIVATION OF THE RÉNYI ENTANGLEMENT ENTROPY SERIES

In this appendix we derive Eq. (2.3) following similar steps to the derivation for the von Neumann entanglement entropy in Sec. II A. As shown in the Supplementary Material, the Rényi entropies can be written in the form

$$\mathcal{S}_n = \frac{1}{1-n}\mathrm{Tr}\{\ln[M^n + (1-M)^n]\}. \tag{A1}$$

We first expand the logarithm in Eq. (A1) as

$$\ln[M^n + (1-M)^n] = -\sum_{j=1}^{\infty} \frac{[(1-M^n) - (1-M)^n]^j}{j}. \tag{A2}$$

By repeated use of the binomial theorem, we obtain

$$\mathrm{Tr}\{\ln[M^n + (1-M)^n]\} = -\sum_{j=1}^{\infty}\sum_{m=0}^{j}\sum_{r=0}^{j-m}\sum_{s=0}^{nm}(-1)^{m+r+s}\frac{1}{j}\binom{j}{m}\binom{j-m}{r}\binom{nm}{s}\mathrm{Tr}(M^{nr+s}). \tag{A3}$$

Before using Eq. (2.16) to write this in terms of the factorial cumulants, we must show that $F_0$ does not contribute, i.e., that the expression vanishes when $nr + s = 0$. Since $n > 1$ and the sum over $m$ vanishes when $r = s = 0$ this is indeed the case. Then we can substitute Eq. (2.16), and moreover use relation between factorial and ordinary cumulants (see Supplementary Material) to get

$$\mathrm{Tr}\{\ln[M^n + (1-M)^n]\} = -\sum_{j=1}^{\infty}\sum_{m=0}^{j}\sum_{r=0}^{j-m}\sum_{s=0}^{nm}\sum_{w=0}^{nr+s}(-1)^{m+r+s+w}\frac{1}{j}\binom{j}{m}\binom{j-m}{r}\binom{nm}{s}\frac{S_1(nr+s,w)}{(nr+s-1)!}C_w. \tag{A4}$$

It is now possible to introduce a cutoff $R$ in the outermost sum and write $\mathrm{Tr}\{\ln[M^n + (1-M)^n]\} = \sum_{k=1}^{2nR}\beta_k(n,R)C_k$, where, after some algebra,

$$\beta_k(n,R) = \frac{1}{1-n}\sum_{r=1}^{R}\sum_{m=0}^{r}\sum_{s=k}^{nr}(-1)^{r+s+nr+nm+k}\frac{1}{r}\binom{R}{r}\binom{r}{m}\binom{nm}{nr-s}\frac{S_1(s,k)}{(s-1)!}. \tag{A5}$$

The coefficients $\beta_k(n, R)$ actually vanish for $k > nR$, so we finally arrive at Eq. (2.4). As a check on the algebra, it is again found that for $R \to \infty$ the coefficients $\beta_k(n, R)$ approach the limiting values derived in the Supplementary Material, although this time only a numerical demonstration is possible.

### APPENDIX B: CONVERGENCE OF THE ENTANGLEMENT ENTROPY SERIES

In this appendix we prove the convergence of series (2.1) for the entanglement entropy, which is relatively simple to show under the assumption that $M$ is finite. In the derivation of Eq. (2.23) the cutoff $K$ originates from the expansion of Eq. (2.9) and hence convergence in the limit $K \to \infty$ corresponds to convergence of the partial sums of Eq. (2.12), i.e., that Eq. (2.12) converges to the correct sum for the $M$ of interest. Now, let us first consider the case where $M$ is a single number $p$ such that $0 < p < 1$. This is the property satisfied by the eigenvalues of $M$ for general $M$ for fermionic systems due to Fermi-Dirac statistics (strictly speaking, $0 \leq p \leq 1$, but at the extremes we understand the entanglement entropy to be zero). The series

$$\mathcal{S} = \sum_{k=1}^{\infty} \frac{p(1-p)^k + p^k(1-p)}{k} \quad \text{(B1)}$$

is absolutely convegent by the ratio test, since $\lim_{k \to \infty} |a_{k+1}/a_k|$ is

$$\lim_{k \to \infty} \frac{[p(1-p)^{k+1} + p^{k+1}(1-p)]/(k+1)}{[p(1-p)^k + p^k(1-p)]/k} = \begin{cases} 1-p & \text{if } 0 < p \leq 1/2, \\ p & \text{if } 1/2 < p < 1. \end{cases} \quad \text{(B2)}$$

In both cases the limit is less than 1 and the series therefore converges. Moreover, each term is positive and by necessity decreasing so that the partial sums approach its limit from below.

If Eq. (2.9) is evaluated in the eigenbasis of $M$ then we can apply the above argument to each eigenvalue, but in general $M$ is a matrix with off-diagonal elements. From the theory of matrices, however, it is known that Eq. (B1) converges for $p$ replaced by a general matrix $M$ if $0 < \sigma(M) < 1$ where $\sigma(M)$ is the spectral radius or maximum eigenvalue of $M$. Therefore the matrix series converges as well.

We can similarly show that Eq. (2.3) for the Rényi entropies converges. In this case the cutoff $R$ arises from the series expansion (A2). Again let us assume first that $M$ consists of the single entry $p$, with $0 < p < 1$. We have

$$\lim_{j \to \infty} \left| \frac{a_{j+1}}{a_j} \right| = |(1-p^n) - (1-p)^n| < 1 \quad \text{(B3)}$$

where $p^n + (1-p)^n$ is positive and $p^n + (1-p)^n < p + (1-p) = 1$ for $n > 1$, so that the series is again absolutely convergent. Application of the matrix spectral radius to $M$ then shows that Eq. (2.3) also converges, and recalling the factor of $1 - n$ left out in Eq. (A2) we see that each term of the series for the Rényi entropies is positive and decreasing so that increasing $K$ improves the approximation from below as for the entanglement entropy.

## APPENDIX C: CUMULANTS IN DMRG

In this appendix we describe how to compute the cumulants of a conserved U(1) charge within DMRG; we show that the fluctuations and higher-order cumulants can be computed with no additional computational effort relative to existing implementations. One can of course do this by computing all possible two-point correlators $\langle \hat{x}_i \hat{x}_j \rangle - \langle \hat{x}_i \rangle \langle \hat{x}_j \rangle$, where $\hat{x} = \hat{S}^z$ or $\hat{x} = \hat{n}$, respectively, inside subsystem $A$; by performing the sum $\mathcal{F}_A = \sum_{i,j \in A} \langle \hat{x}_i \hat{x}_j \rangle - \langle \hat{x}_i \rangle \langle \hat{x}_j \rangle$ one obtains the fluctuations in $A$. It is, however, costly to compute the necessary correlators within DMRG when the system size becomes large. We therefore present a much more elegant way to get the fluctuations directly without any need for computing the correlators: the fluctuations, and indeed all higher-order cumulants, can be extracted simply from the reduced density matrix of the subsystem in the usual DMRG implementation with no additional computational effort.

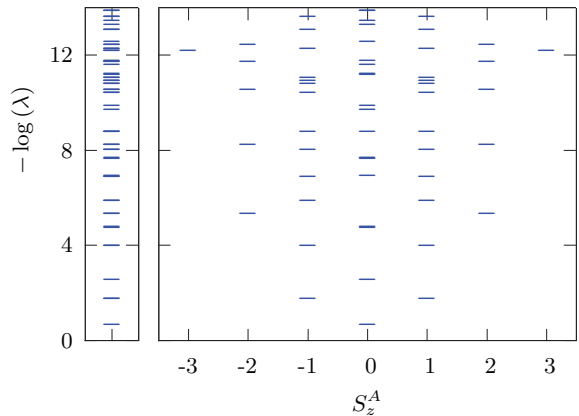

FIG. 18: Eigenvalues $\lambda_i$ of the reduced density matrix for a subsystem $A$, obtained within DMRG. The data is produced for an XXZ spin chain, Eq. (3.20), with $\Delta = 0.2$ and $L = 32$. (Left) Negative logarithm of the eigenvalues, $-\ln \lambda_i$, i.e., the real-space entanglement spectrum. (Right) The same data vs. the spin quantum number $S_z^A$ for subsystem $A$. As expected, the SU(2) multiplets are not present due to broken SU(2) symmetry for $\Delta = 0.2$.

In each DMRG step the full system is split into two blocks, $A$ and $B$, with the reduced density matrices $\rho_A$ and $\rho_B$, respectively. Without loss of generality, we can use the eigenvalues $\lambda_i$ of $\rho_A$ to compute the von Neumann entropy in the Schmidt basis,

$$\mathcal{S}_A = -\sum_i \lambda_i \ln \lambda_i, \quad \text{(C1)}$$

so that the entanglement entropy is a natural byproduct of DMRG, since all $\lambda_i$'s are computed in every DMRG step.

As mentioned in the Introduction, the reduced density matrix is usually stored in block-diagonal format according to the U(1) quantum number (i.e., either the particle number $n$ or the spin projection $S_z$) [38]. In other words, the eigenvalues $\lambda_i$ are in fact labeled and we should write $\lambda_i^{(x)}$ where $x$ is the U(1) quantum number. Now we can write the fluctuations in the Schmidt-basis as

$$\mathcal{F}_A = \sum_x x^2 {\sum_i}' \lambda_i^{(x)} - \left[\sum_x x {\sum_i}' \lambda_i^{(x)}\right]^2 \quad \text{(C2)}$$

where the sum $\sum'$ is restricted to a fixed value of the quantum number $x$. Higher moments are computed analogously while the cumulants are determined from the moments via well-known formulas.

The reader should notice that we refer here to $x$ (e.g., $S_z$) of block $A$ (or $B$) but not to the total $x$ of the whole system. To avoid confusion one could write $x^A$ instead of $x$. Eq. (C2) implemented in the DMRG code provides an efficient way to compute the fluctuations. We present a representative example in Fig. 18 for the eigenvalues $\lambda_i$ of an XXZ spin chain, Eq. (3.20) with $\Delta = 0.2$ and $L = 32$ where the bipartition is made at $L/2$. It is clear that the fluctuations arise quite naturally in DMRG computations.

### APPENDIX D: ILLUSTRATION OF THE DIFFERENCE BETWEEN $\mathcal{F}$ AND $\mathcal{S}^{\text{VB}}$

In order to illustrate the fundamental differences for a SU(2) symmetric model between the fluctuation $\mathcal{F}$ and the VB entanglement entropy $\mathcal{S}^{\text{VB}}$, let us take this simple normalized 4-site singlet state

$$|\Phi_0\rangle = \frac{1}{\sqrt{3}}\left(|\frown \quad \frown\rangle + |\overset{\frown}{\phantom{x}\frown}\rangle\right). \quad \text{(D1)}$$

Here a dimer $\frown$ connects two sites belonging to different sublattices A and B, and is written with the following convention $|\frown\rangle = |\uparrow_A\downarrow_B - \downarrow_A\uparrow_B\rangle/\sqrt{2}$.

The bipartite entanglement shared across a vertical boundary drawn in the middle of the 4 site system is then easy to compute. Regarding VB entanglement entropy, $|\frown | \frown\rangle$ contributes $0 \times \ln 2$ and $|\overset{\frown}{\phantom{x}\frown}\rangle$ contributes $2 \times \ln 2$. Using the definition Eq. (4.9), we simply get

$$\mathcal{S}^{\text{VB}} = \frac{0 + 2\ln(2)/\sqrt{3}}{1/\sqrt{3} + 1/\sqrt{3}} = \ln 2,$$

whereas computing the fluctuation yields

$$\mathcal{F} = \frac{\mathcal{S}^{\text{VB}}}{6\ln 2}, \quad \text{(D2)}$$

which is different from Eq. (4.7).

---

# Supplementary Material for "Bipartite Fluctuations as a Probe of Many-Body Entanglement"

H. Francis Song,[1] Stephan Rachel,[1] Christian Flindt,[2]
Israel Klich,[3] Nicolas Laflorencie,[4] and Karyn Le Hur[1]

[1]*Department of Physics, Yale University, New Haven, CT 06520*
[2]*Département de Physique Théorique, Université de Genève, CH-1211 Genève, Switzerland*
[3]*Department of Physics, University of Virginia, Charlottesville, VA 22904*
[4]*Laboratoire de Physique Théorique, Université de Toulouse, UPS, (IRSAMC), Toulouse, France*
(Dated: September 2, 2011)

In this Supplementary Material we provide detailed calculations for several of the results presented in the main text. These technical details are offered to make the discussions in the main text more transparent and allow interested readers to follow each step of the computations.

### Contents

## I. THE REDUCED DENSITY MATRIX OF NON-INTERACTING FERMIONS IN EQUILIBRIUM

In this section we use the special form of the reduced density matrix for non-interacting fermions in equilibrium to derive the following essential formulas for the von Neumann entanglement entropy $\mathcal{S}$, Rényi entanglement entropies $\mathcal{S}_\alpha$, and the generating function $\chi(\lambda)$ for particle number fluctuations

in a system of non-interacting fermions:

$$\mathcal{S} = -\text{Tr}[M_\Omega \ln M_\Omega + (1 - M_\Omega) \ln(1 - M_\Omega)], \tag{1.1}$$

$$\mathcal{S}_\alpha = \frac{1}{1-\alpha} \text{Tr} \ln[M_\Omega^\alpha + (1 - M_\Omega)^\alpha], \tag{1.2}$$

$$\chi(\lambda) = \det[1 + (e^{i\lambda} - 1)M_\Omega]. \tag{1.3}$$

As explained below, $M_\Omega$ is a Hermitian matrix with dimension equal to the number of fermionic modes in the subsystem $\Omega$ of interest. Eqs. (1.1)-(1.3) apply to any system of non-interacting fermions, but are particularly useful when applied to a *subsystem* of a system of non-interacting fermions. All of the results in this section are known in some form, especially from Ref. [1], but we provide a detailed presentation for clarity and to point out relevant features of the derivation along the way.

### A. Without Pairing

We begin with the simpler case where fermion number is conserved. Non-interacting fermions in any dimension are described by a Hamiltonian of the form

$$\hat{H} = \sum_{i,j} \hat{a}_i^\dagger H_{ij} \hat{a}_j, \tag{1.4}$$

where $\hat{a}_i$ is the fermionic annihilation operator for degree of freedom $i$ (usually $i$ labels a site on the lattice, but more generally may carry a spin index) and $H_{ij}$ is a Hermitian matrix. The Hamiltonian can be diagonalized by a unitary matrix $U$ whose columns are the eigenvectors of the matrix $H$ such that $\Lambda = U^\dagger H U$ is diagonal with matrix elements $\Lambda_{kk'} = \Lambda_k \delta_{kk'}$. Then the Hamiltonian can be written in canonical form as

$$\hat{H} = \sum_k \Lambda_k \hat{b}_k^\dagger \hat{b}_k, \tag{1.5}$$

$$\hat{b}_k = \sum_i U_{ki}^\dagger \hat{a}_i. \tag{1.6}$$

At equilibrium the density matrix of the system is

$$\hat{\rho} = \frac{e^{-\beta \hat{H}}}{\text{Tr}\, e^{-\beta \hat{H}}}, \tag{1.7}$$

and in particular the correlation matrix, or Green's function, at temperature $T = 1/\beta$ is given by

$$M_{ij} = \langle \hat{a}_j^\dagger \hat{a}_i \rangle = \text{Tr}(\hat{\rho} \hat{a}_j^\dagger \hat{a}_i) = \sum_k U_{ik} f(\beta \Lambda_k) U_{kj}^\dagger, \tag{1.8}$$

where

$$f(x) = \frac{1}{e^x + 1} \tag{1.9}$$

is the Fermi occupation function. More simply, we may write

$$M = f(\beta H). \tag{1.10}$$

Clearly $M$ is hermitian. Although we will almost always be concerned with the zero-temperature ground state, i.e., $f(\beta \Lambda_k) = \theta(-\Lambda_k)$, in this section we will keep the discussion general. This will also allow us to briefly explore the behavior of the entanglement entropy and fluctuations at finite temperature in the main text.

Suppose we are interested in a subsystem $\Omega$ of the total system, usually representing a spatial subregion of the lattice. All properties of this subsystem are determined by the reduced density matrix $\hat{\rho}_\Omega$, which may be found using the fact that the reduced density matrix can be written in the thermal form [1]

$$\hat{\rho}_\Omega = \frac{e^{-\beta_\Omega \hat{H}_\Omega}}{\text{Tr } e^{-\beta_\Omega \hat{H}_\Omega}}, \tag{1.11}$$

$$\hat{H}_\Omega = \sum_{i,j \in \Omega} \hat{a}_i^\dagger (H_\Omega)_{ij} \hat{a}_j, \tag{1.12}$$

where the "temperature" $T_\Omega = 1/\beta_\Omega$ of the subsystem can be set to 1 without loss of generality. We note that Eq. (1.11) by itself, without Eq. (1.12), is completely general and applies to *any* system, whether composed of non-interacting fermions or not: it is simply a way to write the positive semi-definite Hermitian density matrix, and the Hamiltonian $\hat{H}_\Omega$ is known in this context as the *entanglement Hamiltonian*. The eigenvalues of $\hat{H}_\Omega$, called the *entanglement spectrum*, have recently been of considerable interest in characterizing topologically-ordered phases [2]. The additional constraint expressed in Eq. (1.12), i.e., that the entanglement Hamiltonian for non-interacting fermions also describes a system of non-interacting fermions, is what makes the following computation possible. Briefly, this is a consequence of Wick's Theorem: since the system is composed of non-interacting fermions and Wick's Theorem must apply for all correlation functions, in particular to correlation functions within subsystem $\Omega$, the reduced density matrix for subsystem $\Omega$ must be of the above form [1].

Since the density matrix described in Eqs. (1.11),(1.12) is identical in form to the original problem, we see that if $u$ plays the role of $U$, $\lambda$ the role of $\Lambda$, and $\hat{c}_q$ the role of $\hat{b}_k$ in the previous calculation, the correlation matrix in subsystem $\Omega$ is given by

$$m_{ij} = \langle \hat{a}_j^\dagger \hat{a}_i \rangle = \text{Tr}(\hat{\rho}_\Omega \hat{a}_j^\dagger \hat{a}_i) = \sum_k u_{ik} f(\lambda_k) u_{kj}^\dagger, \qquad i,j \in \Omega. \tag{1.13}$$

As for $M$, we have that $m = f(h)$. Now, it is clear that $m_{ij}$ must agree with $M_{ij}$ for $i,j$ in subsystem $\Omega$. Moreover, the eigenvalues of $m$ are precisely $f(\lambda_q)$, so that if $\nu_q$ are the eigenvalues of the matrix $M_\Omega$ obtained by restricting (or projecting) $M$ to $A$, then

$$f(\lambda_q) = \nu_q. \tag{1.14}$$

With Eq. (1.14) we can now derive Eqs. (1.1)-(1.3). First, the von Neumann entanglement entropy $\mathcal{S} = -\text{Tr}(\hat{\rho}_\Omega \ln \hat{\rho}_\Omega)$ is simply the ordinary "thermodynamic" entropy of a system of free fermions with energies $\lambda_q$ at inverse temperature $\beta = 1$, which implies

$$\mathcal{S} = \sum_q H_2(f(\lambda_q)) = \sum_q H_2(\nu_q), \tag{1.15}$$

where

$$H_2(x) = -x \ln x - (1-x) \ln(1-x) \tag{1.16}$$

is the binary entropy function. Eq. (1.1) is the matrix form of Eq. (1.15).

The Rényi entropies are usually not considered in thermodynamic applications so we carry out a more explicit computation by writing out the density matrix (1.11) in the basis in which the Hamiltonian is diagonal:

$$\hat{\rho}_\Omega = \prod_q \frac{e^{-\lambda_q \hat{c}_q^\dagger \hat{c}_q}}{1 + e^{-\lambda_q}}. \tag{1.17}$$

In the $q$-occupation basis $|0_q\rangle, |1_q\rangle$ we have explicitly

$$\hat{\rho}_\Omega = \bigotimes_q \begin{pmatrix} 1 - f(\lambda_q) & 0 \\ 0 & f(\lambda_q) \end{pmatrix} = \bigotimes_q \begin{pmatrix} 1 - \nu_q & 0 \\ 0 & \nu_q \end{pmatrix}, \tag{1.18}$$

so that

$$\text{Tr}(\hat{\rho}_\Omega^\alpha) = \prod_q \text{Tr}\begin{pmatrix} (1-\nu_q)^\alpha & 0 \\ 0 & \nu_q^\alpha \end{pmatrix} = \prod_q [\nu_q^\alpha + (1-\nu_q)^\alpha]. \tag{1.19}$$

Therefore

$$\mathcal{S}_\alpha = \frac{1}{1-\alpha} \sum_q \ln[\nu_q^\alpha + (1-\nu_q)^\alpha] \tag{1.20}$$

whose matrix form is Eq. (1.2). We can also confirm that Eq. (1.20) agrees with Eq. (1.15) in the limit $\alpha \to 1$.

The statistics of the number operator in subsystem $\Omega$,

$$\hat{N}_\Omega = \sum_{i \in \Omega} \hat{a}_i^\dagger \hat{a}_i = \sum_q \hat{c}_q^\dagger \hat{c}_q, \tag{1.21}$$

may be computed in a similar fashion. We are interested in the generating function

$$\mathcal{G}(z) = \langle z^{\hat{N}_\Omega} \rangle = \text{Tr}(\hat{\rho}_\Omega z^{\hat{N}_\Omega}), \tag{1.22}$$

from which the standard generating function $\chi(\lambda)$ is obtained by letting $z = e^{i\lambda}$. The factorial generating function $\chi_f(\lambda)$ is similarly given by letting $z = \lambda + 1$. Using the explicit form of the reduced density matrix in Eq. (1.18) we have

$$\mathcal{G}(z) = \text{Tr} \prod_q \frac{e^{-\lambda_q \hat{c}_q^\dagger \hat{c}_q} z^{\hat{c}_q^\dagger \hat{c}_q}}{1 + e^{-\lambda_q}} \tag{1.23}$$

$$= \prod_q \text{Tr}\begin{pmatrix} 1-\nu_q & 0 \\ 0 & \nu_q z \end{pmatrix} \tag{1.24}$$

$$= \prod_q [1 + (z-1)\nu_q], \tag{1.25}$$

which in matrix form is Eq. (1.3) with background charge $Q = 0$ and the substitution $z = e^{i\lambda}$.

It is worth emphasizing that, given the assumptions in Eqs. (1.11),(1.12), the only additional input to the previous computation is the correlation matix of the system restricted to subsystem $\Omega$, i.e., $M_\Omega$. This then determined the fermionic energies $\lambda_q$ in the subsystem, which in turn determine the entanglement entropies and number fluctuations. It is not necessary, in particular, that the correlation function arise from a system of fermions in equilibrium; the procedure for deriving the relevant equations are completely general as long as Wick's theorem applies.

For example, this allows us to apply the same formulas to non-equilibrium problems where the correlation matrix $M$ evolves in time, usually described by a unitary time evolution operator in the space of single-particle modes. The case of a quantum point contact (QPC) was originally described in Ref. [3], though as we shall see, this description used a flawed formula for relating the entanglement entropy to the cumulants of charge transfer.

## B. With Pairing

For completeness we derive Eqs. (1.11),(1.12) in the presence of pairing between fermions, which of course arises in applications involving superconductivity. In this case particle number is not conserved so the interpretation of "number fluctuations" is somewhat problematic, especially for experiments. It does, however, clarify the source of entanglement entropy in such systems so that we present the derivation. Pairing is taken into account by considering the more general quadratic fermionic Hamiltonian

$$\hat{H} = \sum_{i,j} \left[ \hat{a}_i^\dagger A_{ij} \hat{a}_j + \frac{1}{2}(\hat{a}_i^\dagger B_{ij} \hat{a}_j^\dagger + \text{h.c.}) \right], \tag{1.26}$$

where $A$ is a hermitian matrix and $B$ is an anti-symmetric matrix. For simplicity we assume that both $A$ and $B$ are real. To diagonalize the Hamiltonian we perform the transformation [4]

$$\hat{b}_k = \sum_i (U_{ik}\hat{a}_i + V_{ik}\hat{a}_i^\dagger). \tag{1.27}$$

Letting

$$\phi = U + V, \tag{1.28}$$
$$\psi = U - V, \tag{1.29}$$

the transformation (1.27) is both canonical and diagonalizes the Hamiltonian as

$$\hat{H} = \sum_k \Lambda_k \hat{b}_k^\dagger \hat{b}_k + H_0, \qquad H_0 = \frac{1}{2}\text{Tr}(A - \Lambda), \tag{1.30}$$

with $\Lambda_{kp} = \Lambda_k \delta_{kp}$, if the following coupled set of equations are satisfied by the orthonormal matrices $\phi$ and $\psi$:

$$(A + B)\phi = \psi\Lambda, \tag{1.31}$$
$$(A - B)\psi = \phi\Lambda. \tag{1.32}$$

We define the generalized equilibrium correlation matrix

$$G_{ij} = \langle (\hat{a}_i - \hat{a}_i^\dagger)(\hat{a}_j + \hat{a}_j^\dagger) \rangle = \sum_k \psi_{ik}\left(\tanh\frac{\beta\Lambda_k}{2}\right)(\phi^T)_{kj}, \tag{1.33}$$

i.e.,

$$G = \psi\left(\tanh\frac{\beta\Lambda}{2}\right)\phi^T. \tag{1.34}$$

Then

$$G^T G = \phi\left(\tanh^2\frac{\beta\Lambda}{2}\right)\phi^T, \tag{1.35}$$

As in the previous calculation, by Wick's Theorem the reduced density matrix for a subsystem $\Omega$ is the same form as the original Hamiltonian:

$$\hat{\rho}_\Omega = \frac{e^{-\beta_\Omega \hat{H}_\Omega}}{\text{Tr } e^{-\beta_\Omega \hat{H}_\Omega}}, \tag{1.36}$$

$$\hat{H}_\Omega = \sum_{i,j}\left[\hat{a}_i^\dagger (A_\Omega)_{ij}\hat{a}_j + \frac{1}{2}(\hat{a}_i^\dagger (B_\Omega)_{ij}\hat{a}_j^\dagger + \text{h.c.})\right], \tag{1.37}$$

where we can again set $T_\Omega = 1$. The diagonalization procedure is exactly the same for the Hamiltonian of the subsystem as for the original system. Thus if we let $(G^T G)_\Omega$ be the restriction of $G^T G$ to the degrees of freedom in $\Omega$, then the energy levels $\lambda_q$ of the entanglement Hamiltonian $\hat{H}_\Omega$ corresponding to the energy levels $\Lambda_q$ of the original Hamiltonian $\hat{H}$ can be obtained as the eigenvalues $\nu_q^2$ of $(G^T G)_\Omega$ as

$$\nu_q^2 = \tanh^2\frac{\lambda_q}{2}, \tag{1.38}$$

or,

$$f(\lambda_q) = \frac{1 - \nu_q}{2}. \tag{1.39}$$

Comparing to the case without pairing, we obtain all of the same results for the von Neumann and Rényi entanglement entropies by setting

$$M_\Omega = \frac{1 - [(G^T G)_\Omega]^{1/2}}{2}. \tag{1.40}$$

This is of course consistent with the fact that $G = 1 - 2M$ in Eq. (1.33) when $B = 0$, the case without pairing.

One important difference between the case with pairing and the case without, of course, is that the fluctuations of particle number in the subsystem do not represent fluctuations of the number of original particles. The number operator of quasi-particles in the subsystem $\Omega$ is $\hat{N}_\Omega = \sum_q \hat{c}_q^\dagger \hat{c}_q$, where the fermionic operators $\hat{c}_q$ arise from diagonalizing Eq. (1.37). This is the operator being counted by Eq. (1.22). In contrast, the number operator for the original particles is $\hat{N}'_\Omega = \sum_{i \in \Omega} \hat{a}_i^\dagger \hat{a}_i$, and in the presence of a pairing term in the Hamiltonian $\hat{N}_\Omega \neq \hat{N}'_\Omega$.

## II. THE COEFFICIENTS OF THE ENTANGLEMENT ENTROPY–CUMULANTS SERIES IN THE INFINITE-CUTOFF LIMIT

In this section we perform a calculation that generalizes the computation of Ref. [3] to the Rényi entropies and therefore provides, as a limiting case, the von Neumann entropy as well. As noted in the main text, the series resulting from such a derivation is not convergent and therefore cannot be used to approximate the entanglement entropy with a finite number of cumulants. The result, however, does provide an important check on the values of the coefficients of the series when the cutoff is taken to infinity *without taking into account the number of cumulants.* The derivation, moreover, will allow us to point out where this method breaks down.

### 1. Spectral Density Function of $M_\Omega$

Let $\mu(z)$ be the spectral density function of the correlation matrix $M_\Omega$,

$$\mu(z) = \text{Tr}[\delta(z - M_\Omega)] \tag{2.1}$$

$$= \text{Tr}\left(\frac{1}{\pi} \lim_{\epsilon \to 0^+} \text{Im} \frac{1}{z - M_\Omega - i\epsilon}\right) \tag{2.2}$$

$$= \frac{1}{\pi} \text{Im} \lim_{\epsilon \to 0^+} \partial_z \text{Tr}[\ln(z - M_\Omega - i\epsilon)]. \tag{2.3}$$

We recall that $M_\Omega$ has eigenvalues in $z \in [0, 1]$. From Eq. (1.3) we have

$$\ln \chi(\lambda) = \text{Tr}\{\ln[1 + (e^{i\lambda} - 1)M_\Omega]\} - (\text{Tr } Q)i\lambda, \tag{2.4}$$

so that by factoring out $(1 - e^{i\lambda})$ in the appropriate manner we can write

$$\ln \chi(\lambda) = \text{Tr}\left[\ln\left(\frac{1}{1 - e^{i\lambda}} - M_\Omega\right)\right] + (\dim M_\Omega) \ln(1 - e^{i\lambda}) - (\text{Tr } Q)i\lambda. \tag{2.5}$$

With the substitution

$$z = \frac{1}{1 - e^{i\lambda}} \tag{2.6}$$

we therefore obtain

$$\ln \chi(\lambda(z)) = \text{Tr}[\ln(z - M_\Omega)] - (\dim M_\Omega - \text{Tr } Q) \ln z - (\text{Tr } Q) \ln(z - 1). \tag{2.7}$$

Compaing to Eq. (2.3), we see that the spectral density function can be written as

$$\mu(z) = \frac{1}{\pi}\text{Im}[\partial_z \ln \chi(\lambda(z - i0^+))] + (\dim M_\Omega - \text{Tr } Q)\delta(z) + (\text{Tr } Q)\delta(z - 1), \tag{2.8}$$

where we have used $0^+$ as shorthand for taking the limit of $\epsilon \to 0^+$. In inverting Eq. (2.6) to get $\lambda$ in terms of $z$, we have chosen

$$\lambda(z) = -\pi - i\ln\left(\frac{1}{z} - 1\right). \tag{2.9}$$

Although Eq. (2.6) defines $\lambda$ only within multiples of $2\pi$, this turns out to be the correct phase convention.

As an example, consider the simple case where $M_\Omega = 1/2$ and $Q = 0$. The spectral density function in this case is clearly

$$\mu(z) = \delta\left(z - \frac{1}{2}\right). \tag{2.10}$$

We have $\dim M_\Omega = 1$ and $\text{Tr } Q = 0$, so that from Eq. (1.3)

$$\chi(\lambda(z)) = 1 - \frac{1 - e^{i\lambda}}{2} = 1 - \frac{1}{2z}. \tag{2.11}$$

Using Eq. (2.8) and letting $z_0 = z - i0^+$, we get

$$\mu(z) = \frac{1}{\pi}\text{Im}\left[\frac{1}{2z_0(z_0 - 1/2)}\right] + \delta(z) \tag{2.12}$$

$$= \frac{1}{\pi}\text{Im}\left(\frac{1}{z_0 - 1/2} - \frac{1}{z_0}\right) + \delta(z) \tag{2.13}$$

$$= \delta\left(z - \frac{1}{2}\right) \tag{2.14}$$

as expected in Eq. (2.10).

### 2. Rényi Entropies

In terms of the spectral density function, the Rényi entropies in Eq. (1.2) are given by

$$\mathcal{S}_\alpha = \frac{1}{1 - \alpha}\int_0^1 dz\, \mu(z)\ln[z^n + (1-z)^n]. \tag{2.15}$$

Dropping the $\delta$-funcion contributions at $z = 0, 1$ in Eq. (2.8) where the integrand vanishes and performing integration by parts by transferring the derivative, we obtain

$$\mathcal{S}_\alpha = \frac{\alpha}{\pi(\alpha - 1)}\int_0^1 dz\, \frac{z^{n-1} - (1-z)^{n-1}}{z^n + (1-z)^{n-1}}\, \text{Im}[\ln \chi(\lambda(z - i0^+))]. \tag{2.16}$$

We can now neglect the small imaginary part $0^+$. Let

$$u = \frac{1}{2}\ln\left(\frac{1}{z} - 1\right). \tag{2.17}$$

Then, after some simplification,

$$\mathcal{S}_\alpha = -\frac{\alpha}{\pi}\int_{-\infty}^\infty du\, \frac{\tanh(\alpha u) - \tanh u}{\alpha - 1}\, \text{Im}[\ln \chi(-\pi - 2iu)]. \tag{2.18}$$

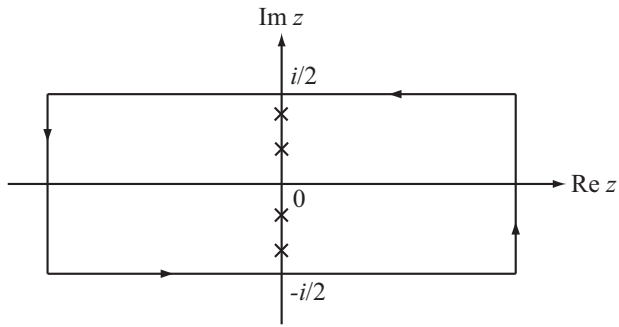

FIG. 1: The contour $C$ used in Eq. (2.23). The poles (crosses) are shown for the case $\alpha = 5$.

In particular, for the von Neumann entropy

$$S = \lim_{\alpha \to 1} S_\alpha = -\frac{1}{\pi} \int_{-\infty}^{\infty} du \, \frac{u}{\cosh^2 u} \, \text{Im}[\ln \chi(-\pi - 2iu)]. \tag{2.19}$$

We can expand the cumulant-generating function as follows:

$$\ln \chi(-\pi - 2iu) = \sum_{k=1}^{\infty} \frac{(-i\pi + 2u)^k}{k!} C_k. \tag{2.20}$$

Now, suppose (contrary to reality) that it is justified to integrate the series term by term, i.e., switch the order of summation and integration, as

$$S_\alpha = \sum_{k=1}^{\infty} \beta_k(\alpha) C_k \tag{2.21}$$

where

$$\beta_k(\alpha) = \frac{\alpha(2\pi)^k}{(1-\alpha)k!} \, \text{Im} \int_{-\infty}^{\infty} du \, [\tanh(\alpha\pi u) - \tanh(\pi u)]\left(u - \frac{i}{2}\right)^k. \tag{2.22}$$

From $\text{Im } z = (z - z^*)/(2i)$ we see that $\beta_k(\alpha) = 0$ for $k$ odd. For $k$ even we can let $z = u - i/2$ and rewrite Eq. (2.22) as the contour integral

$$\beta_k(\alpha) = \frac{\alpha(2\pi)^k}{(1-\alpha)k!} \frac{1}{2i} \int_C dz \, [\tanh(\alpha\pi(z + i/2)) - \coth(\pi z)] z^k \tag{2.23}$$

where $C$ is the rectangular contour shown in Fig. 1. Note that the integrla vanishes on the vertical edges where $|\text{Re } z| \to \infty$. We will compute the integral for $\alpha = n$, $n$ integer, and analytically continue the result to arbitrary $\alpha$.

Treating the case of $n$ even and $n$ odd separately, we find the following:

1. *n odd.* In this case $\tanh(n\pi(z + i/2)) = \coth(n\pi z)$ and the integrand has simple poles at $nz = \pm i, \pm 2i, \ldots, \pm[(n-1)/2]i$. The Residue Thereom then gives

$$\beta_k(n) = \frac{2}{1-n} \frac{1}{k!} \left(\frac{2\pi i}{n}\right)^k \sum_{p=1}^{(n-1)/2} p^k. \tag{2.24}$$

2. *n even.* In this case $\tanh(n\pi(z + i/2)) = \tanh(n\pi z)$ and the integrand has simple poles at $2nz = \pm i, \pm 3i, \ldots, \pm(n-1)$. The Residue Theorem gives

$$\beta_k(n) = \frac{2}{1-n} \frac{1}{k!} \left(\frac{2\pi i}{n}\right)^k \sum_{p=1}^{n/2} \left(p - \frac{1}{2}\right)^k \tag{2.25}$$

The two cases can be combined and analytically continued in $n$ to non-integer values:

$$\beta_k(\alpha) = \frac{2}{\alpha-1}\frac{1}{k!}\left(\frac{2\pi i}{\alpha}\right)^k \zeta\left(-k, \frac{\alpha+1}{2}\right), \tag{2.26}$$

where $\zeta(s,a)$ is the Hurwitz zeta function, cf. Eq. (4.10).

It is instructive to check that Eq. (2.26) reduces to the result of Ref. 3 for $\alpha = 1$. To take the limit as $\alpha \to 1$ we can use l'Hôpital's rule and the identities

$$\partial_a \zeta(s,a) = -s\zeta(s+1,a), \tag{2.27}$$

$$\zeta(1-s) = 2(2\pi)^{-s}\Gamma(s)\zeta(s)\cos\frac{\pi s}{2} \tag{2.28}$$

where $\Gamma(x)$ is the gamma function. Then

$$\lim_{\alpha \to 1}\beta_k(\alpha) = \lim_{\alpha \to 1}\frac{2(2\pi i)^k}{k!}\partial_\alpha\zeta\left(-k, \frac{\alpha+1}{2}\right) \tag{2.29}$$

$$= \frac{(2\pi i)^k}{(k-1)!}\zeta(1-k) \tag{2.30}$$

$$= \begin{cases} 2\zeta(k) & \text{if } k \text{ even}, \\ 0 & \text{if } k \text{ odd}, \end{cases} \tag{2.31}$$

which agrees with the coefficients given in Ref. [3], recalling that for even $k$ we have $2\zeta(k) = (2\pi)^k |B_k|/k!$ where $B_k$ are the Bernoulli numbers.

Of course, the expansion in Eq. (2.20) is *not valid* in the context of integrating Eq. (2.18) term by term, which we have done, because the range of the integral extends along the entire real line. With the exception of gaussian generating functions of the form $\ln \chi(\lambda) = i\mu\lambda - \sigma^2\lambda^2/2$ with only $C_1$ and $C_2$ nonzero (i.e., not an infinite series) all cumulant generating functions occuring in fermionic systems possess singularities in the complex plane so that the series in Eq. (2.20) has a finite radius of convergence. The result is a subtle one: the coefficients obtained from the previous computation are the limiting values of the coefficients of the convergent series derived in the main text (with cutoff), but using the series in this limiting form will result in divergent values for the Rényi entanglement entropies, including the von Neumann entropy.

## III. DIAGONALIZATION OF THE SPIN-1/2 XX CHAIN

In this section we derive the correlation matrix for the Hamiltonian

$$\hat{H}_{\text{XX}} = J\sum_i (\hat{S}_i^x \hat{S}_{i+1}^x + \hat{S}_i^y \hat{S}_{i+1}^y) \tag{3.1}$$

for both periodic boundary conditions (PBCs) and open boundary conditions (OBCs). The computation is very standard, but is provided to make the discussion in the main text more transparent.

The spin-1/2 XX chain is solved by using the Jordan-Wigner transformation [4]

$$\hat{S}_i^+ = \hat{a}_i^\dagger \left[\prod_{j<i}(1-2\hat{a}_j^\dagger \hat{a}_j)\right], \qquad \hat{S}_i^- = \left[\prod_{j<i}(1-2\hat{a}_j^\dagger \hat{a}_j)\right]\hat{a}_i, \qquad \hat{S}_i^z = \hat{a}_i^\dagger \hat{a}_i - \frac{1}{2} \tag{3.2}$$

to map Eq. (3.1) to the free-fermion Hamiltonian

$$\hat{H}_{\text{XX}} = J\sum_{ij}\hat{a}_i^\dagger H_{ij}\hat{a}_j, \qquad H = \frac{J}{2}\begin{pmatrix} & 1 & & & & p \\ 1 & & 1 & & & \\ & \ddots & & \ddots & & \\ & & 1 & & 1 & \\ p & & & & 1 & \end{pmatrix}. \tag{3.3}$$

For PBCs the cyclic term induces a term of the form

$$\hat{S}_L^x \hat{S}_1^x + \hat{S}_L^y \hat{S}_1^y = -\frac{1}{2}(\hat{a}_L^\dagger \hat{a}_1 + \hat{a}_1^\dagger \hat{a}_L) \exp\left(i\pi \sum_j \hat{a}_j^\dagger \hat{a}_j\right), \tag{3.4}$$

and since for the ground state we are interested in the case of zero total magnetization $\sum_j \hat{S}_j^z = 0$, i.e., half-filling of the fermions $\sum_j \hat{a}_j^\dagger \hat{a}_j = L/2$, we have

$$p = \begin{cases} 0 & \text{for OBCs,} \\ 1 & \text{for PBCs, } L \equiv 2 \bmod 4, \\ -1 & \text{for PBCs, } L \equiv 0 \bmod 4. \end{cases} \tag{3.5}$$

We will always assume that $L$ is even. We hope that the dual role of $i$ as an index and as the imaginary unit will not cause confusion.

### A. PBCs

For PBCs the eigenvectors $\phi$ and eigenvalues $J\Lambda$ satisfy

$$\frac{1}{2}\phi_{j-1} - \Lambda \phi_j + \frac{1}{2}\phi_{j+1} = 0, \tag{3.6}$$

subject to the boundary condition $\phi_{L+1} = p\phi_1$. The eigenvalues and corresponding eigenvectors are $\Lambda_k = \cos k$, $\phi_{jk} = e^{ikj}/\sqrt{L}$ where

$$k(n) = \frac{2\pi}{L} \times \begin{cases} n & \text{if } p = 1, \\ n + 1/2 & \text{if } p = -1 \end{cases} \tag{3.7}$$

for $n = -L/2 + 1, \ldots, L/2$. The correlation matrix (1.8) is therefore

$$M_{ij}^{\text{PBC}}(\beta) = \frac{1}{L} \sum_{n=-L/2+1}^{L/2} e^{ik(i-j)} f(\beta J \Lambda_k). \tag{3.8}$$

Although the sum in Eq. (3.8) cannot be carried out analytically for general $L$ and $\beta$, there are two limits of interest where it is possible. First, at zero temperature we have

$$f(\beta J \Lambda_k) \stackrel{\beta J \to \infty}{=} \theta(-\Lambda_k) = \begin{cases} 0 & \text{if } \Lambda_k > 0, \\ 1 & \text{if } \Lambda_k < 0, \end{cases} \tag{3.9}$$

so that Eq. (3.8) becomes, for both $p = \pm 1$,

$$M_{ij}^{\text{PBC}}(\beta J \to \infty) = \delta_{ij} - \frac{1}{2} \frac{\text{sinc}[\pi(i-j)/2]}{\text{sinc}[\pi(i-j)/L]}, \tag{3.10}$$

where

$$\text{sinc } x = \begin{cases} 1 & \text{if } x = 0, \\ \frac{\sin x}{x} & \text{otherwise.} \end{cases} \tag{3.11}$$

In particular, for $L \to \infty$

$$M_{ij}^{\text{PBC}}(\beta J, L \to \infty) = \delta_{ij} - \frac{1}{2}\text{sinc}\left[\frac{\pi}{2}(i-j)\right]. \tag{3.12}$$

Second, at high temperature $\beta J \ll 1$

$$f(\beta J \Lambda_k \ll 1) \simeq \frac{1}{2} - \frac{\beta J}{4}\Lambda_k + O((\beta J)^3), \qquad (3.13)$$

so that the correlation matrix is given by

$$M_{ij}^{\text{PBC}}(\beta J \ll 1) = \frac{1}{2}\delta_{ij} - \frac{\beta J}{4L}\sum_{n=-L/2+1}^{L/2} e^{ik(i-j)}\cos k \qquad (3.14)$$

$$= \frac{1}{2}\left[\delta_{ij} - \frac{\beta J}{4}(\delta_{|i-j|,1} + p\delta_{|i-j|,L-1})\right]. \qquad (3.15)$$

### B. OBCs

For OBCs the eigenvectors $\phi$ and eigenvalues $J\Lambda$ also satisfy Eq. (3.6) but the boundary conditions imply $\phi_0 = \phi_{L+1} = 0$ so that

$$\Lambda_k = \cos k, \qquad \phi_{jk} = \sqrt{\frac{2}{L+1}}\sin jk, \qquad (3.16)$$

where $k(n) = \pi n/(L+1)$ for $n = 1, \ldots, L$. The correlation matrix is therefore

$$M_{ij}^{\text{OBC}}(\beta) = \frac{2}{L+1}\sum_{n=1}^{L}\sin(ik)\sin(jk)f(\beta J\Lambda_k) \qquad (3.17)$$

$$= \frac{1}{L+1}\sum_{n=1}^{L}\cos[(i-j)k]f(\beta J\Lambda_k) - \{(i-j) \to (i+j)\}. \qquad (3.18)$$

At zero temperature we can use Eq. (3.9) to obtain

$$M_{ij}^{\text{OBC}}(\beta J \to \infty) = \frac{1}{2}\delta_{ij} - \frac{1}{2}g(i-j,L)\text{sinc}\left[\frac{\pi}{2}(i-j)\right]$$
$$+ \frac{1}{2}g(i+j,L)\text{sinc}\left[\frac{\pi}{2}(i+j)\right], \qquad (3.19)$$

where the finite-size correction factor is

$$g(j,L) = \frac{1}{8}\left(\frac{\pi j L}{L+1}\right)^2 \frac{\text{sinc}^2\{\pi j L/[4(L+1)]\}}{\text{sinc}\{\pi j/[2(L+1)]\}} \qquad (3.20)$$

and goes to $1 - \cos(\pi j/2)$ for $L \to \infty$. For $L \to \infty$ we therefore have

$$M_{ij}^{\text{OBC}}(\beta J, L \to \infty) = \delta_{ij} - \frac{1}{2}\text{sinc}\left[\frac{\pi}{2}(i-j)\right] + \frac{1}{2}\text{sinc}\left[\frac{\pi}{2}(i+j)\right]. \qquad (3.21)$$

### IV. FLUCTUATION SUMS

In this section we show how to compute spin fluctuations in one dimension,

$$\mathcal{F}(\ell) = \left\langle \left(\sum_{i=1}^{\ell}\hat{S}^z\right)^2\right\rangle - \left\langle \sum_{i=1}^{\ell}\hat{S}^z\right\rangle^2 \qquad (4.1)$$

$$= \sum_{i,j=1}^{\ell}[\langle\hat{S}_i^z\hat{S}_j^z\rangle - \langle\hat{S}_i^z\rangle\langle\hat{S}_j^z\rangle]. \qquad (4.2)$$

In all of the cases considered here the on-site term is $1/4$, $\langle(\hat{S}_i^z)^2\rangle - \langle\hat{S}_i^z\rangle^2 = 1/4$, so that it will be convenient to organize the sums as

$$\mathcal{F}(\ell) = \frac{\ell}{4} + \sum_{\substack{i,j=1 \\ i \neq j}}^{\ell} \rho_{ij} \tag{4.3}$$

where

$$\rho_{ij} = \begin{cases} 0 & \text{if } i = j, \\ \langle\hat{S}_i^z\hat{S}_j^z\rangle - \langle\hat{S}_i^z\rangle\langle\hat{S}_j^z\rangle & \text{otherwise.} \end{cases} \tag{4.4}$$

Before presenting the calculations we list some mathematical formulas and notation needed in the remainder of the section.

### A. Mathematical Formulas

We will make extensive use of the digamma function $\psi(x) = d[\ln \Gamma(x)]/dx$, the logarithmic derivative of the gamma function $\Gamma(x)$. The digamma function has the series representation

$$\psi(x) = -\gamma + \sum_{k=1}^{\infty} \left(\frac{1}{k} - \frac{1}{k+x-1}\right) \tag{4.5}$$

and integral representation

$$\psi(x) = \int_0^\infty dt \left(\frac{e^{-t}}{t} - \frac{e^{-xt}}{1-e^{-t}}\right). \tag{4.6}$$

Some special values of the digamma function and its derivatives are

$$\psi(1) = -\gamma, \qquad \psi\left(\frac{1}{2}\right) = -\gamma - 2\ln 2, \qquad \psi'(1) = \sum_{k=1}^{\infty} \frac{1}{k^2} = \frac{\pi^2}{6}. \tag{4.7}$$

The digamma function is particularly useful in converting sums to a form that can be expanded using the asymptotic expansion

$$\psi(x \to \infty) = \ln x - \frac{1}{2x} - \sum_{n=1}^{\infty} \frac{B_{2n}}{(2n)x^{2n}} \tag{4.8}$$

$$= \ln x - \frac{1}{2x} - \frac{1}{2x^2} + O(x^{-4}), \tag{4.9}$$

where $B_n$ are the Bernoulli numbers.

For sums involving non-integer exponents a useful generalization is the Hurwitz zeta function

$$\zeta(s, a) = \sum_{k=0}^{\infty} \frac{1}{(k+a)^s}, \tag{4.10}$$

which at $a = 1$ reduces to the Riemann zeta function

$$\zeta(s) = \zeta(s, 1) = \sum_{k=1}^{\infty} \frac{1}{k^s}. \tag{4.11}$$

The Hurwitz zeta function has the integral representation

$$\zeta(s, a) = \frac{1}{\Gamma(s)} \int_0^\infty dt \, \frac{t^{s-1}e^{-at}}{1-e^{-t}}. \tag{4.12}$$

The derivative with respect to the second argument is given by

$$\partial_a \zeta(s,a) = -s\zeta(s+1,a). \tag{4.13}$$

We will make use of the asymptotic expansion of the Hurwitz zeta function in its second argument [5]:

$$\zeta(s, a \to \infty) = \frac{1}{(s-1)a^{s-1}} + \frac{1}{2a^s} + \frac{s}{12a^{s+1}} + O(a^{-(s+2)}). \tag{4.14}$$

Double sums that depend only on the difference of the indices can be converted to a single sum as

$$\sum_{i,j=1}^{\ell} f(|i-j|) = f(0)\ell + 2\sum_{k=1}^{\ell}(\ell-k)f(k), \tag{4.15}$$

while sums that depend only on the sum of the indices can be written as

$$\sum_{i,j=1}^{\ell} f(i+j) = \sum_{k=1}^{\ell}[kf(k+1) + (\ell-k)f(k+1+\ell)]. \tag{4.16}$$

$$\sum_{k=1}^{\ell} f(k) = \sum_{k=1}^{\infty}[f(k) - f(k+\ell)]. \tag{4.17}$$

In the following computations we will use the notation

$$\sigma_k = \begin{cases} 1 & \text{if } k \text{ is odd}, \\ 0 & \text{if } k \text{ is even}. \end{cases} \tag{4.18}$$

### B. Spin-1/2 XX Chain

For the spin-1/2 XX chain with zero magnetization $\langle \hat{S}_i^z \rangle = 0$, Wick's Theorem leads to

$$\langle \hat{S}_i^z \hat{S}_j^z \rangle - \langle \hat{S}_i^z \rangle \langle \hat{S}_j^z \rangle = \frac{1}{2}\delta_{ij} - |M_{ij}|^2, \tag{4.19}$$

where the correlation matrix $M_{ij}$ is computed in Sec. III. We therefore carry out Eq. (4.3) with

$$\rho_{ij} = -|M_{ij}|^2, \qquad i \neq j. \tag{4.20}$$

#### 1. PBCs

Here we show how to compute the sum in Eq. (4.3) for [cf. Eq. (3.12)]

$$\rho_{ij} = -\frac{1}{\pi^2} \frac{1}{(i-j)^2} \times \sigma_{i-j}. \tag{4.21}$$

Using Eq. (4.15) the required sum is

$$\mathcal{F}^{\text{PBC}}(\ell) = \frac{\ell}{4} - \frac{2}{\pi^2}(\ell-k)\frac{\sigma_k}{k^2}, \tag{4.22}$$

where, using Eq. (4.17), the sum on the right-hand side can be rewritten as

$$\sum_{k=1}^{\ell}(\ell-k)\frac{\sigma_k}{k^2} = \ell\sum_{k=1}^{\infty}\frac{\sigma_k}{k^2} + \sum_{k=1}^{\infty}\left[-\frac{\sigma_k}{k} + k\frac{\sigma_{k+\ell}}{(k+\ell)^2}\right] \tag{4.23}$$

$$= \ell\sum_{k=1}^{\infty}\frac{\sigma_k}{k^2} + \partial_\alpha\left\{\frac{1}{\alpha}\sum_{k=1}^{\infty}\left[\frac{\sigma_k}{k} - \frac{\sigma_k}{k+\sigma_\ell+\ell/\alpha}\right]\right\}, \tag{4.24}$$

where for the second term in square brackets we have used the trick of converting $\sigma_{k+\ell}$ to $\sigma_k$ by shifting $k$ in the denominator. By writing the sums as (odd terms) = (all terms) − (even terms) we can use Eq. (4.5) to obtain

$$\sum_{k=1}^{\ell}(\ell-k)\frac{\sigma_k}{k^2} = \frac{\pi^2}{8}\ell - \frac{\gamma}{2} + \partial_\alpha\left\{\frac{1}{\alpha}\left[\psi\left(1+\sigma_\ell+\frac{\ell}{\alpha}\right) - \frac{1}{2}\psi\left(1+\frac{\sigma_\ell}{2}+\frac{\ell}{2\alpha}\right)\right]\right\}_{\alpha=1} \tag{4.25}$$

$$= \frac{\pi^2}{8}\ell - \frac{\gamma}{2} - \psi(1+\sigma_\ell+\ell) + \frac{1}{2}\psi\left(1+\frac{\sigma_\ell}{2}+\frac{\ell}{2}\right)$$

$$- \ell\psi'(1+\sigma_\ell+\ell) + \frac{\ell}{4}\psi'\left(1+\frac{\sigma_\ell}{2}+\frac{\ell}{2}\right). \tag{4.26}$$

Expanding for $\ell \gg 1$ using Eq. (4.9) then gives

$$\sum_{k=1}^{\ell}(\ell-k)\frac{\sigma_k}{k^2} = \frac{\pi^2}{8}\ell - \frac{1}{2}(\ln\ell + 1 + \gamma + \ln 2) + O(\ell^{-2}), \tag{4.27}$$

so that

$$\pi^2 \mathcal{F}_{\mathrm{XX}}^{\mathrm{PBC}}(\ell) = \ln\ell + f_1, \qquad f_1 = 1 + \gamma + \ln 2, \tag{4.28}$$

plus $O(\ell^{-2})$ corrections. This result was actually first obtained in Ref. [6] in an attempt to define an effective temperature for the subsystem, though without providing the size of the next-order correction or the corresponding result for OBCs presented in the next section. In the main text, moreover, an alternative method using the theory of Toeplitz matrices is presented for the derivation of Eq. (4.28)

## 2. OBCs

Here we show how to compute the sum in Eq. (4.3) for [cf. Eq. (3.21)]

$$\rho_{ij} = \rho_{ij}^- + \rho_{ij}^+ + \rho_{ij}^{-+}, \tag{4.29}$$

where

$$\rho_{ij}^- = -\frac{1}{\pi^2}\frac{1}{(i-j)^2} \times \sigma_{i-j}, \tag{4.30}$$

$$\rho_{ij}^+ = -\frac{1}{\pi^2}\frac{1}{(i+j)^2} \times \sigma_{i+j}, \tag{4.31}$$

$$\rho_{ij}^{-+} = -\frac{2}{\pi^2}\frac{(-1)^i}{(i^2-j^2)} \times \sigma_{i-j}. \tag{4.32}$$

Comparing to Eq. (4.21), we can write

$$\mathcal{F}_{\mathrm{XX}}^{\mathrm{OBC}}(\ell) = \mathcal{F}_{\mathrm{XX}}^{\mathrm{PBC}}(\ell) + I_+ + I_{-+} \tag{4.33}$$

with

$$I_+ = \sum_{i,j=1}^{\ell} \rho_{ij}^+, \qquad I_{-+} = \sum_{i,j=1}^{\ell} \rho_{ij}^{-+}. \tag{4.34}$$

Using Eq. (4.16) we can write $I^+$ as

$$-\pi^2 I_+ = \sum_{k=1}^{\ell} \left[ k \frac{\sigma_{k+1}}{(k+1)^2} + (\ell-k) \frac{\sigma_{k+1+\ell}}{(k+1+\ell)^2} \right], \tag{4.35}$$

and using Eq. (4.17) we get

$$-\pi^2 I_+ = \sum_{k=1}^{\infty} \left[ (k-1) \frac{\sigma_k}{k^2} + (k-1) \frac{\sigma_k}{(k+2\ell)^2} - 2(k-1) \frac{\sigma_{k+\ell}}{(k+\ell)^2} \right] \tag{4.36}$$

$$= \partial_\alpha \left\{ \frac{1}{\alpha} \sum_{k=1}^{\infty} \left[ -\frac{\sigma_k}{k+1/\alpha-1} - \frac{\sigma_k}{k+(1+2\ell)/\alpha-1} \right. \right.$$

$$\left. \left. + 2 \frac{\sigma_k}{k+\sigma_\ell+(1+\ell)/\alpha-1} \right] \right\}_{\alpha=1}. \tag{4.37}$$

By adding and subtracting $\sigma_k/k$ for each term and noting that (odd terms) = (all terms) − (even terms), we can build the series representation of the digamma function, Eq. (4.5), to obtain

$$-\pi^2 I^+ = \partial_\alpha \left\{ \frac{1}{\alpha} \left[ \psi\left(\frac{1}{\alpha}\right) - \frac{1}{2}\psi\left(\frac{1}{2} + \frac{1}{2\alpha}\right) \right. \right.$$

$$+ \psi\left(\frac{1+2\ell}{\alpha}\right) - \frac{1}{2}\psi\left(\frac{1}{2} + \frac{1+2\ell}{2\alpha}\right)$$

$$\left. \left. - 2\psi\left(\sigma_\ell + \frac{1+\ell}{\alpha}\right) + \psi\left(\frac{1+\sigma_\ell}{2} + \frac{1+\ell}{2\alpha}\right) \right] \right\}_{\alpha=1} \tag{4.38}$$

$$= -\frac{\pi^2}{8} + \frac{\gamma}{2}$$

$$- \psi(1+2\ell) + \frac{1}{2}\psi(1+\ell) + 2\psi(1+\sigma_\ell+\ell) - \psi\left(1+\frac{\sigma_\ell}{2}+\frac{\ell}{2}\right)$$

$$- (1+2\ell)\psi'(1+2\ell) + \frac{1+2\ell}{4}\psi'(1+\ell) + 2(1+\ell)\psi'(1+\sigma_\ell+\ell)$$

$$- \frac{1+\ell}{2}\psi'\left(1+\frac{\sigma_\ell}{2}+\frac{\ell}{2}\right). \tag{4.39}$$

Expanding for $\ell \gg 1$ using Eq. (4.9) then gives

$$\pi^2 I_+ = \left(\frac{\pi^2}{8} - \frac{1+\gamma}{2}\right) - \frac{1}{2}\ln\ell - \frac{3}{4}\frac{1}{\ell} + O(\ell^{-2}). \tag{4.40}$$

Turning now to the cross-term $I_{-+}$ in Eq. (4.33), we use

$$\frac{1}{i^2 - j^2} = -\frac{1}{2i}\left(\frac{1}{j-i} - \frac{1}{j+i}\right) \tag{4.41}$$

to write (note the cancellation of the terms with $i-j =$ even)

$$\pi^2 I_{-+} = \sum_{i=1}^{\ell} \frac{(-1)^i}{i} \left[ \sum_{j=1}^{i-1} \left(\frac{1}{j-i} - \frac{1}{j+i}\right) + \sum_{j=1}^{\ell-i} \left(\frac{1}{j} - \frac{1}{j+2i}\right) \right]. \tag{4.42}$$

The expression in square brackets is

$$\frac{3}{2}\frac{1}{i} + \psi(1+\ell-i) - \psi(1+\ell+i), \tag{4.43}$$

so that we can write

$$\pi^2 I^{-+} = A + B \tag{4.44}$$

with

$$A = \frac{3}{2}\sum_{i=1}^{\ell} \frac{(-1)^i}{i^2}, \tag{4.45}$$

$$B = \sum_{i=1}^{\ell} [\psi(1+\ell-i) - \psi(1+\ell+i)]. \tag{4.46}$$

We evaluate each term in turn. For $A$ in Eq. (4.45) we again rewrite the sum as an infinite series, Eq. (4.17), then use (even terms) − (odd terms) = 2(even terms) − (all terms) for the alternating sum:

$$A = \frac{3}{2}\sum_{i=1}^{\infty} \left[\frac{(-1)^i}{i^2} - \frac{(-1)^{i+\ell}}{(i+\ell)^2}\right] \tag{4.47}$$

$$= \frac{3}{2}\sum_{i=1}^{\infty} \frac{(-1)^i}{i^2} + \frac{3}{2}(-1)^\ell \partial_\alpha \left[\sum_{i=1}^{\infty} \frac{(-1)^i}{i+\alpha}\right]_{\alpha=\ell} \tag{4.48}$$

$$= -\frac{\pi^2}{8} + \frac{3}{2}(-1)^\ell \partial_\alpha \left[\sum_{i=1}^{\infty} \left(-\frac{1}{i+\alpha} + \frac{1}{i+\alpha/2}\right)\right]_{\alpha=\ell}. \tag{4.49}$$

Adding and subtracing $1/i$ in the sum to build the series representation of the digamma function in Eq. (4.5), we obtain

$$A = -\frac{\pi^2}{8} + \frac{3}{2}(-1)^\ell \partial_\alpha \left[\psi(1+\alpha) - \psi\left(1+\frac{\alpha}{2}\right)\right]_{\alpha=\ell} \tag{4.50}$$

$$= -\frac{\pi^2}{8} + \frac{3}{2}(-1)^\ell \left[\psi'(1+\ell) - \frac{1}{2}\psi'\left(1+\frac{\ell}{2}\right)\right]. \tag{4.51}$$

Upon expanding the derivatives of the digamma functions for $\ell \gg 1$ using Eq. (4.9) only the constant term contributes to order $\ell^{-2}$, so that

$$A = -\frac{\pi^2}{8} + O(\ell^{-2}). \tag{4.52}$$

Meanwhile, it has not been possible to give a closed-form expression for $B$ in Eq. (4.46), even in terms of digamma functions. We can, however, directly find the asymptotic behavior for large $\ell$. To this end we first use the integral representation of $\psi(x)$, Eq. (4.6), to write

$$B = -\int_0^\infty dt\, \frac{e^{-(1+\ell)t}}{1-e^{-t}} \sum_{i=1}^{\ell} \left[\frac{(-e^t)^i}{i} - \frac{(-e^{-t})^i}{i}\right]. \tag{4.53}$$

After writing

$$\sum_{i=1}^{\ell} \frac{x^i}{i} = \sum_{i=1}^{\infty} \frac{x^i}{i} - \sum_{i=1}^{\infty} \frac{x^{i+\ell}}{i+\ell} = -\ln(1-x) - \frac{1}{\ell}\sum_{i=1}^{\infty} \frac{x^{i+\ell}}{1+i/\ell}, \tag{4.54}$$

we expand the factor $(1 + i/\ell)^{-1}$ in the second term to get

$$\sum_{i=1}^{\ell} \frac{x^i}{i} = -\ln(1-x) - \sum_{k=0}^{\infty} \frac{(-1)^k}{\ell^{1+k}} x^\ell \sum_{i=1}^{\infty} i^k x^i. \tag{4.55}$$

Let

$$f_k(x) = x^\ell \sum_{i=1}^{\infty} i^k x^i = x^\ell (x\partial_x)^k \left( \sum_{i=1}^{\infty} x^i \right) = x^\ell (x\partial_x)^k \left( \frac{x}{1-x} \right). \tag{4.56}$$

Then we can write $B$ as

$$B = c + \sum_{k=0}^{\infty} \frac{(-1)^k}{\ell^{1+k}} c_k \tag{4.57}$$

with

$$c = \int_0^\infty dt \, \frac{e^{-(1+\ell)t}}{1 - e^{-t}} \ln \frac{1 + e^t}{1 + e^{-t}} \tag{4.58}$$

and

$$c_k = \int_0^\infty dt \, \frac{e^{-(1+\ell)t}}{1 - e^{-t}} [f_k(-e^t) - f_k(-e^{-t})]. \tag{4.59}$$

Before evaluating these terms, we note that some of the previous manipulations are only true *formally* because of the range of $x = e^t$ in the integral of Eq. (4.53). However, the purpose is to exclude terms proportional to $\ell^{-2}$ or smaller, which, as we will see, we have accomplished.

Now, noting that the argument of the logarithm is simply $e^t$, we have from Eqs. (4.12) and (4.14) that

$$c = \zeta(2, 1 + \ell) = \frac{1}{\ell} + O(\ell^{-2}). \tag{4.60}$$

Next,

$$(-1)^{\ell+1} c_0 = \int_0^\infty dt \, \frac{e^{-(1+\ell)t}}{1 - e^{-t}} \left[ \frac{e^{(1+\ell)t}}{1 + e^t} - \frac{e^{-(1+\ell)t}}{1 + e^{-t}} \right] \tag{4.61}$$

$$= \int_0^\infty dt \left[ \frac{e^{-t}}{1 - e^{-2t}} - \frac{e^{-2(1+\ell)t}}{1 - e^{-2t}} \right] \tag{4.62}$$

$$= \frac{1}{2} \int_0^\infty dt \left[ \frac{e^{-t}}{t} - \frac{e^{-(1+\ell)t}}{1 - e^{-t}} \right] - \frac{1}{2} \int_0^\infty dt \left[ \frac{e^{-t}}{t} - \frac{e^{-t/2}}{1 - e^{-t}} \right] \tag{4.63}$$

$$= \frac{1}{2} \left[ \psi(1 + \ell) - \psi\left(\frac{1}{2}\right) \right], \tag{4.64}$$

where we have once again used Eq. (4.6). Expanding for $\ell \gg 1$ with Eq. (4.9), we obtain

$$(-1)^{\ell+1} c_0 = \frac{1}{2} [\psi(1 + \ell) + \gamma + 2 \ln 2] = \frac{1}{2} (\ln \ell + \gamma + 2 \ln 2) + O(\ell^{-1}). \tag{4.65}$$

The next term is

$$(-1)^{\ell+1} c_1 = \int_0^\infty dt \, \frac{e^{-(1+\ell)t}}{1 - e^{-t}} \left[ \frac{e^{(1+\ell)t}}{(1 + e^t)^2} - \frac{e^{-(1+\ell)t}}{(1 + e^{-t})^2} \right] \tag{4.66}$$

$$\tag{4.67}$$

and gives, after a similar calculation,

$$(-1)^{\ell+1} c_1 = \frac{1}{4} \ln \ell + O(\ell^{-2}), \tag{4.68}$$

while $c_k$ for $k > 1$ can be shown to only yield terms of $O(\ell^{-2})$ or smaller. Thus

$$\mathcal{F}_{\text{XX}}^{\text{OBC}}(\ell) = \frac{1}{2} \mathcal{F}_{\text{XX}}^{\text{PBC}}(2\ell) + \frac{1}{2\pi^2 (2\ell)} - \frac{(-1)^\ell}{\pi^2 (2\ell)} [\ln(2\ell) + \gamma + \ln 2]$$
$$+ \frac{(-1)^\ell}{\pi^2 (2\ell)^2} [\ln(2\ell) - \ln 2] \tag{4.69}$$

plus $O(\ell^{-2})$ corrections, recalling the result in Eq. (4.28) for PBCs. As one might expect from conformal field theory arguments [7], up to subleading corrections the OBC result is exactly half the PBC result with the substitution $\ell \to 2\ell$.

### C. Spin-1/2 XXZ Chain (Luttinger Liquids)

Here we show how to compute the sum in Eq. (4.3) for the Luttinger Liquid result

$$\rho_{ij} = -\frac{K}{2\pi^2} \frac{1}{r^2} + \sum_{m=1}^{\infty} 2^{2m^2 K - 2} A_m \frac{(-1)^{mr}}{r^{2m^2 K}}, \qquad r = |i - j|. \tag{4.70}$$

We assume that $1/2 < K \le 1$, which corresponds to anisotropy $0 \le \Delta < 1$ in the spin-1/2 XXZ Hamiltonian where this correlation function is valid. As we will see, the only subleading correction relevant to the fluctuations is the term containing $A_1$. We will, however, keep all of the terms for completeness to show that this is indeed the case. Using the formulas derived here it may also be possible in the future to compute the constant term and demonstrate explicitly the vanishing of an $\ell$-dependent term if the non-universal coefficients $A_m$ are known [8].

Using Eq. (4.15) the sum over Eq. (4.70) becomes

$$\mathcal{F}_{\text{XXZ}}(\ell) = \frac{\ell}{4} - \frac{K}{\pi^2} F_0(\ell) + \sum_{m=1}^{\infty} 2^{2m^2 K - 1} A_m F_m(\ell) \tag{4.71}$$

with

$$F_0(\ell) = \sum_{k=1}^{\ell} (\ell - k) \frac{1}{k^2}, \tag{4.72}$$

$$F_m(\ell) = \sum_{k=1}^{\ell} (\ell - k) \frac{(-1)^{mk}}{k^{2m^2 K}}. \tag{4.73}$$

The usual method of rewriting the finite sums as infinite sums using Eq. (4.17) and converting the resulting sums to the digamma function, Eq. (4.5), gives

$$F_0(\ell) = \ell \sum_{k=1}^{\infty} \frac{1}{k^2} + \sum_{k=1}^{\infty} \left[ -\frac{1}{k} + \frac{k}{(k+\ell)^2} \right] \tag{4.74}$$

$$= \frac{\pi^2}{6} \ell + \partial_\alpha \left\{ \frac{1}{\alpha} \left[ \sum_{k=1}^{\infty} \left( \frac{1}{k} - \frac{1}{k + \ell/\alpha} \right) \right] \right\}_{\alpha=1} \tag{4.75}$$

$$= \frac{\pi^2}{6} \ell + \partial_\alpha \left\{ \frac{1}{\alpha} \left[ \psi\left(1 + \frac{\ell}{\alpha}\right) + \gamma \right] \right\}_{\alpha=1} \tag{4.76}$$

$$= \frac{\pi^2}{6} \ell - \psi(1 + \ell) - \ell \psi'(1 + \ell) - \gamma. \tag{4.77}$$

Expanding for $\ell \gg 1$ using Eq. (4.9), we obtain

$$F_0(\ell) = \frac{\pi^2}{6}\ell - \ln\ell - (1+\gamma) + O(\ell^{-2}). \quad (4.78)$$

For $F_1(\ell)$ we can similarly write

$$F_m(\ell) = \ell\sum_{k=1}^\infty \frac{(-1)^{mk}}{k^{2m^2K}} - \sum_{k=1}^\infty \frac{(-1)^{mk}}{k^{2m^2K-1}} + (-1)^{m\ell}\sum_{k=1}^\infty k\frac{(-1)^{mk}}{(k+\ell)^{2m^2K}} \quad (4.79)$$

$$= \ell\sum_{k=1}^\infty \frac{(-1)^{mk}}{k^{2m^2K}} - \sum_{k=1}^\infty \frac{(-1)^{mk}}{k^{2m^2K-1}}$$

$$- \frac{(-1)^{m\ell}}{2m^2K-1}\partial_\alpha\left[\alpha^{1-2m^2K}\sum_{k=1}^\infty \frac{(-1)^{mk}}{(k+\ell/\alpha)^{2m^2K-1}}\right]_{\alpha=1}. \quad (4.80)$$

We must now treat the case of even and odd $m$ separately for each term. In the first term of Eq. (4.80) for even $m$ the sum is simply the Riemann zeta function, Eq. (4.11), while for odd $m$ we use the identity (even terms) $-$ (odd terms) $=$ 2(even terms) $-$ (all terms) to get

$$\sum_{k=1}^\infty \frac{(-1)^{mk}}{k^{2m^2K}} = \begin{cases}\zeta(2m^2K) & \text{for } m \text{ even,} \\ (2^{1-2m^2K}-1)\zeta(2m^2K) & \text{for } m \text{ odd.}\end{cases} \quad (4.81)$$

In the same way, the second term of Eq. (4.80) can also be evaluated as

$$\sum_{k=1}^\infty \frac{(-1)^{mk}}{k^{2m^2K-1}} = \begin{cases}\zeta(2m^2K-1) & \text{for } m \text{ even,} \\ (2^{2-2m^2K}-1)\zeta(2m^2K-1) & \text{for } m \text{ odd.}\end{cases} \quad (4.82)$$

In the third term of Eq. (4.80), the sum for $m$ even is the Hurwitz zeta function in Eq. (4.10),

$$\sum_{k=1}^\infty \frac{(-1)^{mk}}{(k+\ell/\alpha)^{2m^2K-1}} = \zeta\left(2m^2K-1, 1+\frac{\ell}{\alpha}\right), \quad (m \text{ even}), \quad (4.83)$$

so that

$$\frac{1}{2m^2K-1}\partial_\alpha\left[\alpha^{1-2m^2K}\sum_{k=1}^\infty \frac{(-1)^{mk}}{(k+\ell/\alpha)^{2m^2K-1}}\right]_{\alpha=1}$$

$$= -\zeta(2m^2K-1, 1+\ell) - \frac{\ell}{2m^2K-1}\partial_\alpha\zeta\left(2m^2K-1, 1+\frac{\ell}{\alpha}\right)\bigg|_{\alpha=1} \quad (4.84)$$

$$= -\zeta(2m^2K-1, 1+\ell) + \ell\zeta(2m^2K, 1+\ell) \quad (4.85)$$

where in the last line we have used Eq. (4.13) for the derivative of the Hurwitz zeta function. By expanding for $\ell \gg 1$ using Eq. (4.14) it can be seen that for even $m$ (smallest $m = 2$ and $K > 1/2$) the sum does not contribute to the overall result up to $O(\ell^{-2})$ corrections.

Meanwhile, for $m$ odd we simply separate the even and odd terms to write

$$\sum_{k=1}^\infty \frac{(-1)^{mk}}{(k+\ell/\alpha)^{2m^2K-1}}$$

$$= 2^{2m^2K-1}\left[\zeta\left(2m^2K-1, 1+\frac{\ell}{2\alpha}\right) - \zeta\left(2m^2K-1, \frac{1}{2}+\frac{\ell}{2\alpha}\right)\right], \quad (m \text{ odd}). \quad (4.86)$$

Then

$$\frac{1}{2m^2K-1}\partial_\alpha\left[\alpha^{1-2m^2K}\sum_{k=1}^{\infty}\frac{(-1)^{mk}}{(k+\ell/\alpha)^{2m^2K-1}}\right]_{\alpha=1}$$
$$= 2^{1-2m^2K}\left[-\zeta\left(2m^2K-1,1+\frac{\ell}{2}\right)+\zeta\left(2m^2K-1,\frac{1}{2}+\frac{\ell}{2}\right)\right.$$
$$\left.+\frac{\ell}{2}\zeta\left(2m^2K,1+\frac{\ell}{2}\right)-\frac{\ell}{2}\zeta\left(2m^2K,\frac{1}{2}+\frac{\ell}{2}\right)\right], \quad (4.87)$$

where we have again made use of the derivative formula (4.13). In this case, expanding for $\ell \gg 1$ using Eq. (4.14) gives corrections smaller than $O(\ell^{-2})$ for all $m$ except $m = 1$, in which case

$$\frac{1}{2m^2K-1}\partial_\alpha\left[\alpha^{1-2m^2K}\sum_{k=1}^{\infty}\frac{(-1)^{mk}}{(k+\ell/\alpha)^{2m^2K-1}}\right]_{\alpha=1} = \frac{1}{4}\frac{1}{\ell^{2K}}+O(\ell^{-2}), \quad m=1. \quad (4.88)$$

Combining the previous results, we find

$$F_m(\ell) = \begin{cases} (2^{1-2m^2K}-1)\zeta(2m^2K)\ell - (2^{2-2m^2K}-1)\zeta(2m^2K-1) \\ \quad -\frac{(-1)^\ell}{4}\frac{1}{\ell^{2K}} & m=1, \\ (2^{1-2m^2K}-1)\zeta(2m^2K)\ell - (2^{2-2m^2K}-1)\zeta(2m^2K-1) & m \text{ odd}, m>1, \\ \zeta(2m^2K)\ell - \zeta(2m^2K-1) & m \text{ even}. \end{cases}$$
$$+ O(\ell^{-2}), \quad (4.89)$$

so that

$$\pi^2 \mathcal{F}_{\text{XXZ}}(\ell) = K\ln\ell + f_{\text{LL}} - \frac{\pi^2 A_1}{2^{3-2K}}\frac{(-1)^\ell}{\ell^{2K}} \quad (4.90)$$

plus $O(\ell^{-2})$ corrections, where

$$\frac{f_{\text{LL}}}{\pi^2} = K\frac{1+\gamma}{\pi^2} - \sum_{m=1,3,5,\ldots}(2-2^{2m^2K-1})\zeta(2m^2K-1)A_m$$
$$- \sum_{m=2,4,6,\ldots}2^{2m^2K-1}\zeta(2m^2K-1)A_m \quad (4.91)$$

and we have dropped a linear term $a_{\text{LL}}\ell$ on the assumption that

$$\frac{a_{\text{LL}}}{\pi^2} = \left(\frac{1}{4}-\frac{K}{6}\right) + \sum_{m=1,3,5,\ldots}(1-2^{2m^2K-1})\zeta(2m^2K)A_m$$
$$+ \sum_{2,4,6,\ldots}2^{2m^2K-1}\zeta(2m^2K)A_m \quad (4.92)$$
$$= 0. \quad (4.93)$$

If all of the coefficients $A_m$ are known this assumption can, in principle, be demonstrated explicitly. We can check, for example, that the first few terms are correct at the free-fermion point $\Delta = 0$ where

$$K = 1, \quad A_1 = \frac{1}{2\pi^2}, \quad A_{m\geq 2} = 0, \quad (4.94)$$

which follows from comparing Eq. (4.70) with Eq. (4.21). Then in Eq. (4.92) clearly $a_{\text{LL}} = 0$, while in Eq. (4.91)

$$f_{\text{LL}} = 1 + \gamma - \frac{1}{2}\lim_{x\to 1}(2-2^{2x-1})\zeta(2x-1) = 1 + \gamma + \ln 2, \quad (4.95)$$

consistent with Eq. (4.28). In the next section a more elaborate example with slightly different exponents is given.

## D. Haldane-Shastry Chain

Here we show how to compute the sum in Eq. (4.3) for [9–11]

$$\rho_{ij} = \frac{1}{4}(-1)^r \frac{\text{Si}(\pi r)}{\pi r}, \qquad r = |i - j|. \tag{4.96}$$

Before doing an exact calculation we may observe that $\rho_{ij}$ has the asymptotic expansion

$$\rho_{ij} = -\frac{1}{4\pi^2}\frac{1}{r^2} + \frac{(-1)^r}{8r} - \frac{1}{4}\sum_{m=1}^{\infty}\frac{(-1)^m(2m)!}{(\pi r)^{2m+2}}. \tag{4.97}$$

Comparison to Eq. (4.70) shows that the leading behavior of the fluctuations can be obtained by letting

$$K = \frac{1}{2}, \qquad A_1 = \frac{1}{4} \tag{4.98}$$

in Eq. (4.90), which gives

$$\pi^2 \mathcal{F}_{\text{HS}}(\ell) = \frac{1}{2}\ln\ell + f_{\text{HS}} - \frac{\pi^2}{16}\frac{(-1)^\ell}{\ell} \tag{4.99}$$

plus $O(\ell^{-2})$ corrections. The value $K = 1/2$ is expected from the fact that the Haldane-Shastry model is in the same universality class as the isotropic Heisenberg chain, for which $K = 1/2$ at the renormalization group fixed point. The difference is that the correlation functions of the Haldane-Shastry model do not have logarithmic corrections due to marginal (in the renormalization group sense) operators.

For the Haldane-Shastry model we can also show explicitly that the "volume" term $a_{\text{HS}}\ell$ corresponding to Eq. (4.92) vanishes, and moreover compute the numerical value of the constant term corresponding to Eq. (4.91). Taking the difference in exponents between Eq. (4.70) and Eq. (4.96) into account, we find

$$\frac{a_{\text{HS}}}{\pi^2} = \frac{1 - \ln 2}{4} - \frac{1}{2}\sum_{m=1}^{\infty}\frac{(-1)^{m+1}(2m-2)!}{\pi^{2m}}\zeta(2m) \tag{4.100}$$

and

$$\frac{f_{\text{HS}}}{\pi^2} = \frac{1}{8} + \frac{1+\gamma}{2\pi^2} + \frac{1}{2\pi^2}\sum_{m=1}^{\infty}\frac{(-1)^m(2m)!}{\pi^{2m}}\zeta(2m+1). \tag{4.101}$$

Both sums appear to be divergent since the Riemann zeta function $\zeta(x) \to 1$ as $x \to \infty$ so that terms grow larger in magnitude: this is not surprising since we summed an asymptotic expansion. Nevertheless, we can show that the following divergent summation methods produce the correct expressions for the coefficient of the linear term (zero) and constant term, as compared to numerical results from DMRG presented in the main text.

First, recalling that $(-1)^{m+1}(2m)!\zeta(2m)/\pi^{2m} = 2^{2m-1}B_{2m}$ where $B_n$ are the Bernoulli numbers, we can write Eq. (4.100) as

$$\frac{a_{\text{HS}}}{\pi^2} = \frac{1 - \ln 2}{4} - \frac{1}{2}\sum_{m=1}^{\infty}\frac{1}{(2m)(2m-1)}\frac{B_{2m}}{(1/2)^{2m-1}} \tag{4.102}$$

$$= \frac{1 - \ln 2}{4} - \frac{1}{2}\sum_{m=1}^{\infty}\left[\int_{1/2}^{\infty}dx\,\frac{B_{2m}}{(2m)x^{2m}}\right]. \tag{4.103}$$

Reversing the sum and integral and comparing to the series representation of the digamma function, Eq. (4.8), we see that

$$\frac{a_{\text{HS}}}{\pi^2} = \frac{1 - \ln 2}{4} - \frac{1}{2} \int_{1/2}^{\infty} dx \left[ \ln x - \frac{1}{2x} - \psi(x) \right]. \tag{4.104}$$

Using the original definition of the digamma function, $\psi(x) = d[\ln \Gamma(x)]/dx$, and $\ln \Gamma(x) \to (x - 1/2) \ln x - x + (1/2) \ln(2\pi)$ for $x \to \infty$, we arrive at the expected result

$$\frac{a_{\text{HS}}}{\pi^2} = \frac{1 - \ln 2}{4} - \frac{1}{2} \left[ \left( x - \frac{1}{2} \right) \ln x - x - \ln \Gamma(x) \right]_{1/2}^{\infty} \tag{4.105}$$

$$= \frac{1 - \ln 2}{4} - \frac{1}{2} \left( \frac{1 - \ln 2}{2} \right) \tag{4.106}$$

$$= 0. \tag{4.107}$$

Meanwhile, the series in Eq. (4.101) can be carried out using Borel summation. Let

$$y(z) = \sum_{k=1}^{\infty} y_k z^k \tag{4.108}$$

with

$$y_k = \begin{cases} i^k k! \zeta(k+1) & \text{for } k \text{ even,} \\ 0 & \text{for } k \text{ odd.} \end{cases} \tag{4.109}$$

The desired value is then

$$\frac{f_{\text{HS}}}{\pi^2} = \frac{1}{8} + \frac{1 + \gamma + y(1/\pi)}{2\pi^2}. \tag{4.110}$$

The Borel transform of $y(z)$ is

$$\mathcal{B}y(t) = \sum_{k=1}^{\infty} \frac{y_k}{k!} t^k = \sum_{k=1}^{\infty} \zeta(2k+1)(it)^{2k}. \tag{4.111}$$

There is a useful identity, essentially the Taylor expansion of the digamma function at $z = 1$, that relates the digamma function with the Riemann zeta function:

$$\sum_{k=1}^{\infty} \zeta(k+1)(-z)^k = -\gamma - \psi(1+z). \tag{4.112}$$

The even terms can be singled out by using the alternating factor $(-1)^k$:

$$\mathcal{B}y(t) = -\gamma - \frac{1}{2}[\psi(1+it) + \psi(1-it)]. \tag{4.113}$$

The Borel sum of $y(z)$ is thus

$$y(z) = \int_0^{\infty} dt\, e^{-t}[\mathcal{B}y(tz)] \tag{4.114}$$

$$= -\gamma \int_0^{\infty} dt\, e^{-t} - \frac{1}{2} \int_0^{\infty} dt\, [\psi(1+itz) + \psi(1-itz)] \tag{4.115}$$

$$= -\gamma - \frac{1}{2} \int_0^{\infty} dt\, [\psi(1+itz) + \psi(1-itz)]. \tag{4.116}$$

We can now write Eq. (4.110) as

$$\frac{f_{\text{HS}}}{\pi^2} = \frac{1}{8} + \frac{1-A}{2\pi^2} \qquad (4.117)$$

where

$$A = \frac{1}{2}\int_0^\infty dt\, e^{-t}\left[\psi\left(1 + i\frac{t}{\pi}\right) + \psi\left(1 - i\frac{t}{\pi}\right)\right]. \qquad (4.118)$$

It is worth rewriting the constant $A$ to keep all expressions real. Using the integral representation of $\psi(x)$, Eq. (4.6), and switching the order of integration, we obtain

$$A = \int_0^\infty dx \int_0^\infty dt\, e^{-t}\left(\frac{e^{-x}}{x} - \frac{e^{-x}}{1 - e^{-x}}\cos\frac{xt}{\pi}\right) \qquad (4.119)$$

$$= \int_0^\infty dx\, e^{-x}\left[\frac{1}{x} - \frac{1}{1 - e^{-x}}\frac{1}{1 + (x/\pi)^2}\right]. \qquad (4.120)$$

Numerical integration of Eq. (4.120) yields

$$\frac{f_{\text{HS}}}{\pi^2} \simeq 0.197217. \qquad (4.121)$$

## V. THE RELATION BETWEEN FACTORIAL AND ORDINARY CUMULANTS

In this section we derive the general relation between factorial and ordinary cumulants, which is not new but to our knowledge is not well-known. The ordinary cumulants, or simply "cumulants," of a probability distribution $P(X)$ are defined as the coefficients $C_n$ in the expansion of the cumulant generating function

$$\ln\chi(\lambda) = \sum_{n=1}^\infty \frac{(i\lambda)^n}{n!}C_n, \qquad (5.1)$$

where $\chi(\lambda)$ is the characteristic function

$$\chi(\lambda) = \langle e^{i\lambda X}\rangle = \sum_X e^{i\lambda X}P(X). \qquad (5.2)$$

Thus

$$C_n = (-i\partial_\lambda)^n \ln\chi(\lambda)|_{\lambda=0}. \qquad (5.3)$$

Meanwhile, the factorial cumulants of the same probability distribution are defined as the coefficients $F_n$ in the expansion of the factorial cumulant generating function

$$\ln\chi_f(z) = \sum_{n=1}^\infty \frac{z^n}{n!}F_n, \qquad (5.4)$$

where $\chi_f(z)$ is the factorial moment generating function

$$\chi_f(z) = \langle(z+1)^X\rangle = \sum_X (z+1)^X P(X). \qquad (5.5)$$

Therefore

$$F_n = \partial_z^n \ln\chi_f(z)|_{z=0}. \qquad (5.6)$$

We see that the two generating functions $\chi(\lambda)$ and $\chi_f(z)$ are related by $z + 1 = e^{i\lambda}$, i.e.,

$$\chi_f(z) = \chi(-i\ln(z+1)), \tag{5.7}$$

$$\chi(\lambda) = \chi_f(e^{i\lambda} - 1). \tag{5.8}$$

The two relations (5.7) and (5.8) can be used to express the two kinds of cumulants in terms of each other. First, to express the factorial cumulants in terms of ordinary cumulants we use Eq. (5.7) and the series representations (5.1) and (5.4) to obtain

$$\sum_{n=1}^{\infty} \frac{z^n}{n!} F_n = \sum_{k=1}^{\infty} \frac{[\ln(z+1)]^k}{k!} C_k. \tag{5.9}$$

The RHS of Eq. (5.9) can be rewritten using the identity [12]

$$\sum_{n=k}^{\infty} \frac{s(n,k)}{n!} z^n = \frac{[\ln(z+1)]^k}{k!} \tag{5.10}$$

where $s(n,k)$ are the *signed* Stirling numbers of the first kind, so that

$$\sum_{n=1}^{\infty} \frac{z^n}{n!} F_n = \sum_{k=1}^{\infty} \sum_{n=k}^{\infty} \frac{s(n,k)}{n!} z^n C_k. \tag{5.11}$$

Switching the order of the sums on the RHS gives

$$\sum_{n=1}^{\infty} \frac{z^n}{n!} F_n = \sum_{n=1}^{\infty} \frac{z^n}{n!} \left[ \sum_{k=1}^{n} s(n,k) C_k \right], \tag{5.12}$$

and by equating coefficients we obtain

$$F_n = \sum_{k=1}^{n} s(n,k) C_k. \tag{5.13}$$

Similarly, to express the ordinary cumulants in terms of factorial cumulants we first write, using Eq. (5.8),

$$\sum_{n=1}^{\infty} \frac{(i\lambda)^n}{n!} C_n = \sum_{k=1}^{\infty} \frac{(e^{i\lambda}-1)^k}{k!} F_k. \tag{5.14}$$

The identity [12]

$$\sum_{n=k}^{\infty} \frac{S(n,k)}{n!} (i\lambda)^n = \frac{(e^{i\lambda}-1)^k}{k!} \tag{5.15}$$

where $S(n,k)$ are the Stirling numbers of the second kind allows us to rewrite the RHS of Eq. (5.14) to get

$$\sum_{n=1}^{\infty} \frac{(i\lambda)^n}{n!} C_n = \sum_{k=1}^{\infty} \sum_{n=k}^{\infty} \frac{S(n,k)}{n!} (i\lambda)^n F_k, \tag{5.16}$$

and switching the order of sums gives

$$\sum_{n=1}^{\infty} \frac{(i\lambda)^n}{n!} C_n = \sum_{n=1}^{\infty} \frac{(i\lambda)^n}{n!} \left[ \sum_{k=1}^{n} S(n,k) F_k \right]. \tag{5.17}$$

Equating coefficients, we then obtain

$$C_n = \sum_{k=1}^{n} S(n,k) F_k. \qquad (5.18)$$

Eq. (5.18) can also be obtained from Eq. (5.13) by noting that $s(n,k)$ and $S(n,k)$ are inverses of each other when considered as square matrices.

For convenience, in the main text we use the notation $S_1(n,k) = |s(n,k)| = (-1)^{n-k} s(n,k)$ for the *unsigned* Stirling numbers of the first kind.

---